%% file: TOP-19-009_temp.tex
\pdfoutput=1
\documentclass[11pt,twoside,a4paper,cmspaper,final,collab]{cms-tdr}
\def\svnVersion{6602742}\def\svnDate{2021/12/15}\def\cmsCernNoTag{CERN-EP-2021-124}\def\cmsCernDate{\today}\def\cmsMessage{Published in the Journal of High Energy Physics as \href{http://dx.doi.org/10.1007/JHEP12(2021)161}{\doi{10.1007/JHEP12(2021)161}.}}
\begin{document}\cmsNoteHeader{TOP-19-009}

\newlength\cmsTabSkip\setlength{\cmsTabSkip}{1ex}
\newcommand{\cmsTable}[1]{\resizebox{\textwidth}{!}{#1}}

\newcommand{\pxmiss}{\ensuremath{p_{x}^{\text{miss}}}\xspace}
\newcommand{\pymiss}{\ensuremath{p_{y}^{\text{miss}}}\xspace}
\newcommand{\PFrelIso}{\ensuremath{I_{\text{rel}}}\xspace}
\newcommand{\mtop}{\ensuremath{m_{\PQt}}\xspace}
\newcommand{\ptL}{\ensuremath{p_{\mathrm{T}, l}}\xspace}
\newcommand{\vecptL}{\ensuremath{\vec{p}_{\mathrm{T}, l}}\xspace}
\newcommand{\pxL}{\ensuremath{p_{x, l}}\xspace}
\newcommand{\pyL}{\ensuremath{p_{y, l}}\xspace}
\newcommand{\pzL}{\ensuremath{p_{z, l}}\xspace}
\newcommand{\ptNu}{\ensuremath{p_{\mathrm{T}, \nu}}\xspace}
\newcommand{\vecptNu}{\ensuremath{\vec{p}_{\mathrm{T}, \nu}}\xspace}
\newcommand{\pxNu}{\ensuremath{p_{x, \nu}}\xspace}
\newcommand{\pyNu}{\ensuremath{p_{y, \nu}}\xspace}
\newcommand{\pzNu}{\ensuremath{p_{z, \nu}}\xspace}
\newcommand{\wjets}{\ensuremath{\PW{+}\text{jets}}\xspace}
\newcommand{\zjets}{\ensuremath{\PZ{+}\text{jets}}\xspace}
\newcommand{\gjets}{\ensuremath{\PGg{+}\text{jets}}\xspace}
\newcommand{\vjets}{\ensuremath{\PV{+}\text{jets}}\xspace}
\newcommand{\ljets}{\ensuremath{l{+}\text{jets}}\xspace}
\newcommand{\statprof}{\ensuremath{\,(\text{stat}{+}\text{prof})}\xspace}

\cmsNoteHeader{TOP-19-009}
\title{Measurement of the top quark mass using events with a single reconstructed top quark in \texorpdfstring{$\Pp\Pp$}{pp} collisions at \texorpdfstring{$\sqrt{s} = 13\TeV$}{sqrt(s) = 13 TeV}}

\date{\today}

\abstract{
A measurement of the top quark mass is performed using a data sample enriched with single top quark events produced in the $t$ channel. The study is based on proton-proton collision data, corresponding to an integrated luminosity of 35.9\fbinv, recorded at $\sqrt{s} = 13\TeV$ by the CMS experiment at the LHC in 2016. Candidate events are selected by requiring an isolated high-momentum lepton (muon or electron) and exactly two jets, of which one is identified as originating from a bottom quark. Multivariate discriminants are designed to separate the signal from the background. Optimized thresholds are placed on the discriminant outputs to obtain an event sample with high signal purity. The top quark mass is found to be $172.13^{+0.76}_{-0.77}\GeV$, where the uncertainty includes both the statistical and systematic components, reaching sub-\GeVns{{}} precision for the first time in this event topology. The masses of the top quark and antiquark are also determined separately using the lepton charge in the final state, from which the mass ratio and difference are determined to be $0.9952^{+0.0079}_{-0.0104}$ and $0.83^{+1.79}_{-1.35}\GeV$, respectively. The results are consistent with $CPT$ invariance.
}

\hypersetup{
pdfauthor={CMS Collaboration},
pdftitle={Measurement of the top quark mass using events with a single reconstructed top quark in pp collisions at sqrt(s)=13 TeV },
pdfsubject={CMS},
pdfkeywords={CMS,  top quark, single top, top quark mass, CPT invariance}}

\maketitle

\section{\label{sec:introduction} Introduction}

The mass of the top quark, \mtop, is an important parameter of the standard model (SM) of particle physics.
Its precise measurement is of profound importance, both for theory and experiment.
On the theory side, it constitutes a major input to global electroweak (EW) fits~\cite{Haller:2018nnx,Durieux:2019rbz}, used to verify the self-consistency of the SM.
It is also directly related to the stability of the EW vacuum, because among all known elementary particles it has the largest contribution in terms of radiative corrections to the Higgs boson mass~\cite{topHiggs,Degrassi:2012ry}.
From the experimental perspective, it also provides a benchmark for the identification and calibration of heavy-flavor jets~\cite{BTV-16-002} arising from bottom or charm quarks, both in resolved and boosted topologies~\cite{Aaboud:2018psm}.

The majority of LHC results on \mtop~\cite{Aaboud:2018zbu,Khachatryan:2015hba,Sirunyan:2018gqx,Sirunyan:2018mlv,Sirunyan:2019rfa,Sirunyan:2019zvx,Sirunyan:2019jyn,Sirunyan:2018goh} have been obtained with top quark pair (\ttbar) events.
Such events are predominantly produced via gluon-gluon fusion (90\%) along with a subdominant contribution from quark-antiquark annihilation (10\%). 
The most precise \mtop measurements of $172.08 \pm 0.48$ and $172.44 \pm 0.49\GeV$ are reported by the ATLAS and CMS experiments, respectively, in \ttbar events by combining results based on data recorded at $\sqrt{s}=7$ and $8\TeV$~\cite{Aaboud:2018zbu,Khachatryan:2015hba}. 
Current models for color reconnection (CR)~\cite{Christiansen:2015yqa,Argyropoulos:2014zoa} indicate that it can have a potentially large impact on the measured \mtop values~\cite{Sirunyan:2018gqx,Sirunyan:2018mlv}.
Measurements based on complementary event topologies are important to better understand such systematic effects, as well as to possibly reduce the impact of dominant sources of uncertainty via a combination of results.

\begin{figure*}[!htb]
\centering
\includegraphics[width=0.35\textwidth]{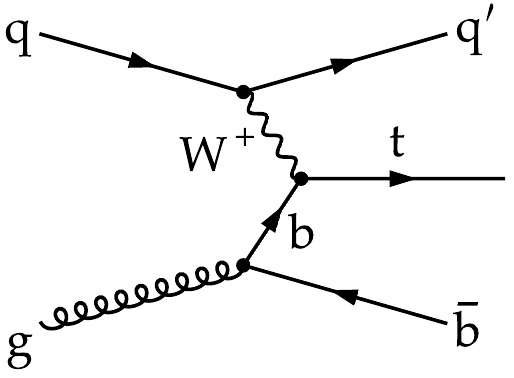}\hspace{2cm}
\includegraphics[width=0.35\textwidth]{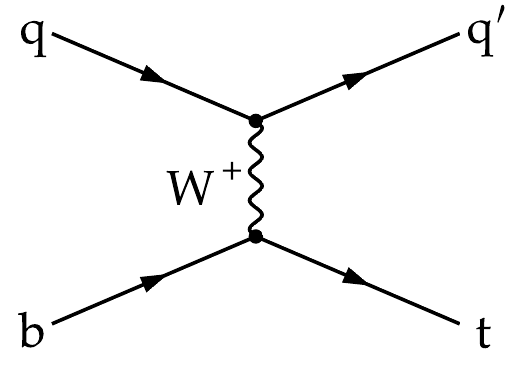}\\
\caption{\label{fig:SingleTopProcesses} Feynman diagrams of the $t$-channel single top quark production at LO corresponding to four- (left) and five-flavor (right) schemes.
At NLO in perturbative QCD, the left diagram is also part of the five-flavor scheme.}
\end{figure*}

Top quarks can be singly produced at the LHC through charged-current EW interactions via the exchange of a \PW boson. 
At leading order (LO) in the SM, single top quark production can be realized in three modes, depending on the virtuality of the \PW boson involved in the process, namely the $t$ channel (spacelike), the \PQt{}\PW channel (on-shell), and the $s$ channel (timelike).
The $t$-channel diagram shown in Fig.~\ref{fig:SingleTopProcesses} constitutes the dominant process for single top quark production in proton-proton (pp) collisions at the LHC, with a total cross section of $217^{+9}_{-8}\unit{pb}$ at $\sqrt{s}=13\TeV$ calculated at next-to-LO (NLO)~\cite{Hathor1,Hathor2} in perturbative quantum chromodynamics (QCD).
Within uncertainties, the measured cross section~\cite{Sirunyan:2018rlu} agrees with this prediction. 

The $t$-channel single top quark production occurs at a lower energy scale ($\geq 170\GeV$) compared to \ttbar ($\geq 340\GeV$), offering a partially independent event sample for measurement in a complementary region of phase space.
Furthermore, it enhances the range of available measurements with systematic sources that are partially uncorrelated from those for \ttbar events.
The $t$-channel process involves CR only between the top quark and the proton from which the \PQb quark or the gluon in the initial state arises, not to the whole event, unlike in the case of \ttbar~\cite{Argyropoulos:2014zoa}.
Such alternative measurements provide an important validation for the modeling of nonperturbative QCD processes in Monte Carlo (MC) simulations, which could be a source of large systematic effects.
The unique production mode and event topology may also shed light on the interpretation of the \mtop parameter~\cite{Butenschoen:2016lpz}, since, as opposed to \ttbar events, the single top quark process probes energy scales down to the production threshold of about 170\GeV.

The signature of $t$-channel single top quark production comprises an isolated high-momentum charged lepton, a neutrino, which results in an overall transverse momentum \pt imbalance, a light-quark jet, often produced in the forward direction, and a jet arising from the hadronization of a bottom quark (\PQb jet) from the top quark decay.
The second \PQb jet, arising in the production process via gluon splitting, as shown in Fig.~\ref{fig:SingleTopProcesses} (left), has a softer \pt spectrum and a broader pseudorapidity $\eta$ distribution compared to the \PQb jet originating from the top quark, and thus frequently escapes detection.
Hence, candidate events are required to contain one muon or electron and exactly two jets, of which one is \PQb tagged, in the final state.

In this paper, a measurement of \mtop is reported based on data recorded at $\sqrt{s} = 13\TeV$ by the CMS experiment in 2016 and corresponding to an integrated luminosity of 35.9\fbinv~\cite{CMS:2021xjt}.
The study uses a sample enriched in $t$-channel single top quark events selected via a multivariate analysis (MVA) technique, and supersedes the earlier CMS measurement at $8\TeV$~\cite{CMS_topMass_SingleTop_8TeV}.
The masses of the top quark and antiquark are also measured separately using events with positively and negatively charged leptons in the final state, respectively.
The mass ratio and difference from these measurements are used to test $CPT$  invariance. 
Violation of $CPT$  symmetry would imply a nonlocal field theory~\cite{Greenberg:2002uu}, signaling physics beyond the SM.
Tabulated results are provided in HEPData~\cite{hepdata}.

The paper is organized as follows.
The CMS detector is briefly discussed in Section~\ref{sec:detector}.
Section~\ref{sec:simulation} describes the simulation samples used to model signal and background events.
Section~\ref{sec:selection} provides an overview of the reconstruction and identification of physics objects, as well as of the selection of candidate events.
A method relying on sideband (SB) data used to estimate the QCD multijet background is described in Section~\ref{sec:bkgmodel}.
Section~\ref{sec:mva} discusses the MVA technique designed to distinguish the $t$-channel single top quark signal from the \ttbar, EW, and QCD backgrounds.
The maximum-likelihood (ML) fit used to extract the value of \mtop from its reconstructed distribution is explained in Section~\ref{sec:topmass}.
In Section~\ref{sec:syst} we describe various systematic sources affecting the measurement and their individual contributions.
Results and overall impact are explained in Section~\ref{sec:result}, and Section~\ref{sec:summary} summarizes the main findings.

\section{\label{sec:detector} The CMS detector}

The central feature of the CMS apparatus is a superconducting solenoid of 6\unit{m} diameter, providing an axial magnetic field of 3.8\unit{T}.
Within the solenoid volume lie a silicon pixel and microstrip tracker, a lead tungstate crystal electromagnetic calorimeter (ECAL), and a brass and plastic scintillator hadron calorimeter, each composed of a barrel and two endcap sections.
Forward calorimeters, based on steel absorbers with quartz fibers, extend the fiducial coverage provided by the barrel and endcap detectors up to $\abs{\eta} = 5$.
Muons are measured in the range $\abs{\eta} < 2.4$, with detection planes made using three technologies: drift tubes, cathode strip chambers, and resistive plate chambers, embedded in the steel flux-return yoke outside the solenoid.
A detailed description of the CMS detector, together with a definition of the coordinate system used and relevant kinematic variables, can be found in Ref.~\cite{Chatrchyan:2008zzk}.

Events of interest are selected with a two-tiered trigger system~\cite{Khachatryan:2016bia}.
The first level~\cite{Sirunyan_2020}, composed of custom hardware processors, uses information from the calorimeters and muon detectors to select events at a rate of 100\unit{kHz}.
The second level, known as the high-level trigger, comprises a farm of processors running a version of the full event reconstruction software optimized for fast processing at a rate of about 1\unit{kHz}. 

\section{\label{sec:simulation} Simulation of events}

We simulate signal and background events to NLO QCD accuracy using either the \POWHEG~\cite{Nason:2004rx,Frixione:2007vw,Alioli:2010xd} or \MGvATNLO 2.2.2~\cite{Alwall:2014hca} generator.
The signal process~\cite{Alioli:2009je} is simulated with \POWHEG 2.0 in the four-flavor scheme (4FS), where \PQb quarks are produced via gluon splitting as shown in Fig.~\ref{fig:SingleTopProcesses} (left).
This scheme is expected to yield a more accurate description of the kinematic distributions of $t$-channel events than the five-flavor scheme (5FS)~\cite{Frederix:2012dh,Aaboud:2017pdi,Sirunyan:2019hqb} indicated in Fig.~\ref{fig:SingleTopProcesses} (right).
For the normalization of signal samples, we employ the 5FS predictions calculated using \textsc{hathor} 2.1~\cite{Hathor1,Hathor2} at NLO in perturbative QCD.
This is because with the inclusion of both gluon- and \PQb-quark-initiated diagrams, the 5FS cross section calculation yields a more accurate value than that of 4FS, which is based on the gluon-initiated diagram only.

The \ttbar background~\cite{Alioli:2011as} is simulated using \POWHEG 2.0 and is normalized to the cross section calculated with \textsc{top++} 2.0~\cite{Czakon:2011xx} at next-to-NLO (NNLO) in perturbative QCD, including soft gluon resummation at next-to-next-to-leading-log (NNLL) accuracy. 
We use \POWHEG 1.0 to simulate the production of single top quarks in association with \PW bosons (\PQt{}\PW) in the 5FS~\cite{Re:2010bp}, and normalize the events to a prediction providing approximate NNLO accuracy~\cite{Kidonakis:2010ux,Kidonakis:2013zqa}.
The $s$-channel contribution is modeled using \MGvATNLO 2.2.2 in the 4FS with up to one additional parton and with the FxFx merging scheme~\cite{Frederix:2012ps} and normalized to a cross section calculated at NLO QCD accuracy in 5FS by \textsc{hathor} 2.1~\cite{Hathor1,Hathor2}.
While the \mtop value is set to 172.5\GeV in the nominal simulation samples of signal and top quark background events, a number of such samples generated with alternate \mtop hypotheses ranging between 169.5 and 178.5\GeV are considered for the purpose of mass calibration (Section~\ref{sec:topmass}).
In all these samples the top quark width is set to its nominal value of 1.31\GeV.

Events with \PW and \PZ bosons produced in association with jets (referred to as \vjets) are simulated at NLO accuracy using \MGvATNLO 2.2.2 and the FxFx merging scheme.
Predictions calculated with \FEWZ 3.1~\cite{Gavin:2010az,Gavin:2012sy,Li:2012wna} are employed for the normalization of these two processes. 
The \wjets events are simulated with zero, one, and two additional partons exclusively, in order to retain a sufficient number of events surviving the selection criteria, while the \zjets events are generated with up to two additional partons inclusively.
Contributions from \PW{}\PW, \PW{}\PZ, and \PZ{}\PZ (collectively referred to as \PV{}\PV) processes are simulated at NLO with \MGvATNLO 2.2.2 with the FxFx merging scheme. 

For all samples, \PYTHIA 8.212~\cite{Sjostrand:2014zea} is used to simulate parton showering and hadronization.
We model the underlying event (UE) activities with the tune CUETP8M1~\cite{Khachatryan:2015pea} for all samples except for \ttbar, where we use the tune \textsc{CUETP8M2T4}~\cite{CMS-PAS-TOP-16-021}, as it provides a more accurate description of the kinematic distributions of the top quark pair and jets in \ttbar events.
The parton distribution functions (PDFs) predicted by NNPDF3.0 NLO~\cite{Ball:2014uwa} are used in all simulations.
The cross sections of simulated signal and background processes are listed in Table~\ref{tab:sample_Xsec}. 

All generated events undergo a full simulation of the detector response using a model of the CMS detector implemented in \GEANTfour~\cite{Agostinelli:2002hh}.
Additional \Pp{}\Pp interactions within the same or nearby bunch crossings (pileup) included in the simulation are reweighted such that the corresponding multiplicity distribution agrees with that observed in data.

\begin{table*}[!htb]
\centering
\caption{\label{tab:sample_Xsec}Summary of signal and background simulations discussed in Section~\ref{sec:simulation} with their respective cross sections.}
\cmsTable{
\begin{tabular}{c|c|c|c}
Process & Simulation & Cross section (\unit{pb}) & Accuracy\\
\hline
Single top $t$-channel, \PQt{} & \POWHEG 2.0 4FS & 136 & NLO, estimated using \textsc{hathor} 2.1~\cite{Hathor1, Hathor2} in 5FS \\
Single top $t$-channel, \PAQt{} & \POWHEG 2.0 4FS & 81 & NLO, estimated using \textsc{hathor} 2.1~\cite{Hathor1, Hathor2} in 5FS \\
\hline
Single top $s$-channel, $\PQt{}\ +\ \PAQt{}$ & \MGvATNLO 2.2.2 4FS & 10 & NLO, estimated using \textsc{hathor} 2.1~\cite{Hathor1, Hathor2} in 5FS \\
\PQt{}\PW{} (\PAQt{}\PW{}) & \POWHEG 1.0 5FS & 36 (36) & Approximate NNLO~\cite{Kidonakis:2012db} \\
\ttbar & \POWHEG 2.0 & 832 & NNLO + NNLL, estimated using \textsc{top++} 2.0 ~\cite{Czakon:2011xx} \\
\hline
 \PW{}($\to l\nu$)+ 0 jet & \MGvATNLO 2.2.2 & 50132 & NNLO, estimated using \FEWZ 3.1~\cite{Gavin:2010az,Gavin:2012sy,Li:2012wna} \\
 \PW{}($\to l\nu$)+ 1 jet & \MGvATNLO 2.2.2 & 8426 & NNLO, estimated using \FEWZ 3.1~\cite{Gavin:2010az,Gavin:2012sy,Li:2012wna} \\
 \PW{}($\to l\nu$)+ 2 jets & \MGvATNLO 2.2.2 & 3173 & NNLO, estimated using \FEWZ 3.1~\cite{Gavin:2010az,Gavin:2012sy,Li:2012wna} \\
 \PZ{}($\to ll$)+ jets, ($m_{ll} > 50\GeV$) & \MGvATNLO 2.2.2 & 5765 & NNLO, estimated using \FEWZ 3.1~\cite{Gavin:2010az,Gavin:2012sy,Li:2012wna}\\
 \PW{}\PW{}$\to l\nu$\PQq{}\PQq{} & \MGvATNLO 2.2.2 & 46 & NLO \\
 \PW{}\PZ{}$\to l\nu$\PQq{}\PQq{} & \MGvATNLO 2.2.2 & 11 & NLO \\
 \PZ{}\PZ{}$\to ll$\PQq{}\PQq{} & \MGvATNLO 2.2.2 & 3 & NLO \\  
\end{tabular}}
\end{table*}

\section{\label{sec:selection} Event reconstruction}

\subsection{\label{ssec:evtSel} Event selection}

Events in the muon final state are selected using a trigger that requires at least one isolated muon with $\pt > 24\GeV$ and $\abs{\eta} < 2.4$.
For the electron final state, the trigger requires the presence of at least one isolated electron with $\pt > 32\GeV$ and $\abs{\eta} < 2.1$.

Events with at least one reconstructed \Pp{}\Pp interaction vertex are retained for further analysis.
The vertex must be reconstructed from at least four tracks that have a longitudinal distance $\abs{d_{z}}<24\cm$ and a radial distance $\abs{d_{xy}}< 2 \unit{cm}$ from the nominal interaction point.
If multiple vertices are found in an event, the one with the largest value of summed $\pt^2$ of physics objects is taken as the primary \Pp{}\Pp interaction vertex.
The objects are the jets, clustered with the tracks assigned to the vertex as inputs, and the associated missing transverse momentum \ptvecmiss, taken as the negative of the \ptvec sum of those jets.

The particle-flow algorithm~\cite{PFAlgo}, which combines information from various subdetectors, is used to reconstruct the individual particles. 
Muon candidates must have at least one hit in the muon detector and a minimum of five hits in the silicon microstrip tracker.
They are then reconstructed by a global fit to the combined information from the tracker and muon detector. Selected muons must have $\pt > 26\GeV$ and $\abs{\eta} < 2.4$.
Electron candidates are reconstructed~\cite{reco_el} from good quality tracks in the tracker, matched to clusters in the ECAL.
They are identified by applying dedicated selection criteria on nine variables related to tracking and shower shape.
Electrons are required to pass the tight identification criteria~\cite{Sirunyan:2020ycc} corresponding to an average efficiency of approximately 70\% and have $\pt > 35\GeV$ and $\abs{\eta} < 2.1$, while those falling into the gap between the ECAL barrel and endcap regions ($1.44 < \abs{\eta} < 1.57$) are rejected.
The relative isolation (\PFrelIso) for a muon (electron) candidate is calculated by summing the transverse energy deposited by photons and charged and neutral hadrons within a cone of size 
$\Delta R=\sqrt{\smash[b]{(\Delta\eta)^{2}+(\Delta\phi)^{2}}} < 0.4$ $(0.3)$ around its direction, corrected for contributions from pileup~\cite{Sirunyan:2018}, divided by its \pt.
The transverse energy is defined as $E\sin\theta$, where $E$ is the energy and $\theta$ is the polar angle of the energy deposit.
The muon and electron candidates are required to pass the criterion $\PFrelIso < 0.06$.

Events containing additional muons (electrons) with $\pt > 10$ (15)\GeV and $\abs{\eta} < 2.4$ (2.5) are rejected.
In such cases, the criteria on lepton isolation are relaxed to $\PFrelIso < 0.2$ for muons and $\PFrelIso < 0.18$ (0.16) for electrons in the barrel (endcap) ECAL.
We apply \pt- and $\eta$-dependent scale factors in simulation for both selected and vetoed leptons to correct for observed differences in the lepton reconstruction efficiencies with data.

Jets are reconstructed using the anti-\kt clustering algorithm~\cite{anti-kt} with a distance parameter of 0.4, as implemented in the \FASTJET package~\cite{Cacciari:2011ma}.
The effect of additional tracks and calorimetric energy deposits from pileup on the jet momentum is mitigated by discarding tracks identified to be originating from pileup vertices, as well as by applying an offset correction to account for residual neutral pileup contributions~\cite{CMS-PAS-JME-16-003,Cacciari:2007fd}.
Loose identification criteria~\cite{CMS-PAS-JME-16-003} are applied to suppress jets arising from spurious sources, such as electronics noise in the calorimeters. 
Energy corrections are derived from simulation to bring the measured average response of jets to that of particle-level jets.
In-situ measurements of the momentum balance in dijet, \gjets, \zjets, and QCD multijet events are used to account for any residual differences in the jet energy scale (JES) between data and simulation~\cite{Khachatryan:2016kdb}.
In this analysis, jets are required to have $\pt > 40\GeV$ and $\abs{\eta} < 4.7$.

A combined MVA tagging algorithm~\cite{BTV-16-002} is used to identify \PQb jets, which are required to have $\pt > 40\GeV$ and $\abs{\eta} < 2.4$.
The efficiency to correctly identify \PQb jets is about 55\% at the chosen working point, while the misidentification probability is 0.1\% for light-quark or gluon jets and 6\% for charm jets.
Simulated events are corrected using dedicated scale factors that account for the differences in the \PQb tagging efficiencies and misidentification probabilities when compared to data.
Candidate events must contain exactly two jets.
For the signal sample selection, one of these jets is required to be \PQb{}-tagged and the other must not satisfy the \PQb tagging criterion (referred to as the ``untagged" jet in the following discussion).

To suppress the background from QCD multijet processes, we require the transverse mass of the charged lepton plus neutrino system, defined as
\begin{linenomath}
\begin{equation}
\label{eq:mtw}
\mT = \sqrt{(\ptL + \ptmiss)^{2} - (\pxL + \pxmiss)^{2} - (\pyL + \pymiss)^{2}},
\end{equation}
\end{linenomath} 
to exceed $50\GeV$.
Here, \ptmiss is the magnitude of \ptvecmiss, which is the negative of the \ptvec sum of all reconstructed particle-flow objects in an event.
The energy scale corrections applied to jets are propagated to \ptvecmiss~\cite{Sirunyan:2019kia}.
The variables \pxmiss and \pymiss denote the $x$ and $y$ components of \ptvecmiss, respectively.
The symbol \ptL represents the magnitude of \ptvec of the charged lepton and \pxL (\pyL) is its $x$ ($y$) component.

\subsection{\label{ssec:evtCat} Event categories}

\begin{figure*}[!htb]
\centering
\includegraphics[width=0.45\textwidth]{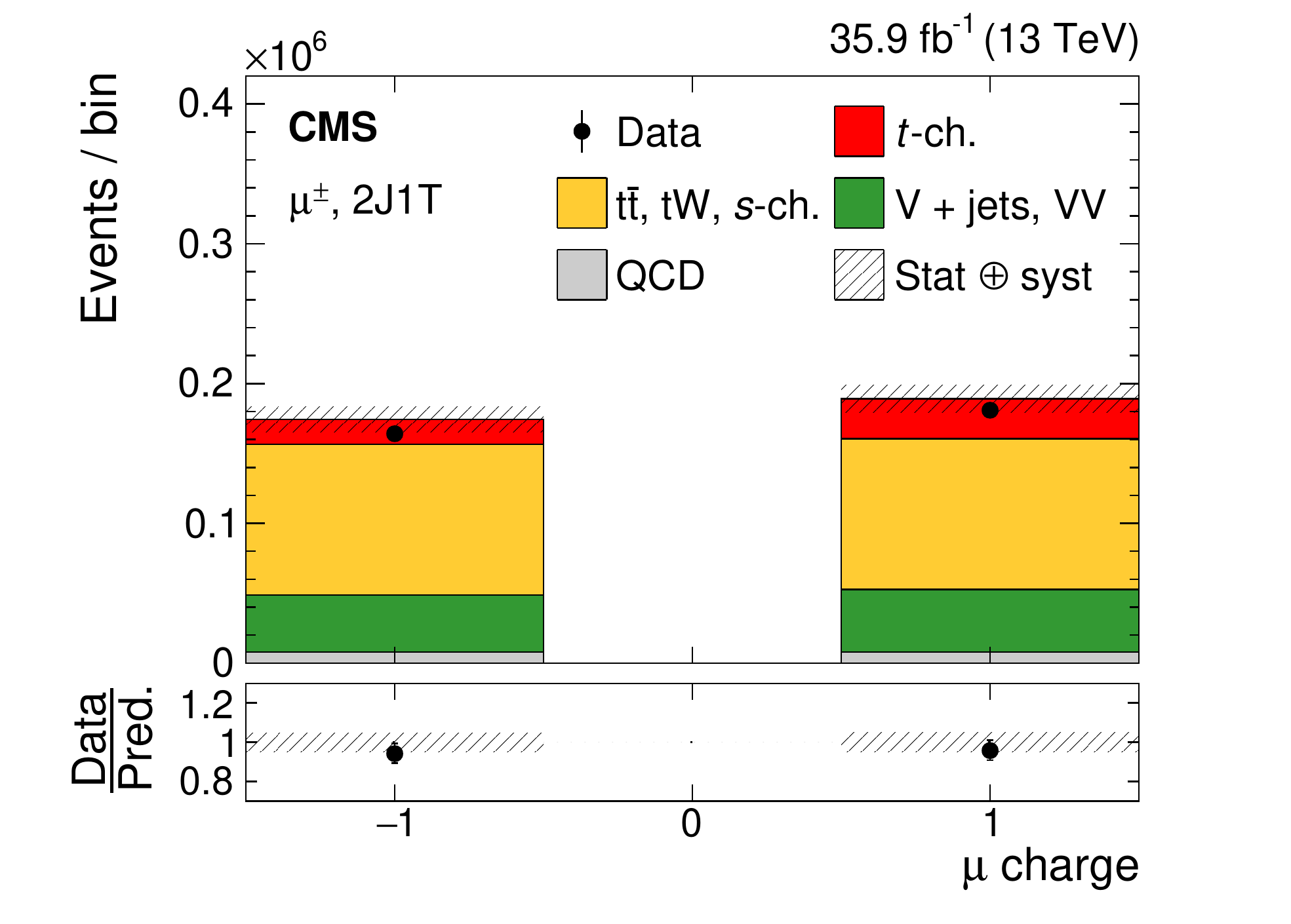}
\includegraphics[width=0.45\textwidth]{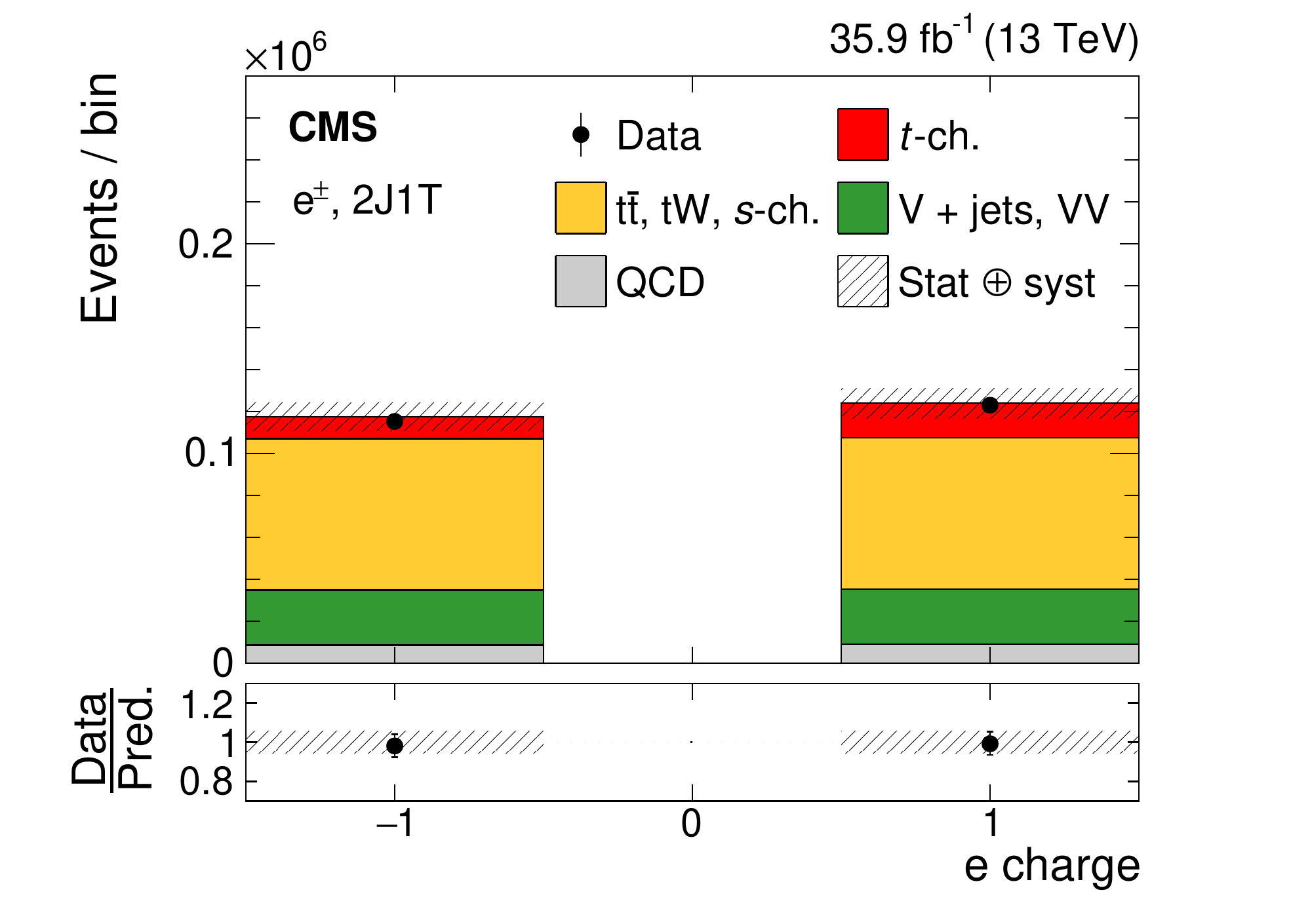}\\
\caption{\label{fig:lepCharge} Event yields corresponding to positively and negatively charged muons (left) and electrons (right) in the final states obtained for the 2J1T category from data (points) and from simulated signal and background processes (colored histograms).
The vertical bars on the points show the statistical uncertainty in the data, and the cross-hatched bands denote the quadrature sum of the statistical and systematic uncertainties in the predictions.
The lower panels show the ratio of the data to the predictions.}
\end{figure*}

The selected events are divided into two categories, depending on the number of jets, $n$, and number of \PQb{}-tagged jets, $m$, (labeled $n$J$m$T).
The 2J1T category has the largest contribution from signal events and is referred to as the signal category.  
Besides this, we use the 2J0T category, where both the jets do not satisfy the \PQb tagging criterion, to validate the estimation of the QCD multijet background contribution in data.
A similar approach was used in earlier CMS measurements of the inclusive and differential $t$-channel single top quark cross sections~\cite{Sirunyan:2018rlu,Sirunyan:2019hqb}.
The event yields in the 2J1T category after applying all selection criteria are shown in Fig.~\ref{fig:lepCharge} for the muon (left) and electron (right) final states.
The yields are shown separately for events with positively and negatively charged leptons. 
The contribution from the QCD multijet background is determined from data as described in Section~\ref{sec:bkgmodel}.
For other processes, the event yields are obtained from simulation.

\subsection{\label{ssec:top4v} Reconstruction of the top quark}

The four-momentum of the top quark (and hence its mass) is reconstructed from the momenta of its decay products: the charged lepton, the reconstructed neutrino, and the \PQb-tagged jet.
The momenta of the lepton and \PQb-tagged jet are measured, while the transverse momentum of the neutrino \ptNu is inferred from \ptvecmiss.
Assuming energy-momentum conservation at the $\PW\to l\nu$ vertex and setting the \PW boson mass $m_{\PW}$ to 80.4\GeV~\cite{Zyla:2020zbs}, the longitudinal momentum of the neutrino \pzNu can be calculated from the following constraint:
\begin{linenomath}
\begin{equation}
\label{eq:mW_quad}
m_{\PW}^{2} = \left(E_{l}+\sqrt{(\ptmiss)^{2}+\pzNu^{2}}\right)^{2} - (\vecptL+\ptvecmiss)^{2} - (\pzL+\pzNu)^{2},
\end{equation}
\end{linenomath} 
where $E_{l}$ is the lepton energy and $\pzL$ is the $z$ component of its momentum.
The above equation generally leads to two possible solutions for \pzNu, given by
\begin{linenomath}
\begin{equation}
\label{eq:pz_Nu}
\pzNu = \frac{\Lambda \pzL}{\ptL^{2}} \pm \frac{1}{\ptL^{2}} \sqrt{\Lambda^{2} \pzL^{2} - \ptL^{2} [E_{l}^{2} (\ptmiss)^{2} - \Lambda^{2} ] }, 
\end{equation}
\end{linenomath} 
where, $\Lambda = (m_{\PW}^{2}/2) + \vecptL \cdot \ptvecmiss$.
If both solutions have real values, the one with the smaller magnitude is retained~\cite{Single-top-CDF, Single-top-D0}.
This choice yields a better consistency between the inferred and true values of \pzNu in simulation.
In the case of complex solutions, the \ptNu values are modified so that the radical in Eq.~(\ref{eq:pz_Nu}) becomes zero, while still fulfilling Eq.~(\ref{eq:mW_quad}).
Setting the radical equal to zero, we get two pairs of possible solutions for \pxNu and \pyNu.
Out of the two pairs, the one resulting in a $\vecptNu$ with a lower $\abs{\Delta\phi}$ value with respect to \ptvecmiss in the event is chosen.

The reconstructed \mtop distribution after event selection for the signal and background is shown in Section~\ref{sec:mva}.
Potential inadequacy in the determination of \pzNu leads to a softer reconstructed \pzNu spectrum compared to the true spectrum in simulation.
This in turn leads to a mismatch of the reconstructed \mtop with the true value used in simulation.
A calibration is applied to the reconstructed \mtop value in order to compensate for this difference, and the related uncertainty is considered as a source of systematic uncertainty, as discussed in Sections~\ref{sec:topmass} and \ref{sec:syst}, respectively.

\section{\label{sec:bkgmodel} Estimation of the QCD multijet background}

The QCD multijet production has a large cross section in \Pp{}\Pp collisions but a tiny acceptance in the phase space used in this analysis.
Therefore, a very large sample of simulated QCD multijet events would be needed in order to retain a sufficient event yield surviving our selection criteria to ensure a reliable description of this background.
In the absence of simulated event samples of the required size, an alternative approach is followed by defining an SB in data enriched in QCD multijet events.
The SB is obtained by requiring the selected muon to have $0.2 < \PFrelIso < 0.5$ and the selected electron to fail the tight identification criteria.
The underlying assumption here is that the description of kinematic variables for QCD multijet events in the SB is similar to that in the signal region.
We have verified this assumption using simulated samples.
Shapes for QCD multijet events are derived by subtracting the total non-QCD contribution from data in this SB.
As such, the SB data contain 93 (70)\% QCD multijet events for the muon (electron) final state.
The QCD multijet contribution in the signal region is estimated by means of a binned ML fit to the \mT distribution with two components: QCD and non-QCD.
As the QCD multijet background has a larger contribution in the 2J0T category, this category is used to validate the above method.
The procedure is then applied in the signal-enriched 2J1T category where the background estimation is performed separately for positively and negatively charged leptons, as well as inclusive of the lepton charge in the final state.

\begin{figure*}[!htb]
\centering
\includegraphics[width=0.45\textwidth]{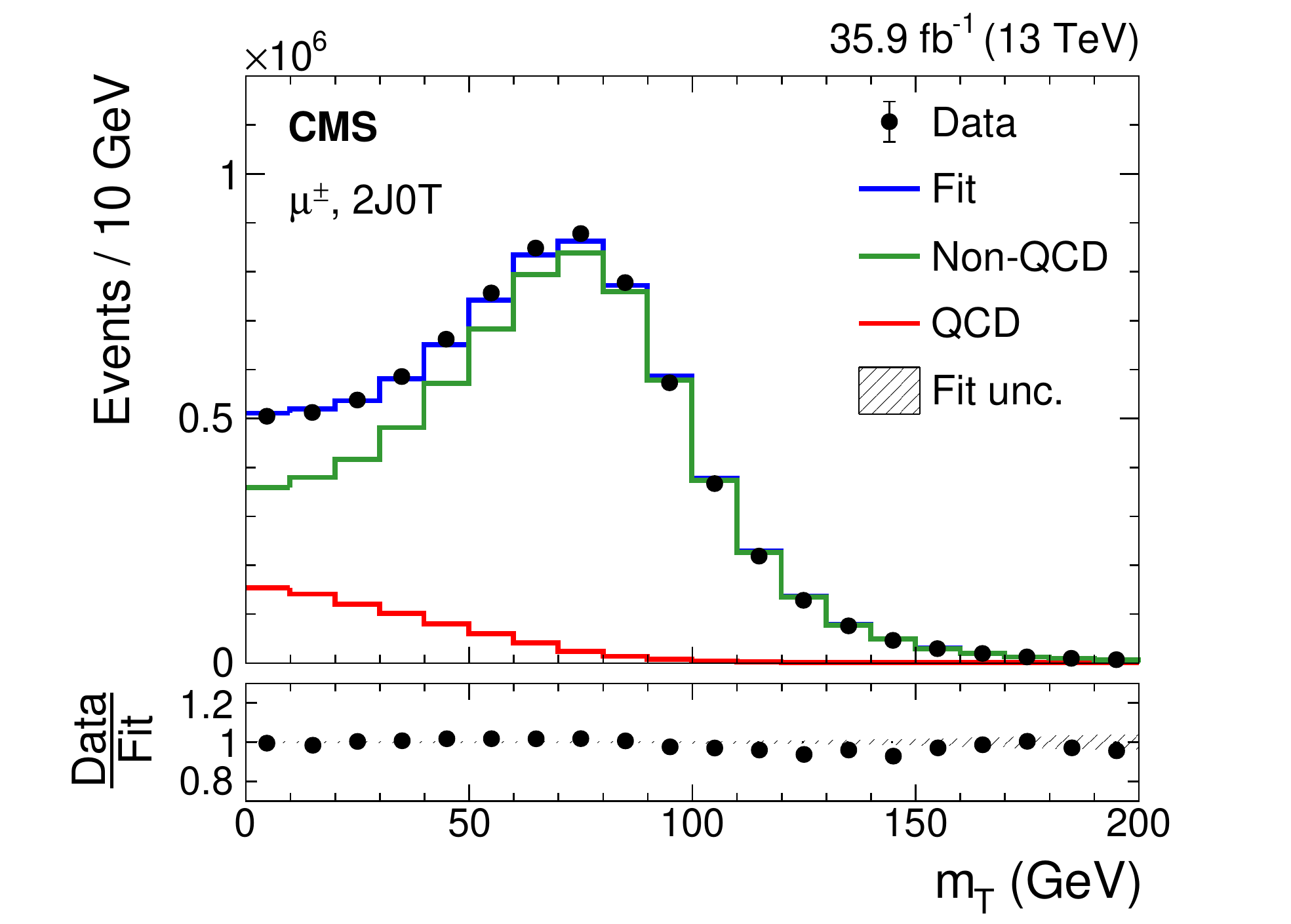}
\includegraphics[width=0.45\textwidth]{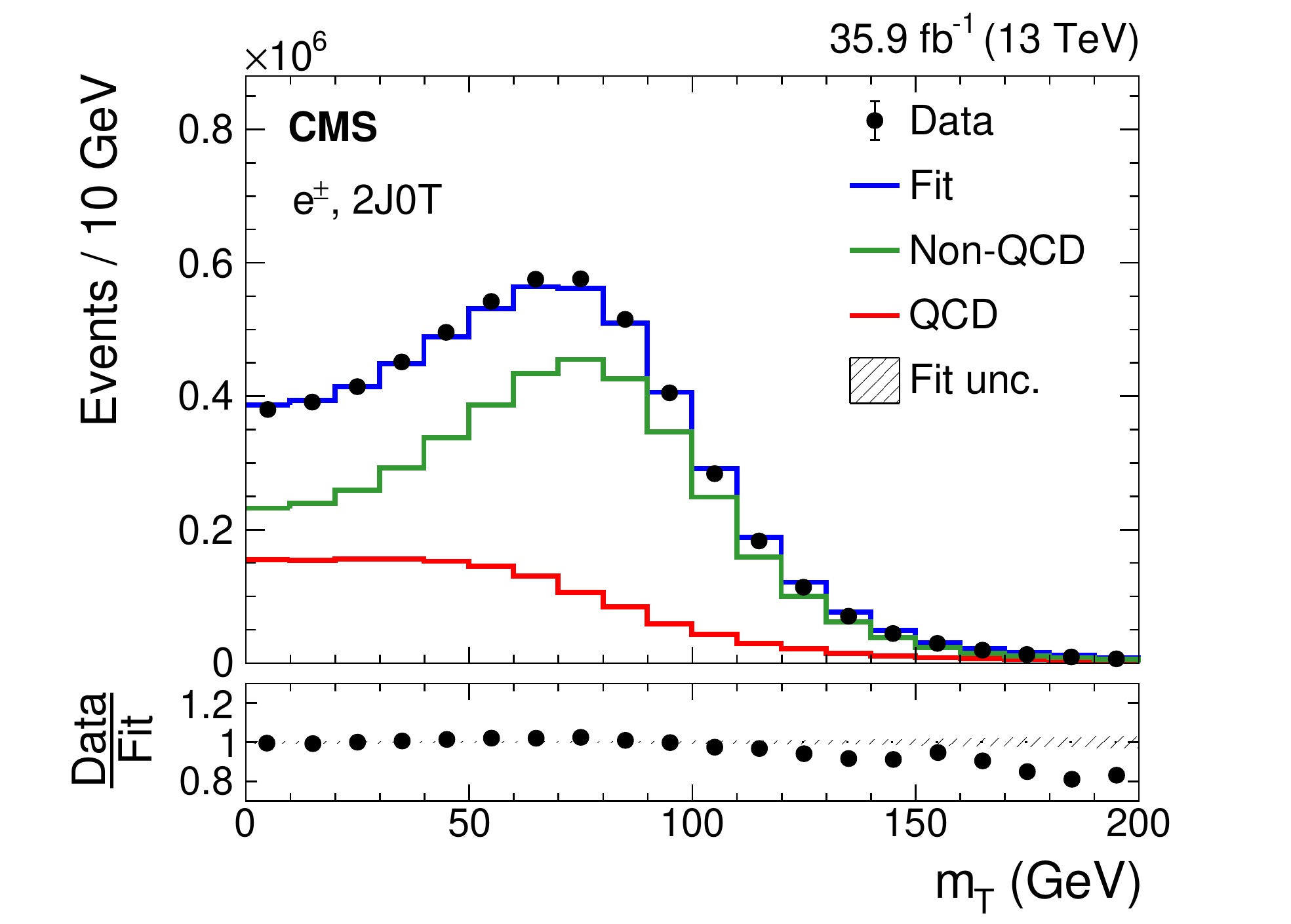}\\
\includegraphics[width=0.45\textwidth]{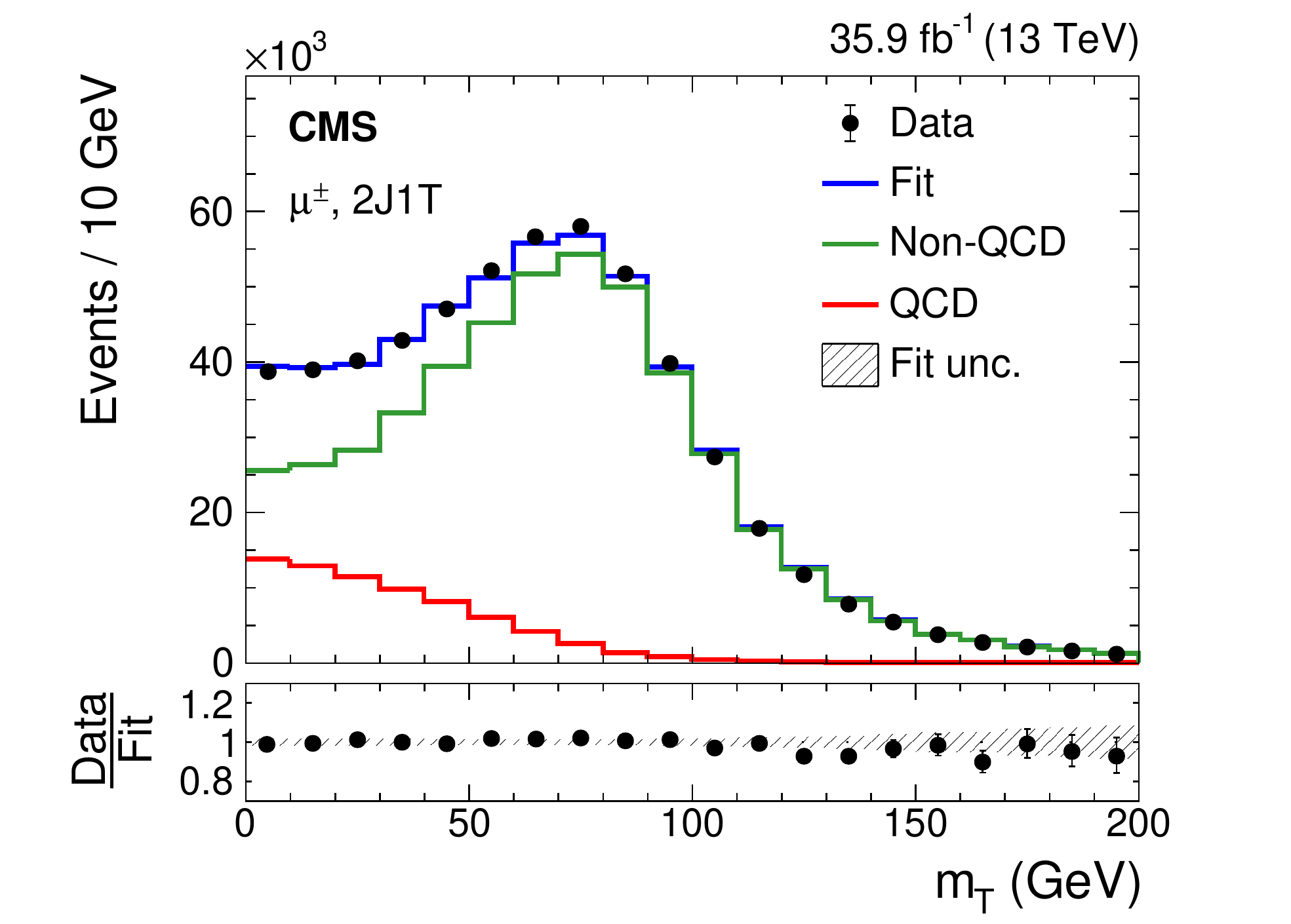}
\includegraphics[width=0.45\textwidth]{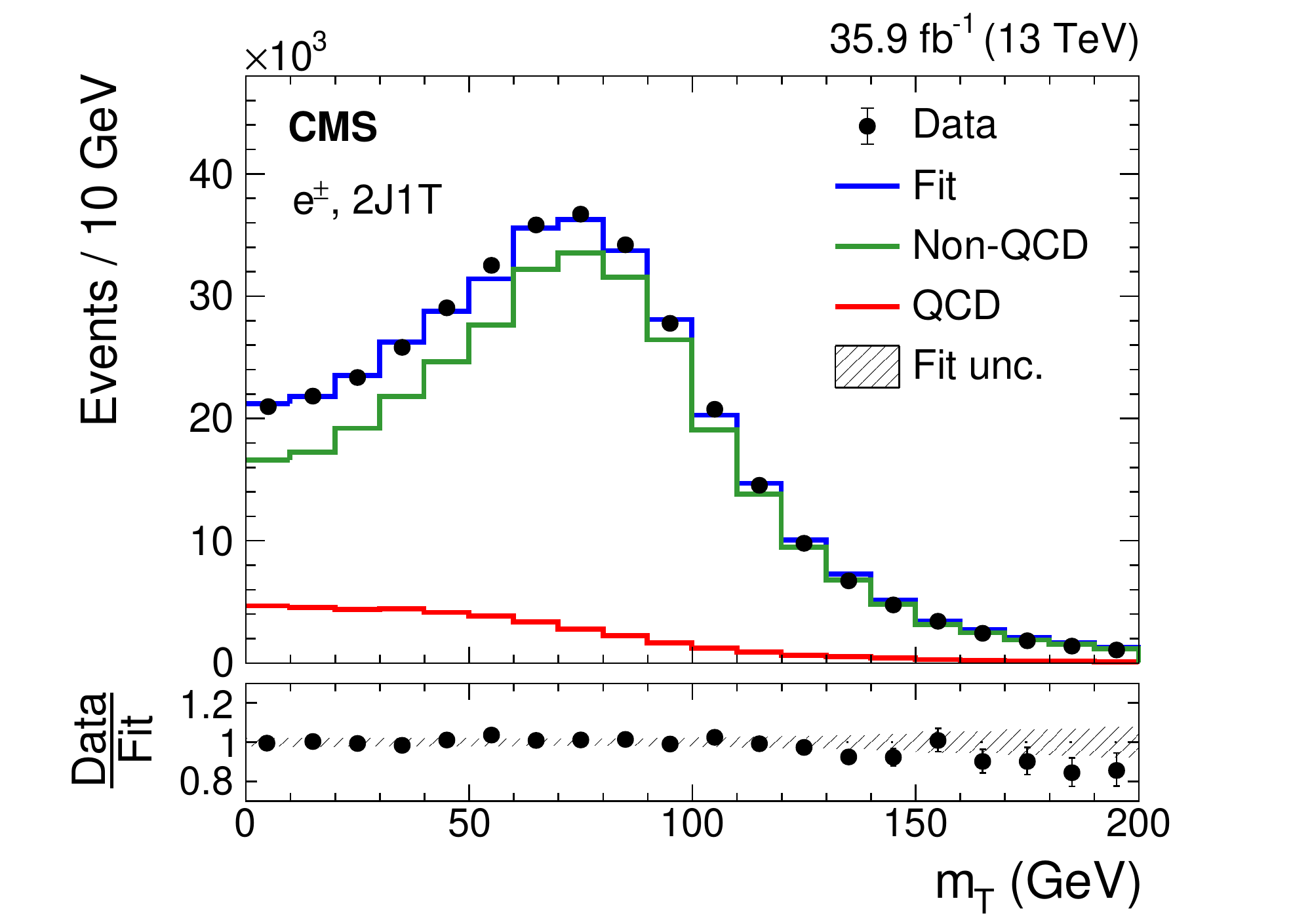}\\
\caption{\label{fig:mtwfit} Projections of the fit results onto the \mT distributions for the muon (left) and electron (right) final states in the 2J0T (upper) and 2J1T (lower) categories for the data (points) and the various components of the fit (colored lines).
The lower panels show the ratio of the data to the fit predictions.
The bands represent the uncertainty in the \mT distribution predicted by the fit.}
\end{figure*}

Figure~\ref{fig:mtwfit} shows the projections of the fit results onto the \mT distributions in the 2J0T (upper) and 2J1T (lower) categories for the muon (left) and electron (right) final states.
To account for possible differences between QCD multijet \mT shapes obtained from the SB and that in the signal region, a separate 50\% systematic uncertainty is assigned to that background rate and shape (see Section~\ref{sec:syst-exp}).
This propagated uncertainty is twice the maximum difference between the data and the prediction found in the tail of the \mT distributions.

\section{\label{sec:mva} Multivariate analysis}

A number of variables are combined into an MVA discriminant to separate $t$-channel single top quark events as the signal from the rest as the background.
The latter is the sum total of top quark (\ttbar, \PQt{}\PW, and $s$-channel), EW (\vjets and \PV{}\PV), and QCD multijet events selected in the 2J1T category.
All background contributions are obtained with simulation samples except for the QCD multijet where the SB data are used, as described in Section~\ref{sec:bkgmodel}.
We develop two boosted decision trees (BDTs) using the \textsc{tmva} package~\cite{Hocker:2007ht} with the variables listed in Table~\ref{tab:mvaInputs} as inputs for the muon and electron final states in the 2J1T category, respectively.

\begin{figure*}
\centering
\includegraphics[width=0.45\textwidth]{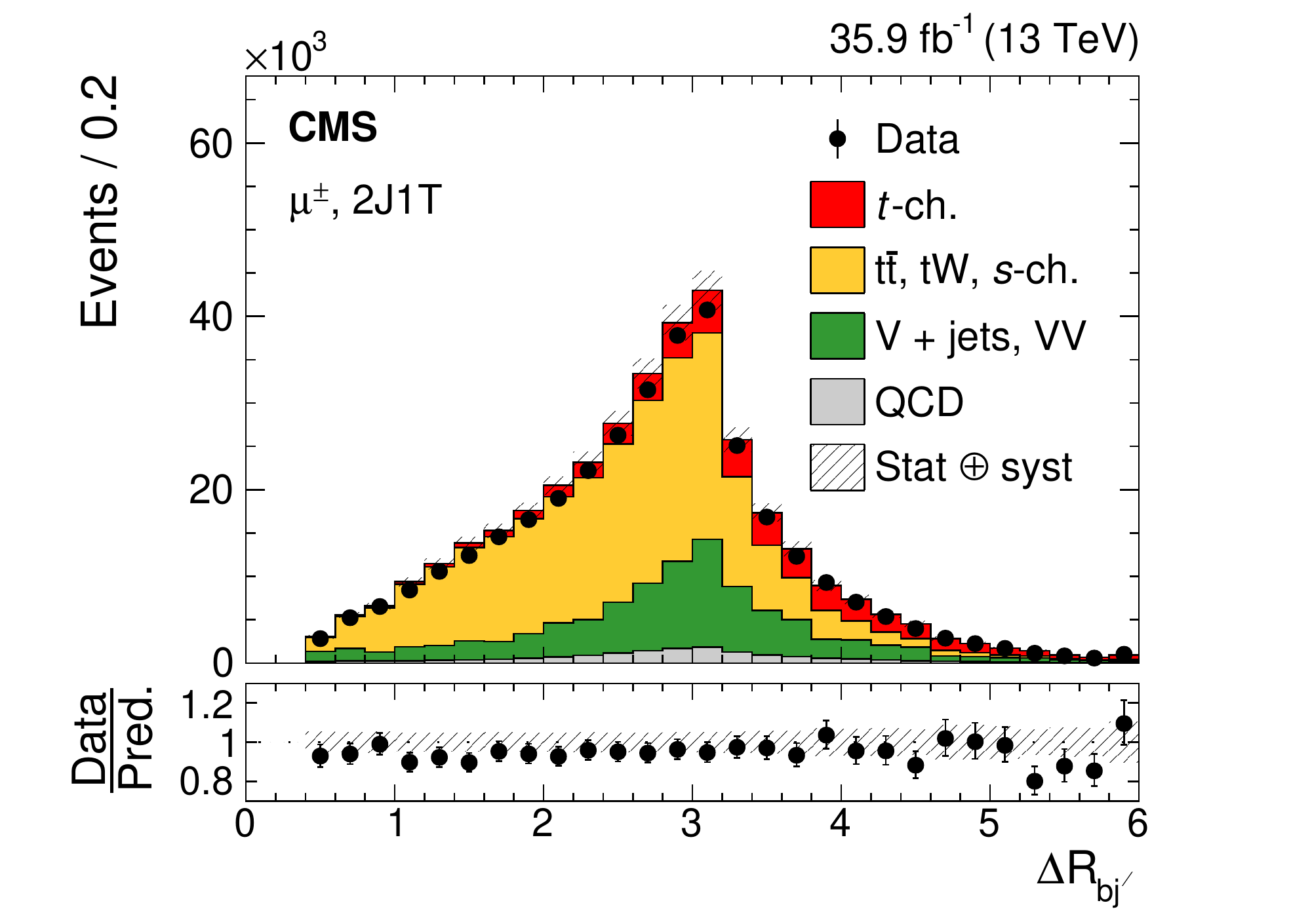}
\includegraphics[width=0.45\textwidth]{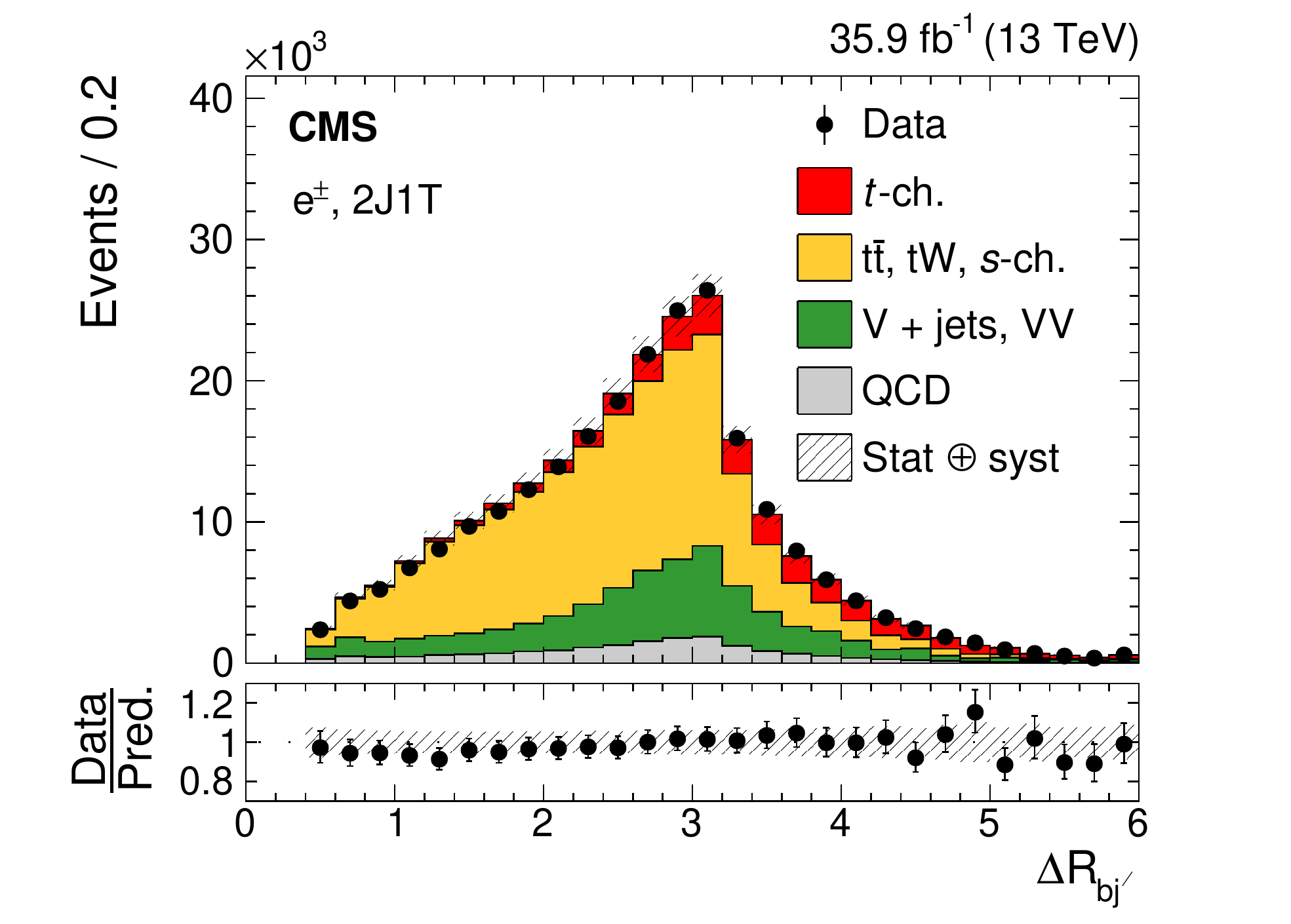}\\
\includegraphics[width=0.45\textwidth]{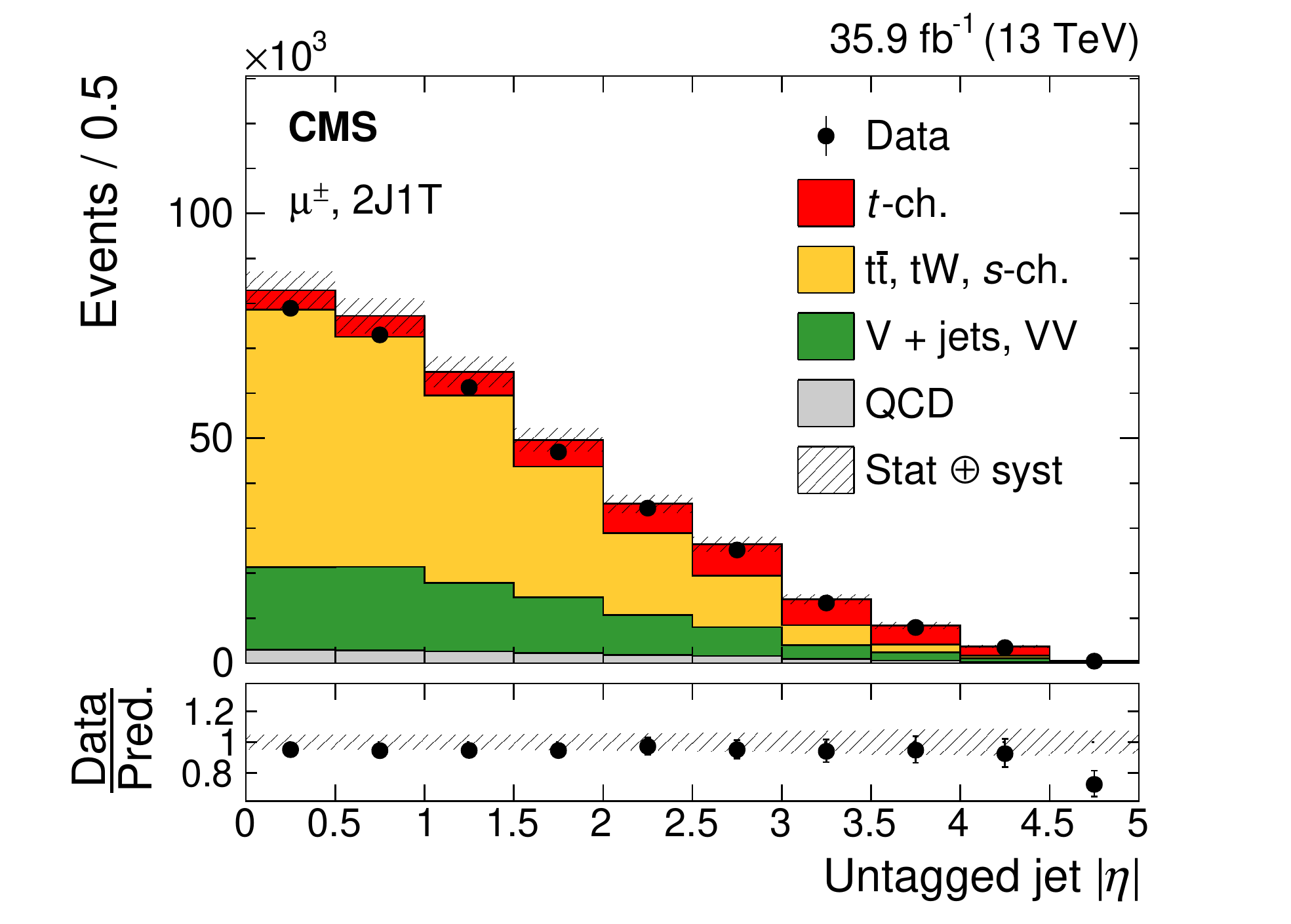}
\includegraphics[width=0.45\textwidth]{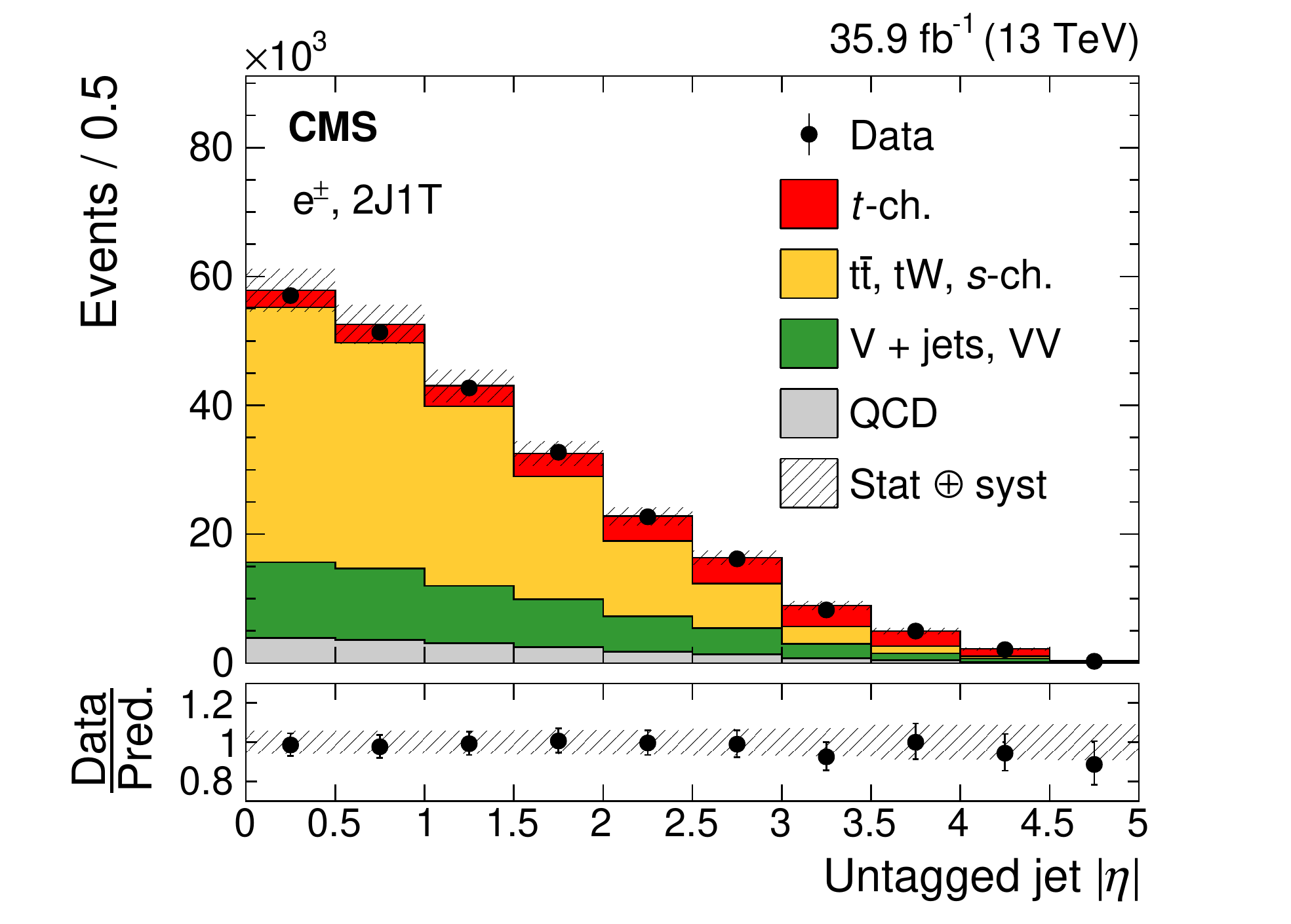}\\
\includegraphics[width=0.45\textwidth]{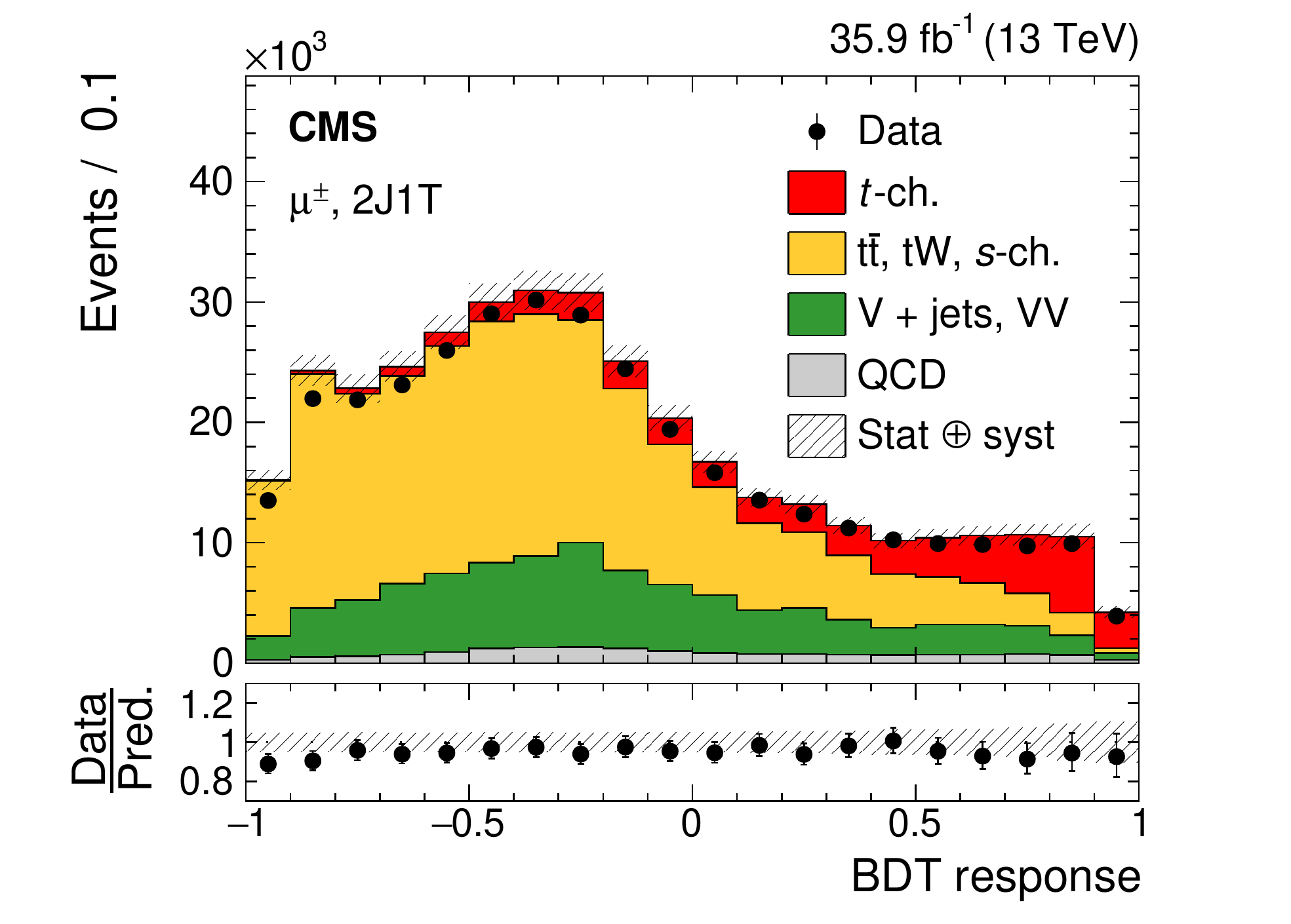}
\includegraphics[width=0.45\textwidth]{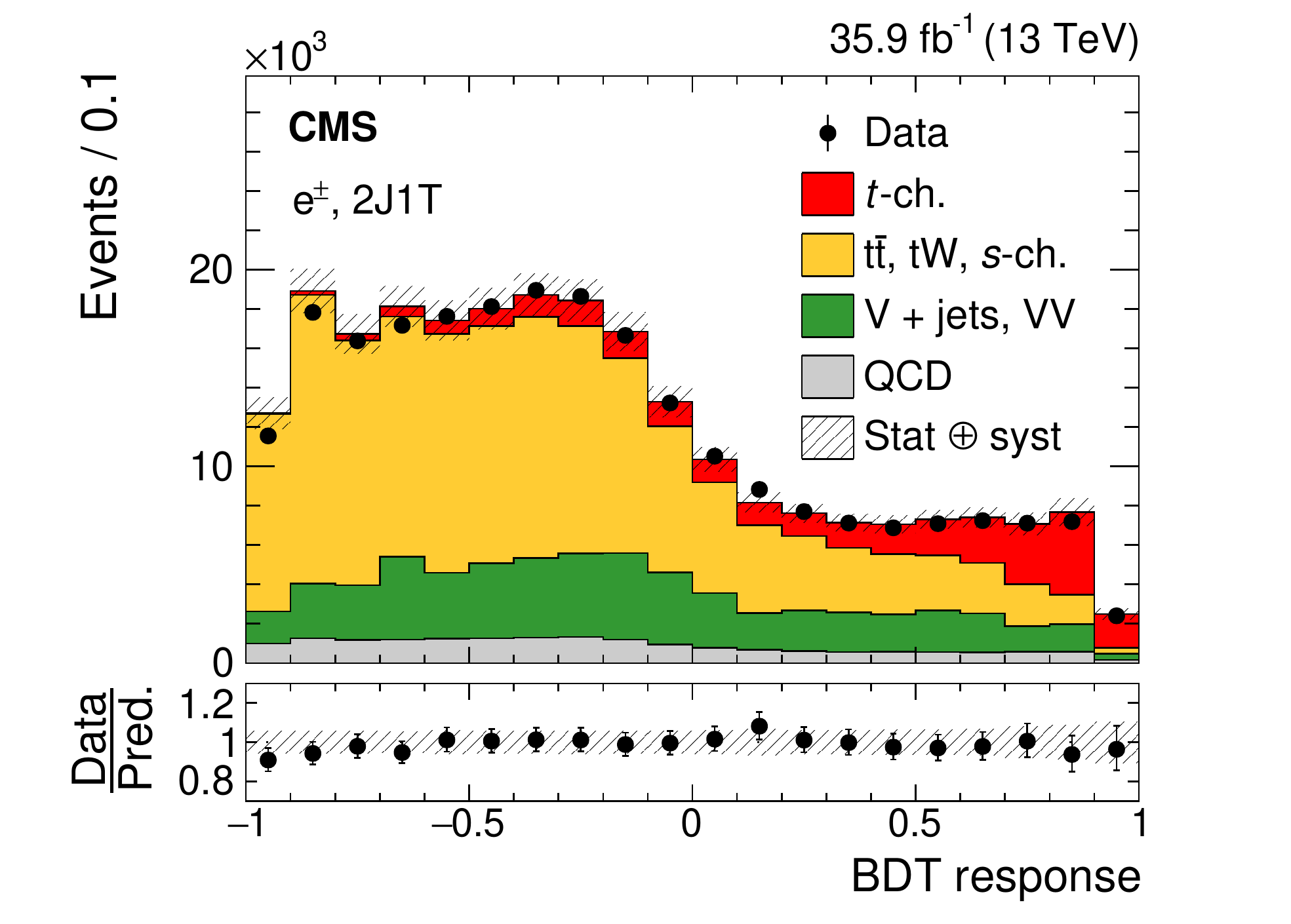}\\
\caption{\label{fig:inputVars} Data and simulation comparison for $\Delta R_{\PQb\mathrm{j^{\prime}}}$ (upper row), untagged jet $\abs{\eta}$ (middle row), and BDT response (lower row) in the 2J1T category for the muon (left) and electron (right) final states.
The lower panels show the ratio of the data to simulation predictions.
The bands indicate the statistical and systematic uncertainties added in quadrature.
The last bin in each of the upper-row plots includes the overflow.}
\end{figure*}

Figure~\ref{fig:inputVars} shows a comparison of the data and simulation for the two highest-ranked variables in Table~\ref{tab:mvaInputs}. 
The correlations among the variables are also taken into account by using the decorrelation method available in \textsc{tmva} during evaluation of the BDT response.
The modeling of the other input variables in data and simulation, along with their correlations in signal and background events before and after applying the decorrelation, are presented in Appendix~\ref{sec:suppl_MVA_Inputs}.

The input variables to the BDTs are chosen keeping in mind the following aspects:
\begin{itemize}
\item{good separation power to discriminate signal from background;} 
\item{low correlation with the reconstructed \mtop.}
\end{itemize}
The variables listed in Table~\ref{tab:mvaInputs} constitute a minimal set satisfying the above conditions in the muon and electron final states, and are selected as the BDT input variables.
They are ranked in decreasing order of their separation power (\textit{SP}) in the respective final states.
The \textit{SP} of a given variable $x$ is defined as
\begin{linenomath}
\begin{equation}
\label{eq:sep_pow}
\textit{SP}^{2} = \frac{1}{2}\int\frac{[\hat{X}_{S}(x) - \hat{X}_{B}(x)]^2}{\hat{X}_{S}(x) + \hat{X}_{B}(x)}\ \rd x, 
\end{equation}
\end{linenomath} 
where $\hat{X}_{S}$ ($\hat{X}_{B}$) denotes the probability density of $x$ in the signal (background) event category.

\begin{table*}[!htb]
\centering
\caption{\label{tab:mvaInputs} BDT input variables ranked in decreasing order of their \textit{SP} values for the muon and electron final states in the 2J1T category.}   
\resizebox{\textwidth}{!}{
\begin{tabular}{cccl}
 & \multicolumn{2}{c}{Rank} & \\
 Variable & Muon & Electron & Description\\
 \hline
 $\Delta R_{\PQb\mathrm{j^{\prime}}}$ &  1 & 1 & Angular separation in ($\eta$, $\phi$) space between the \PQb-tagged and untagged jets \\
 Untagged jet $\abs{\eta}$ ($\abs{\eta_{\mathrm{j}^{\prime}}}$) & 2 & 2 & Absolute pseudorapidity of the untagged jet \\
 $m_{\PQb\mathrm{j^{\prime}}}$ & 3 & 3 & Invariant mass of the system comprising the \PQb-tagged and untagged jets \\
 \multirow{2}{*}{$\cos{ \theta^{*}}$} & \multirow{2}{*}{4} & \multirow{2}{*}{4} & Cosine of the angle between the lepton and untagged jet in the rest frame\\ 
 &  &  & of the top quark \\
 \mT & 5 & 5 & Transverse mass as defined in Eq.~(\ref{eq:mtw}) \\
 FW1 & \NA & 6 & First-order Fox--Wolfram moment~\cite{FWM,Bernaciak:2012nh} (electron final state)\\ 
 $\abs{\Delta\eta_{l\PQb}}$ & 6 & 7 & Absolute pseudorapidity difference between the lepton and \PQb-tagged jet \\
 $\pt^{\PQb} + \pt^{\mathrm{j}^{\prime}}$ & 7 & 8 & Scalar sum of the \pt of the \PQb-tagged and untagged jets \\
 $\abs{\eta_{l}}$ & 8 & \NA & Absolute pseudorapidity of the lepton (muon final state) \\
\end{tabular}}
\end{table*}

\begin{figure*}
\centering
\includegraphics[width=0.45\textwidth]{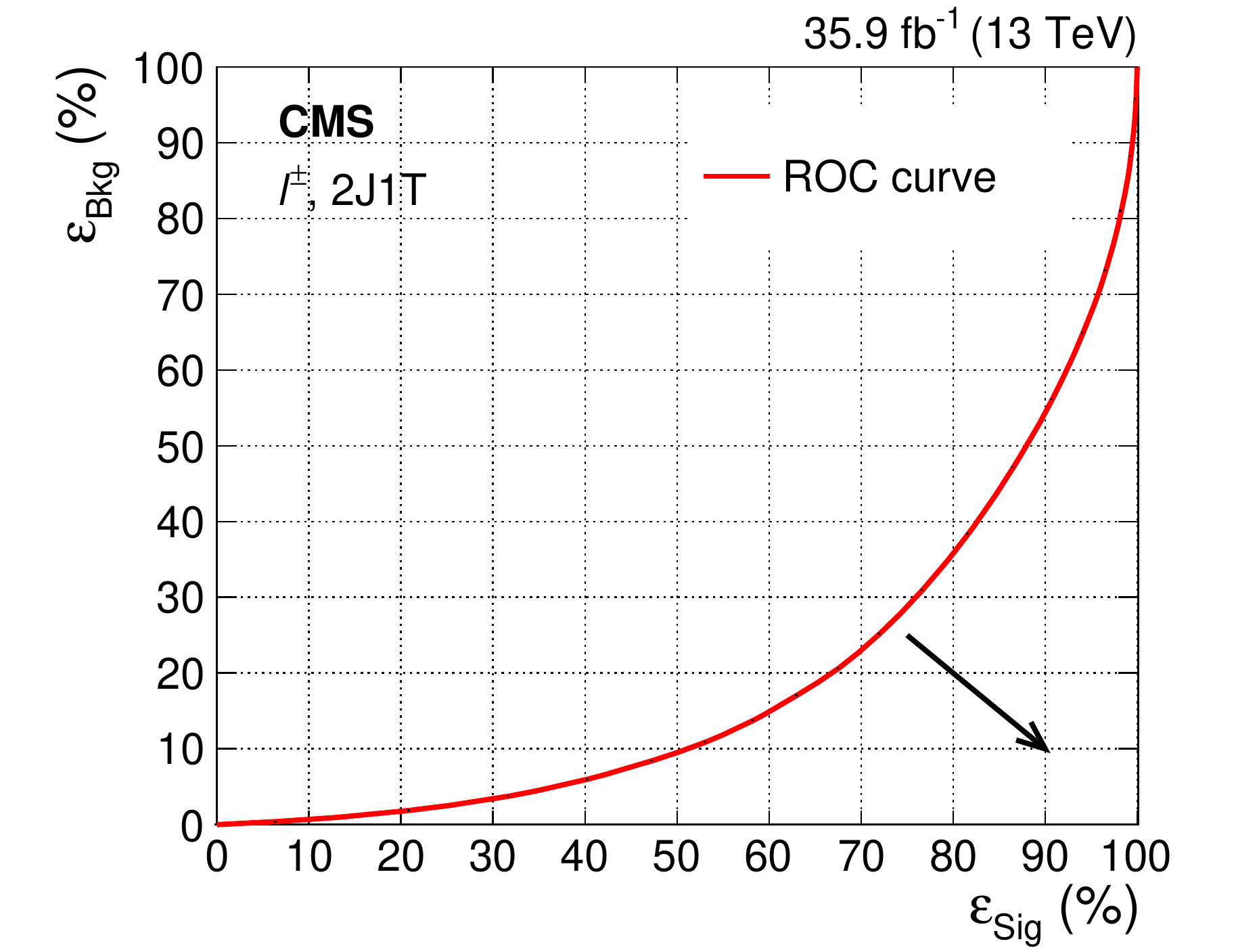}
\includegraphics[width=0.45\textwidth]{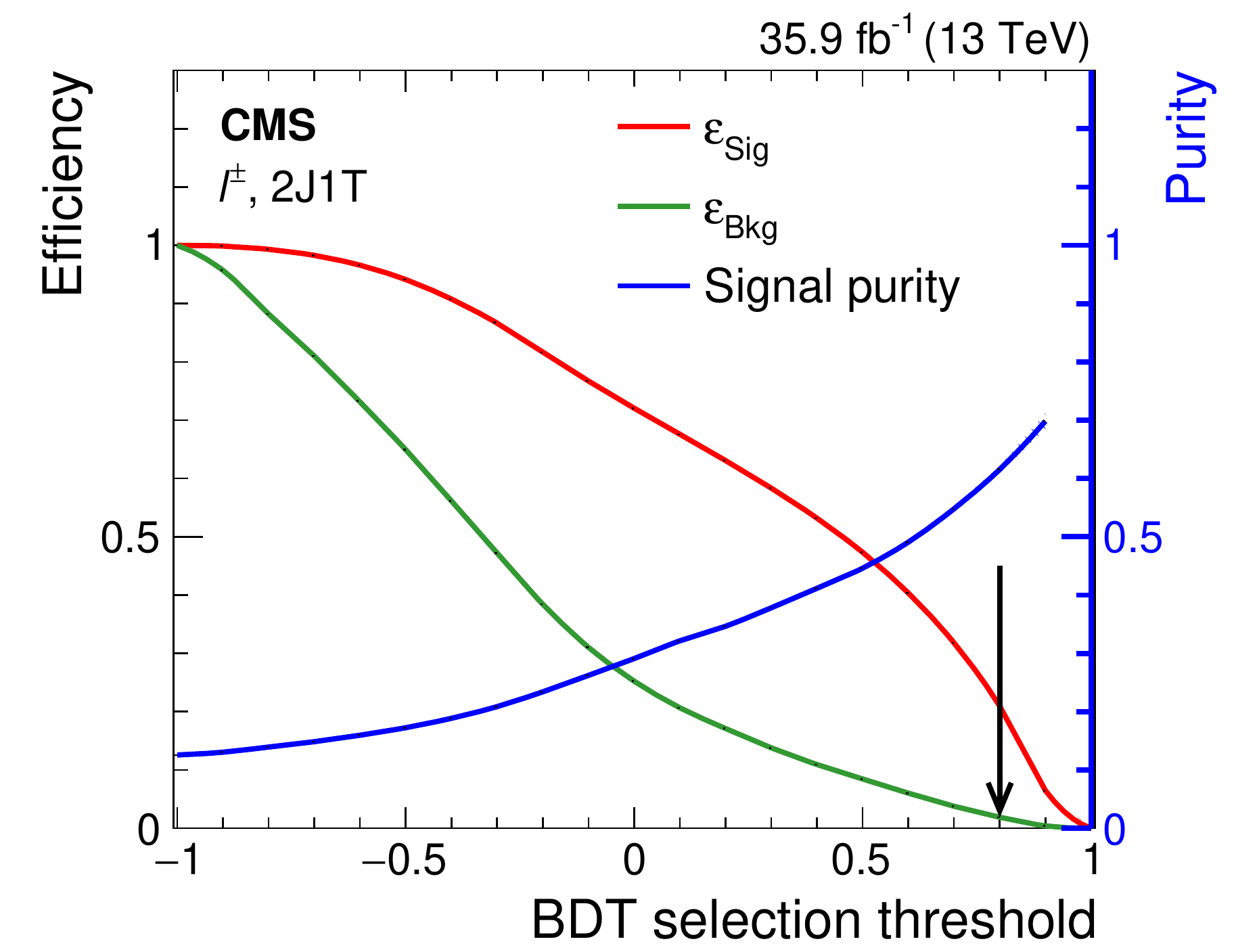}\\
\includegraphics[width=0.45\textwidth]{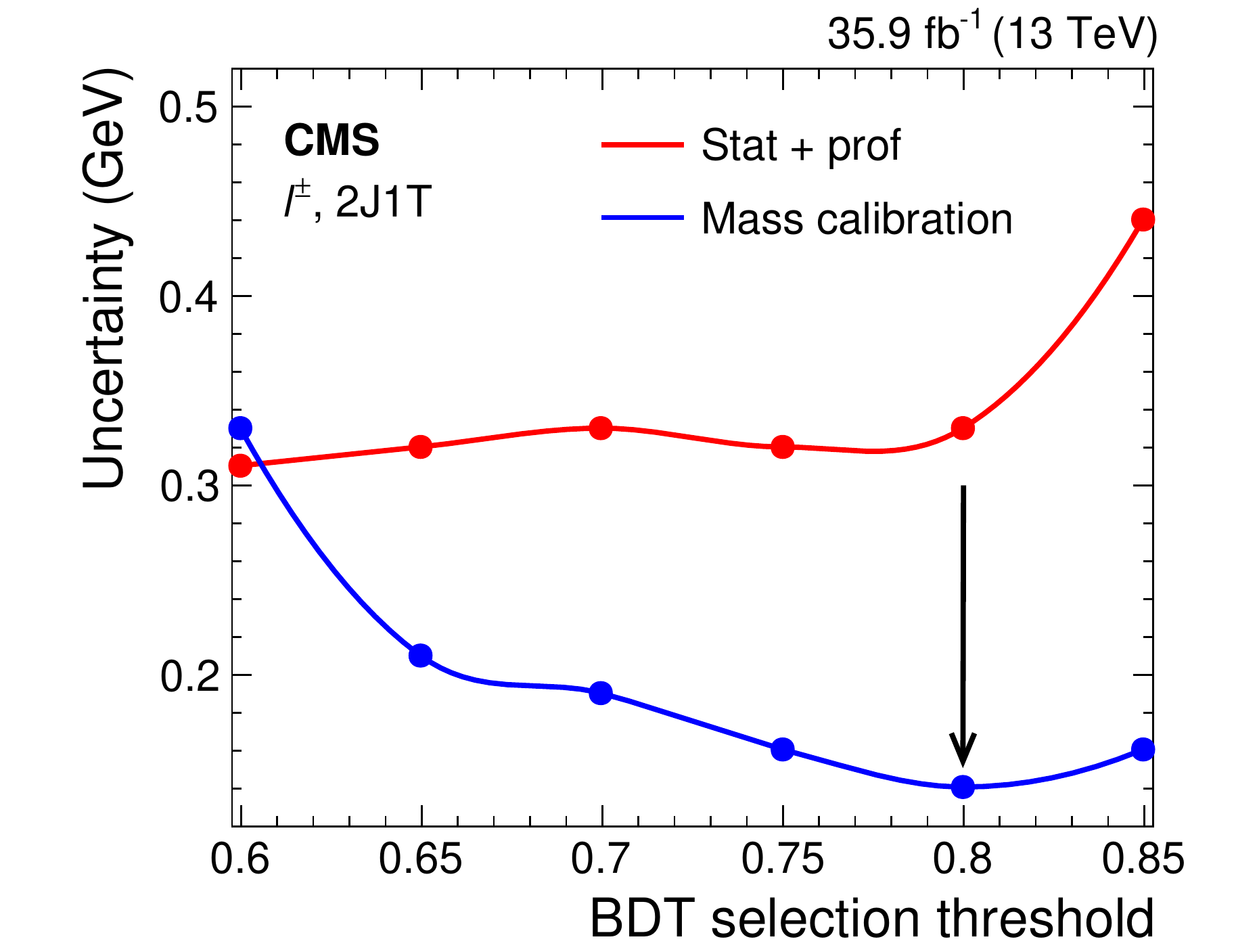}
\includegraphics[width=0.49\textwidth]{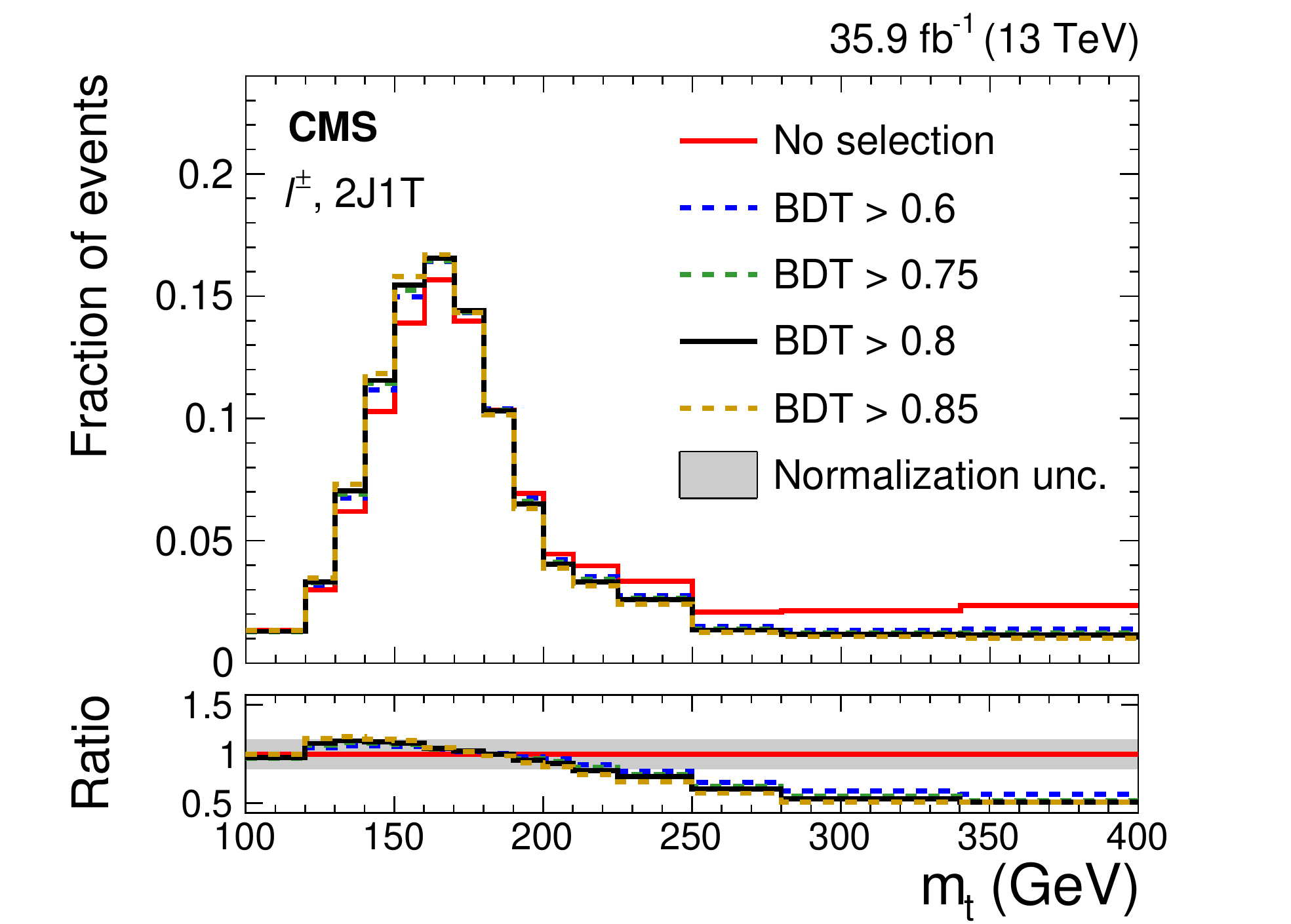}\\
\caption{\label{fig:bdt_cut_optimization} Combined BDT performance in terms of the ROC curve (upper left), and the signal and background efficiencies and signal purity as functions of the BDT selection threshold (upper right) in the muon and electron final states.
The uncertainty in \mtop from the combined statistical and profiled systematic components (red curve), and from the mass calibration (blue curve) as a function of the BDT selection threshold (lower left).
Arrows on the plots indicate the region of better separation (upper left) and optimized selection criteria (upper right and lower left).
A comparison of the reconstructed \mtop shapes from simulated signal events for different selection thresholds on the BDT response is shown on the lower right.
The lower panel shows the ratio relative to the case where no selection (red) is applied, with the grey band denoting the normalization uncertainty around the red curve.}
\end{figure*}

During training, signal and background events are weighted according to their relative contribution in the 2J1T category.
In total, 400 decision trees with a depth of three layers per tree are combined into a forest.
The adaptive boosting algorithm~\cite{Freund:1997xna} implemented in \textsc{tmva} is used with the learning rate and minimum node size set to 40\% and 1\%, respectively.
The BDT setup is checked for overtraining by dividing the MC samples into two independent subsamples of equal size, one for training and the other to validate its performance. 
Signal and background events are picked randomly to set up the training and validation subsamples in order to avoid bias.
We perform Kolmogorov--Smirnov (KS) tests to compare the BDT output distribution between the two subsamples, and obtain KS probabilities ranging between 57 and 93\% for the signal and background events.
These results confirm no significant overtraining.
Therefore, we combine the training and validation subsamples for further studies and evaluate the BDT responses for all signal and background events  according to the probabilities obtained from training in the respective final states.
The data-to-simulation comparisons of the BDT response distributions are shown in Fig.~\ref{fig:inputVars}, separately for the muon (left) and electron (right) final states in the 2J1T category.  
The performance of the BDTs is quantified via a combined receiver-operator-characteristic (ROC) curve, as shown in Fig.~\ref{fig:bdt_cut_optimization} (upper left).
The area between the ROC curve and the horizontal axis is ${\approx}0.16$. 
A smaller area indicates a better separation between the signal and background.
The correlation between the reconstructed \mtop and BDT response is about $-13$\% in simulated signal events.

To obtain an event sample enriched in $t$-channel single top quark events, selection thresholds on the BDT responses are optimized by studying the signal and background efficiencies, together with the signal purity after selection, as shown in Fig.~\ref{fig:bdt_cut_optimization} (upper right).
Ideally, high signal purity with high yield or efficiency is desirable for a precise \mtop measurement in $t$-channel single top quark events.
In reality, however, the signal efficiency and purity have opposite trends with increasing selection thresholds. 
The optimal working point is determined by studying the combined contribution of the statistical and profiled systematic uncertainties (Section~\ref{sec:syst}) along with that of the mass calibration, in the \mtop value as a function of the BDT thresholds applied to both muon and electron final states in simulated events.
These uncertainties are evaluated by means of an ML fit discussed in Section~\ref{sec:topmass}, based on pseudo-experiments derived from simulated events.
The result of this study is presented in Fig.~\ref{fig:bdt_cut_optimization} (lower left).
The starting point for the optimization is chosen to be 0.6, where about 50\% signal purity is observed for both final states.
The mass calibration uncertainty reaches its minimum when the selection threshold is at 0.8 and thus offers an optimum point where the relative impact of background contamination is at its lowest.
Beyond this point, both uncertainties start to increase rapidly with higher threshold values due to the depletion of the signal events.
The criterion BDT response $>$0.8 is chosen, which yields about 65 (60)\% signal purity in the muon (electron) final state.

\begin{figure*}
\centering
\includegraphics[width=0.45\textwidth]{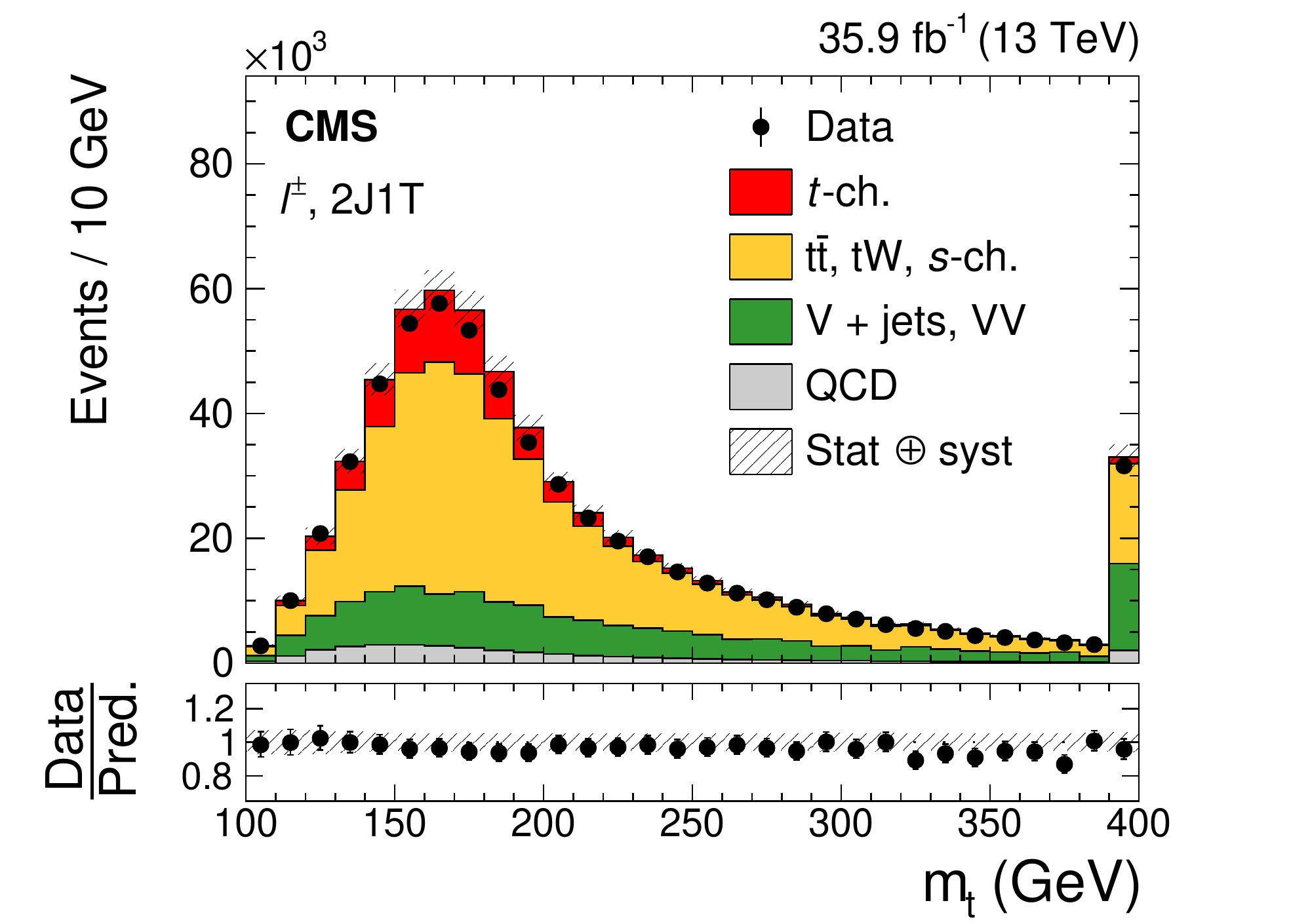}
\includegraphics[width=0.45\textwidth]{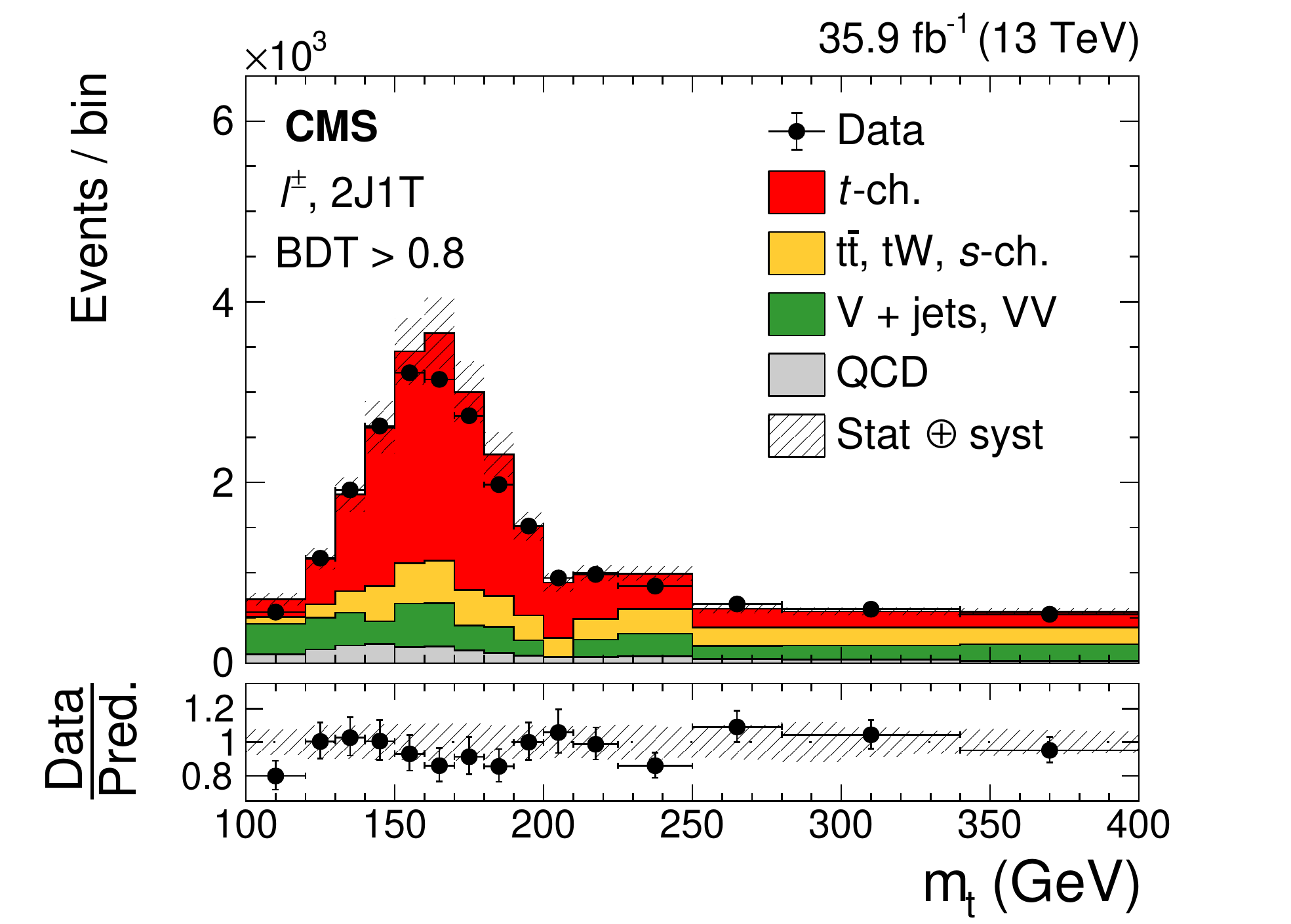}\\
\includegraphics[width=0.45\textwidth]{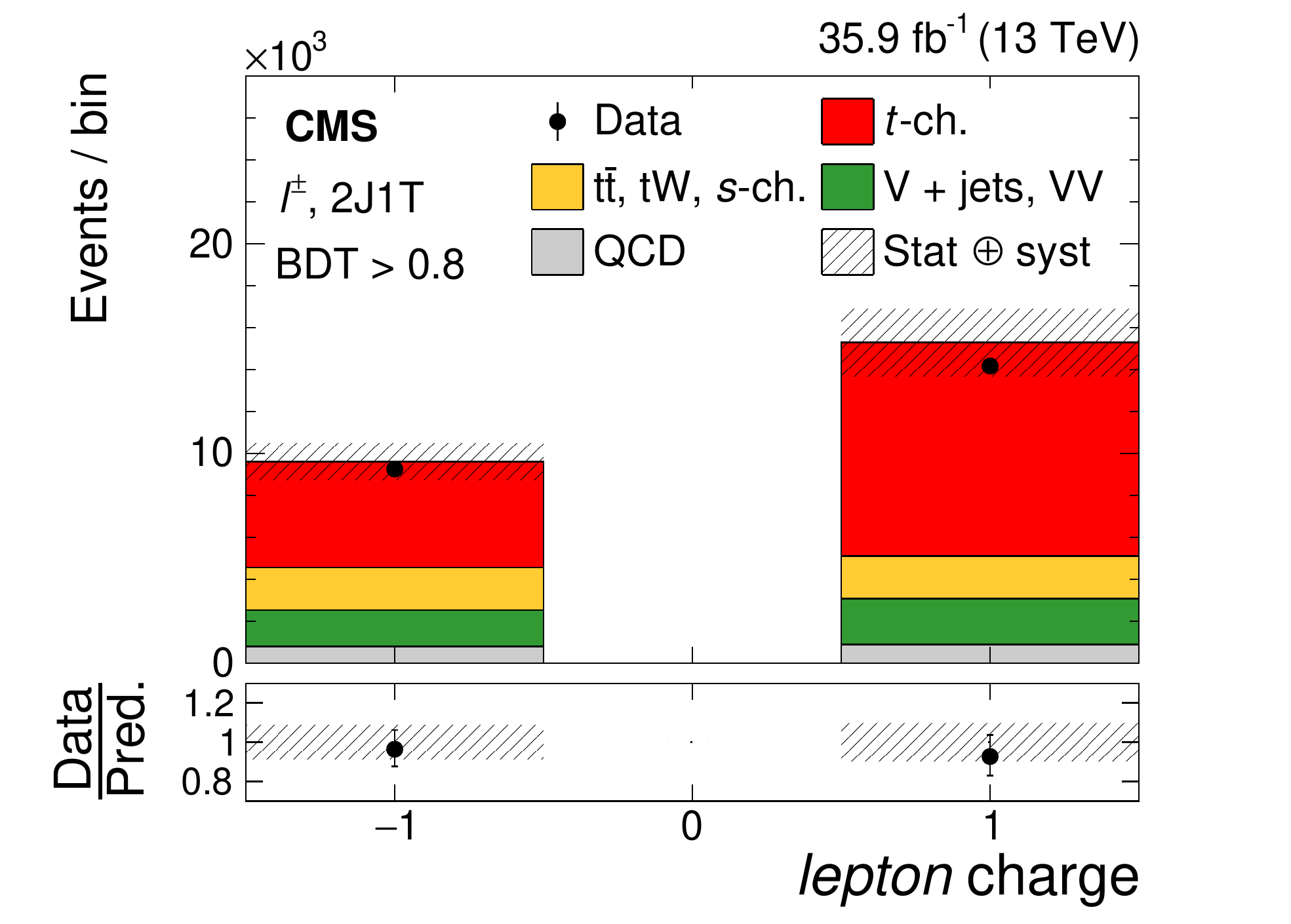}
\includegraphics[width=0.45\textwidth]{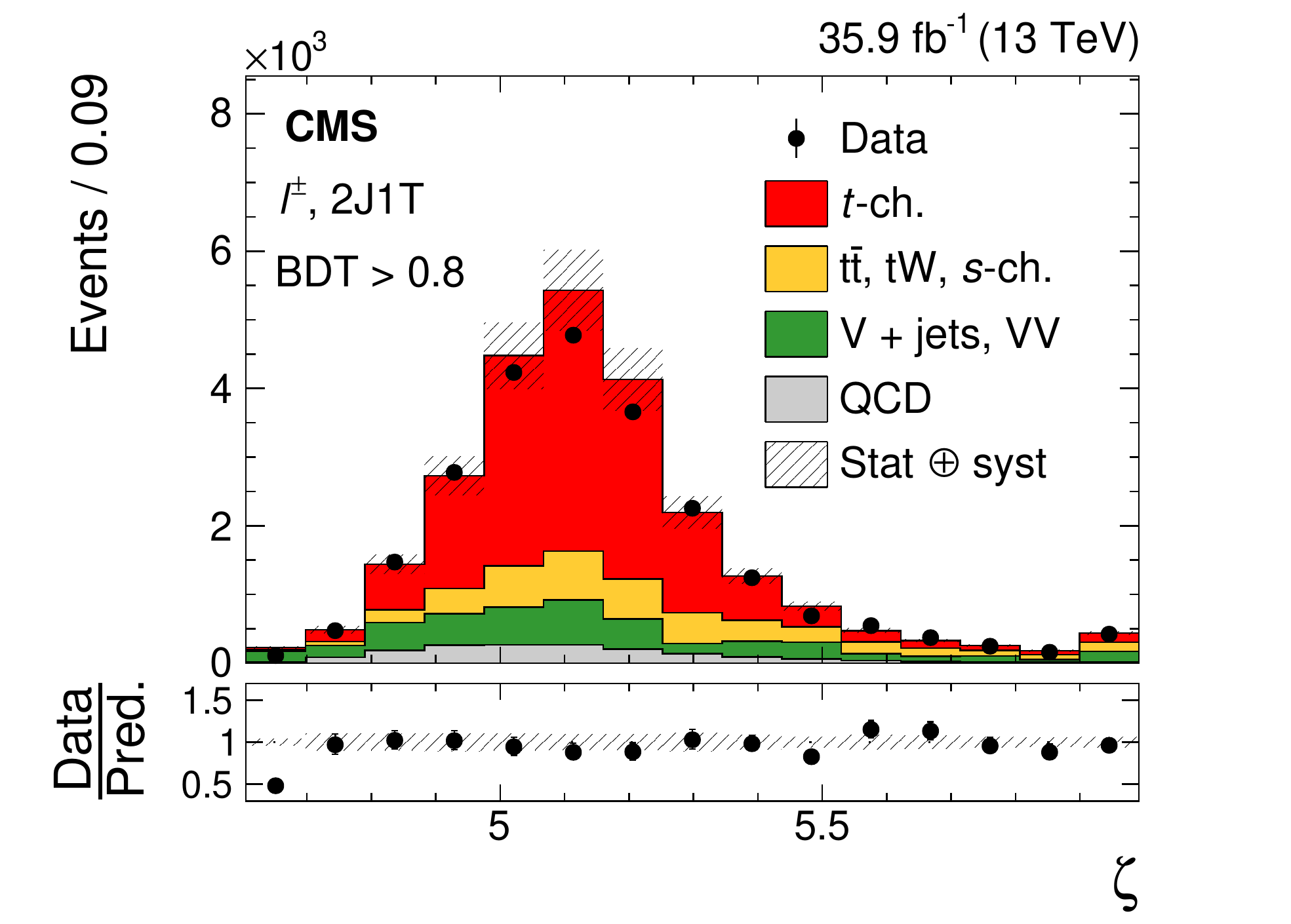}\\
\caption{\label{fig:mt_bdt} Reconstructed \mtop distributions before (upper-left) and after (upper-right) applying the optimized BDT selection from data (points) and simulation (colored histograms) are shown in the upper row.
The lower-left plot shows the event yields per lepton charge in data and simulation after optimized BDT selection. 
Data-to-simulation comparison of the fit variable $\zeta\ =\ln(\mtop/\ 1\GeV)$ inclusive of lepton charge after optimized BDT selection is presented in the lower right plot.
The horizontal bars in the upper-right plot indicate the variable bin width.
The first and last bins include the underflow and overflow, respectively, for each plot.
The bands denote a quadrature sum of the statistical and systematic uncertainties in the prediction.
The lower panels show the ratio of the data to the prediction. 
The deviation from unity seen in the first bins of the upper-right and lower-right ratio plots arises because of significantly less underflow events in the data compared to the simulation.}
\end{figure*}

Because of the low correlation between the BDT response and the reconstructed \mtop in the signal events, a selection on the former does not alter the \mtop distribution significantly, as shown in Fig.~\ref{fig:bdt_cut_optimization} (lower right).
The differences in signal \mtop shapes, obtained with various selection criteria on the BDT responses, are mostly covered by the normalization uncertainty around the peak.
This is evident in the lower panel of the plot where a comparison of shape ratios obtained with different selection thresholds relative to the one without any selection on the BDT response is shown.
The differences, observed in the higher tails of the \mtop distributions with and without the optimized BDT selection, are covered by the signal shape variation discussed in Section~\ref{sec:syst-mod}.
The \mtop distribution in data and the simulated signal and background before and after the application of the BDT selection criteria are compared in Fig.~\ref{fig:mt_bdt}. 

\section{\label{sec:topmass} Extraction of the top quark mass}
The high skewness of the \mtop distribution, coupled with the low background rate after the BDT selection, poses a considerable challenge in obtaining appropriate analytic shapes for both the signal and background.
Instead, a suitable alternative is found using the variable $\zeta = \ln(\mtop/\ 1\GeV)$.
The natural logarithm significantly reduces the skewness of the \mtop distribution since more extreme values to the right of the peak are pulled in, and those to the left are stretched further away~\cite{Gaur:2013uou}. 
The transformed probability density functions are well-behaved since this is a monotonic one-to-one mapping. 

The $\zeta$ distributions obtained from the muon and electron final states are considered in a simultaneous ML fit~\cite{Verkerke:2003ir}.
The fit is separately performed on events with a positively charged lepton ($l^{+}$), negatively charged lepton ($l^{-}$), and inclusive of the lepton charge ($l^{\pm}$) in the final state.
The inclusion of the QCD multijet background as an additional component in the fit would require a reliable parameterization, which turns out to be challenging in our case.
Instead, this background contribution is subtracted from data before the fit.
The QCD-subtracted binned $\zeta$ distribution is described by a analytic model $F(\zeta)$ for each final state ($l=\PGm$ or $\Pe$).
The total likelihood is given as
\begin{linenomath}
\begin{equation}
\label{eq:Likelihood}
\mathcal{L}_{\text{tot}} = \prod_{l}\mathcal{L}_{l}\ \ \ \text{with}\ \ \ \mathcal{L}_{l} = \prod_{i,\ j}\ \mathcal{P}[N^{\text{obs}}_{i,\ l}\ | \ F_{l}(\zeta;\ \zeta_{0},\ f_{j})]\ \Theta(f_{j}),
\end{equation}
\end{linenomath} 
where $i$ is the bin index, $\zeta_{0}$ is the peak position of the $\zeta$ distribution, $\mathcal{P}$ denotes the probability of the analytic model,  $F_{l}(\zeta;\ \zeta_{0},\ f_{j})$ to describe the observed $\zeta$ distribution, and $\Theta$ is the penalty term to account for the nuisance parameters $f_{j}$.
These parameters are defined for the rates of various event components $j$, namely signal, \ttbar, and EW backgrounds, as
\begin{linenomath}
\begin{equation}
\label{eq:sf}
f_{j} = \frac{N_{j}^{\text{obs}}}{N_{j}^{\text{exp}}}, \ \ j \in \{\text{sig},\ \ttbar,\ \mathrm{EW} \},  
\end{equation}
\end{linenomath} 
where $N_{j}^{\text{obs}}$ ($N_{j}^{\text{exp}}$) is the observed (expected) yield for the event component $j$.
We express $F_{l}(\zeta;\ \zeta_{0},\ f_{j})$ as
\begin{linenomath}
\begin{equation}
\label{eq:Model}
F_{l}(\zeta;\ \zeta_{0},\ f_{j}) = f_{\text{sig}} F_{\text{sig}}(\zeta;\ \zeta_{0})\ +\ f_{\ttbar} F_{\ttbar}(\zeta;\ \zeta_{0})\ +\ f_{\mathrm{EW}} F_{\mathrm{EW}}(\zeta),
\end{equation}
\end{linenomath} 
where $F_{\text{sig}}$, $F_{\ttbar}$, and $F_{\mathrm{EW}}$ represent the analytic shapes for the signal, \ttbar, and EW background, respectively.
Small contributions (6\%) from the \PQt{}\PW and $s$-channel single top quark processes are absorbed into the significantly larger (94\%) \ttbar component in forming what we call the \ttbar background above.

The $F_{\text{sig}}$ shape is parameterized with a sum of an asymmetric Gaussian ($\zeta_{0}$) function as the core and a Landau~\cite{Fanchiotti:2003yp} function to model the higher tail.
The $F_{\ttbar}$ shape is described by a Crystal Ball ($\zeta_{0}$) function~\cite{CrystalBallRef}.
The $F_{\mathrm{EW}}$ shape describes contributions from the \vjets and diboson processes.
It is modeled with a Novosibirsk function~\cite{Ikeda:1999aq}.

The parameter $\zeta_{0}$ of the combined signal and \ttbar background shapes is the parameter of interest (POI), and is allowed to float in the fit.
We extract \mtop from the best-fit $\zeta_{0}$ value.
Parameters, except for the POI, that alter the signal and background shapes are fixed to their estimated values during the fit. 
These are obtained by fitting individual models to the respective distributions of simulated signal and background events in the muon and electron final states.
Out of the shape parameters, the ones having large correlations with the POI are varied up to three standard deviations about their estimated values.
The resulting uncertainty is considered as a separate systematic uncertainty in the measured mass.
The nuisance parameters $f_{\text{sig}}$, $f_{\ttbar}$, and $f_{\mathrm{EW}}$ are constrained in the fit using log-normal priors with 15, 6, and 10\% widths, respectively.
The constraint on $f_{\text{sig}}$ takes into account the uncertainty in the cross section of the inclusive $t$-channel single top quark production measured at $\sqrt{s} = 13\TeV$~\cite{Sirunyan:2018rlu}.
The constraint on $f_{\ttbar}$ is driven by the uncertainty in the predicted \ttbar production cross section~\cite{Czakon:2013goa}.
The constraint on $f_{\mathrm{EW}}$ is relaxed to around three times the uncertainty in the measured \vjets cross sections~\cite{Sirunyan:2017wgx,Sirunyan:2018cpw} in order to account for mismodeling of heavy-flavor jet production in simulation, as well as to cover the uncertainties due to the renormalization ($\mu_{\mathrm{R}}$) and factorization ($\mu_{\mathrm{F}}$) scales and PDF in the EW background.
A similar approach was used in the measurement of the inclusive $t$-channel single top quark production cross section~\cite{Sirunyan:2018rlu}. 

\begin{figure*}
\centering
\includegraphics[width=0.45\textwidth]{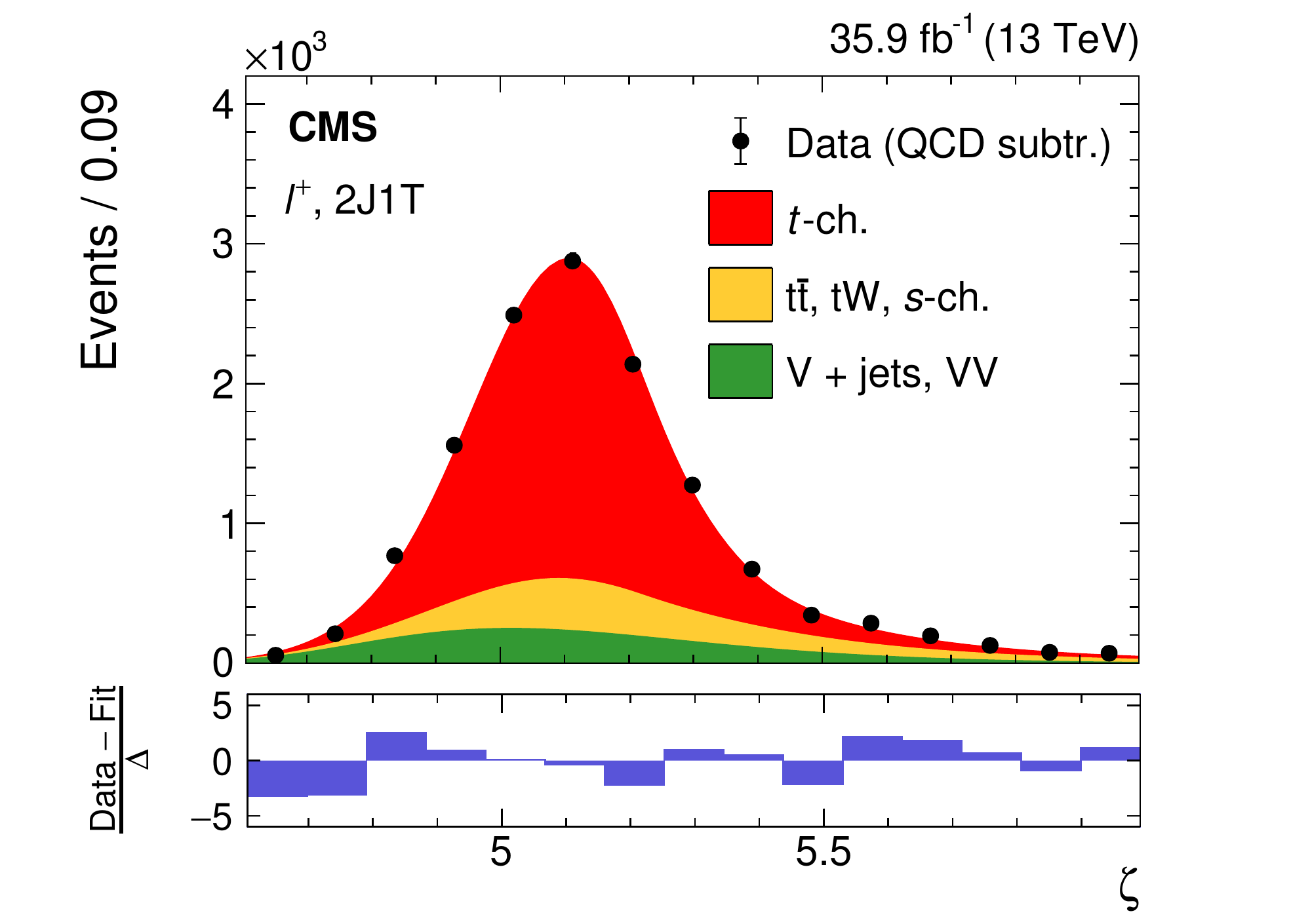}
\includegraphics[width=0.45\textwidth]{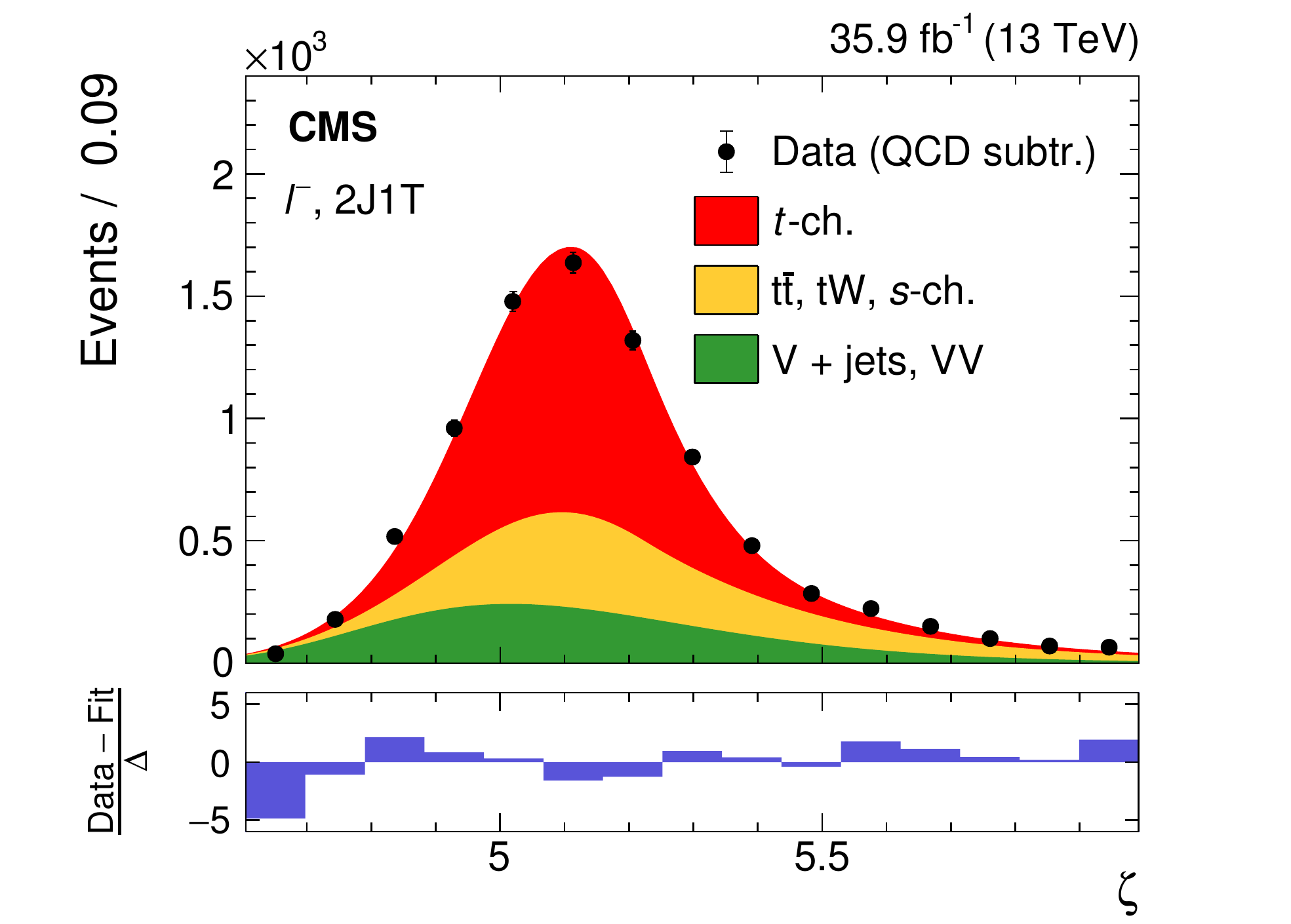}\\
\includegraphics[width=0.45\textwidth]{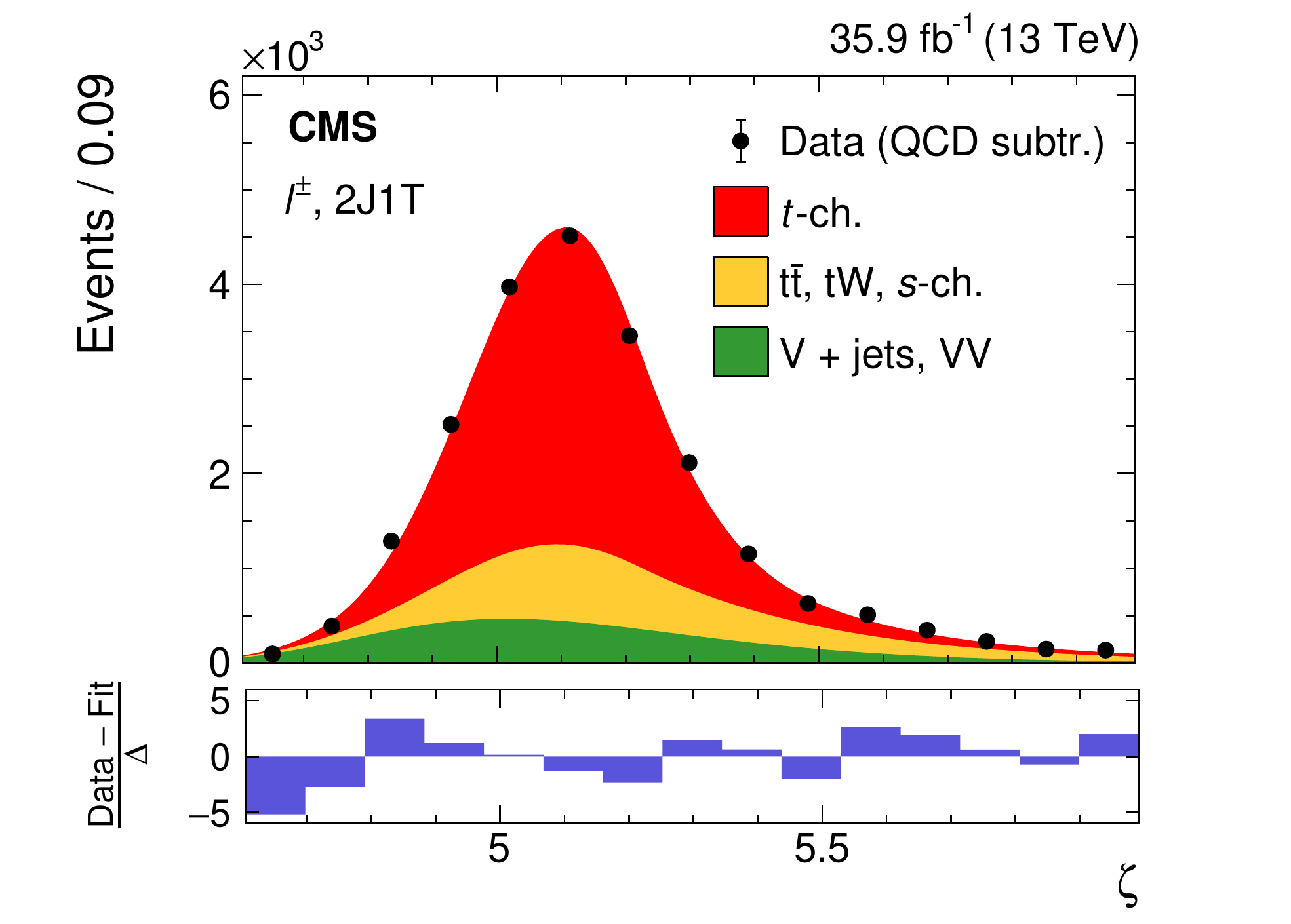}\\
\caption{\label{fig:massfit} Projections of fit results onto the $\zeta\ = \ln(\mtop/\ 1\GeV)$ distributions for the $l^{+}$ (upper left), $l^{-}$ (upper right), and $l^{\pm}$ (lower row) final states for signal and background processes compared to data in the 2J1T category with the optimized BDT selection applied.
The lower panels show the pulls that are determined using the bin contents of the data  distributions (after QCD subtraction) and the $F(\zeta)$ values evaluated at the center of the bins.
The fitted $\zeta$ distributions for each final state are combined in these plots for a comparison with data for the three different cases based on the lepton charge.}
\end{figure*}

Projections of the fit results onto the $\zeta$ distributions for the $l^{+}$, $l^{-}$, and $l^{\pm}$ cases are shown in Fig.~\ref{fig:massfit}. 
The lower panels in Fig.~\ref{fig:massfit} show the normalized residuals or pulls, which are defined as the difference between the distribution in data and the one predicted by the fit, divided by the uncertainty $\Delta = \sqrt{\smash[b]{\Delta^{2}_{\text{data}} - \Delta^{2}_{\text{fit}}}}$.
Here, $\Delta_{\text{data}}$ is the Poisson uncertainty in the data and $\Delta_{\text{fit}}$ is the uncertainty in the fit that includes both the statistical and profiled systematic components. 
Most of the pull values lie within $\pm$2 for all three cases, with the maximum deviations occurring in the first $\zeta$ bin because of significantly less underflow events in the data than in the simulation.

We validate the fit model given in Eq.~(\ref{eq:Model}) by applying it to a control sample defined by $-0.2 <\text{BDT}<0.8$, which is dominated by \ttbar events.
The resulting best-fit values of \mtop and the nuisance parameters are found to agree within the uncertainties with that obtained from the signal-enriched region ($\text{BDT}>0.8$). 
The scan of the profile likelihood ratio with the POI, together with the correlation among the fit parameters in the signal-enriched region for the $l^{\pm}$ case, are presented in Appendix~\ref{sec:suppl_fit}.
The fit consistency is checked by performing pseudo-experiments based on the $F(\zeta)$ model with profiled systematic uncertainties only.
We do not observe any bias in the fit parameters, and find the corresponding pulls follow a Gaussian distribution having a mean and width consistent with $0$ and $1.0$, respectively, within their uncertainties.

The linearity of the best-fit mass ($m_{\text{fit}}$) is checked against different \textit{true} mass ($m_{\text{true}}$) hypotheses using dedicated simulation samples for signal and \ttbar processes.
A calibration is performed by applying a suitable offset to $m_{\text{fit}}$ to account for the differences relative to $m_{\text{true}}$.
The difference between $m_{\text{fit}}$ and $m_{\text{true}}$ can be attributed to the inadequacy in the determination of \pzNu in the signal process discussed in Section~\ref{ssec:top4v}.
Details about the associated systematic uncertainty due to the mass calibration are discussed in Section~\ref{sec:syst-exp}.

\section{\label{sec:syst} Systematic uncertainties}

We consider several sources of systematic uncertainties.
They can be broadly classified into two categories as follows.
\begin{enumerate}
\item{\textit{Profiled (prof)}: The signal and background normalizations are profiled by including them as nuisance parameters in the ML fit.
The impacts of the profiled systematic sources are obtained directly from the fit via correlations between the POI and the nuisance parameters.}
\item{\textit{Externalized (ext)}: All other uncertainty sources are externalized, i.e., the ML fit is repeated with varied $\zeta$ shapes.
Thus, full variations of the shapes are considered due to these sources.
The impacts of these sources are calculated by taking the difference between the offset corrected best-fit values of \mtop corresponding to the nominal and varied shapes.
The largest shift relative to the nominal result is quoted as the uncertainty for a particular systematic source in a conservative approach, unless otherwise specified.
The total uncertainty due to the externalized systematic sources in the central value is obtained by separately combining the positive and negative shifts.} 
\end{enumerate} 
The externalized systematic sources can be further divided into two subcategories, namely experimental and modeling.

\subsection{\label{sec:syst-exp} Experimental uncertainties}

\begin{figure*}
\centering
\includegraphics[width=0.45\textwidth]{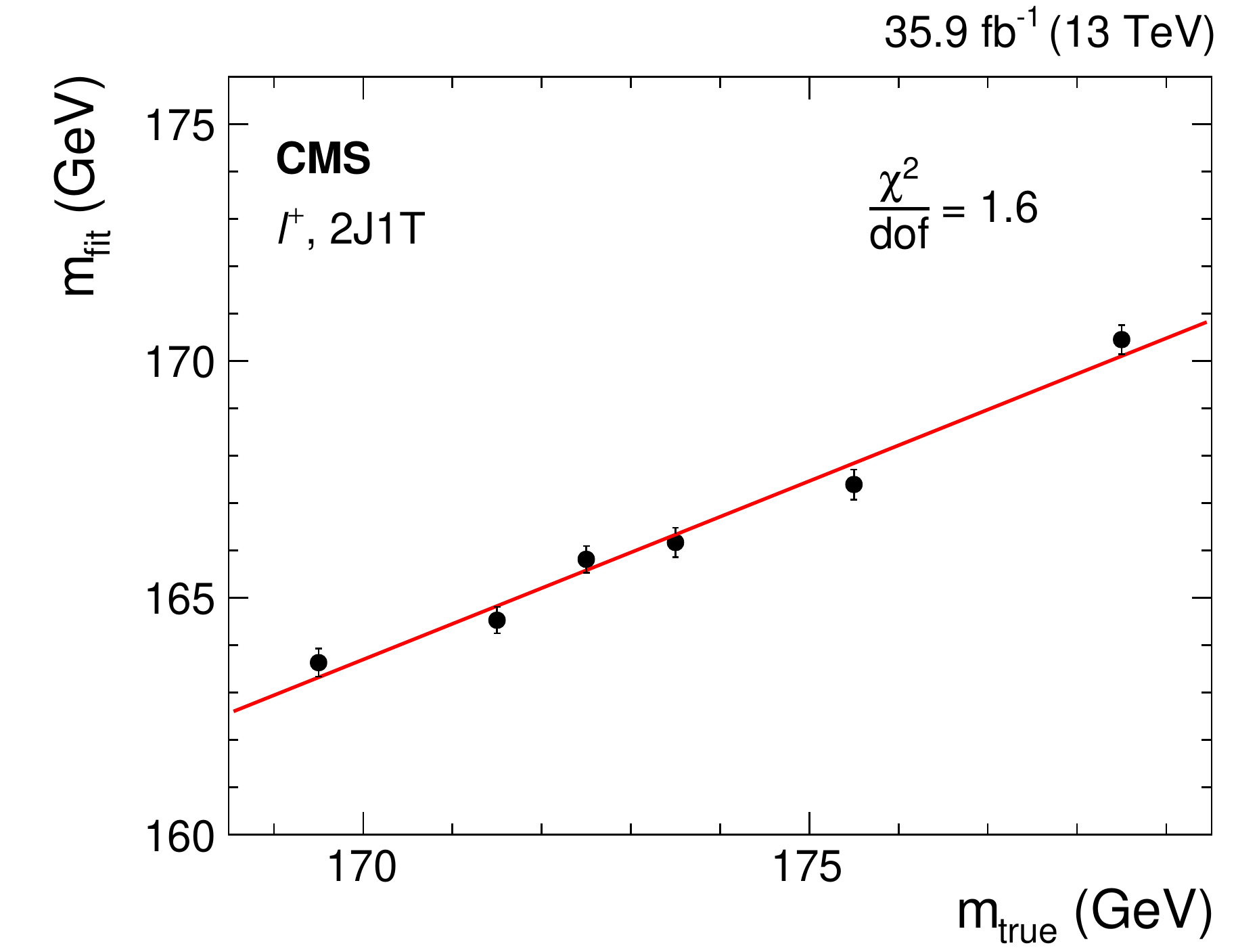}
\includegraphics[width=0.45\textwidth]{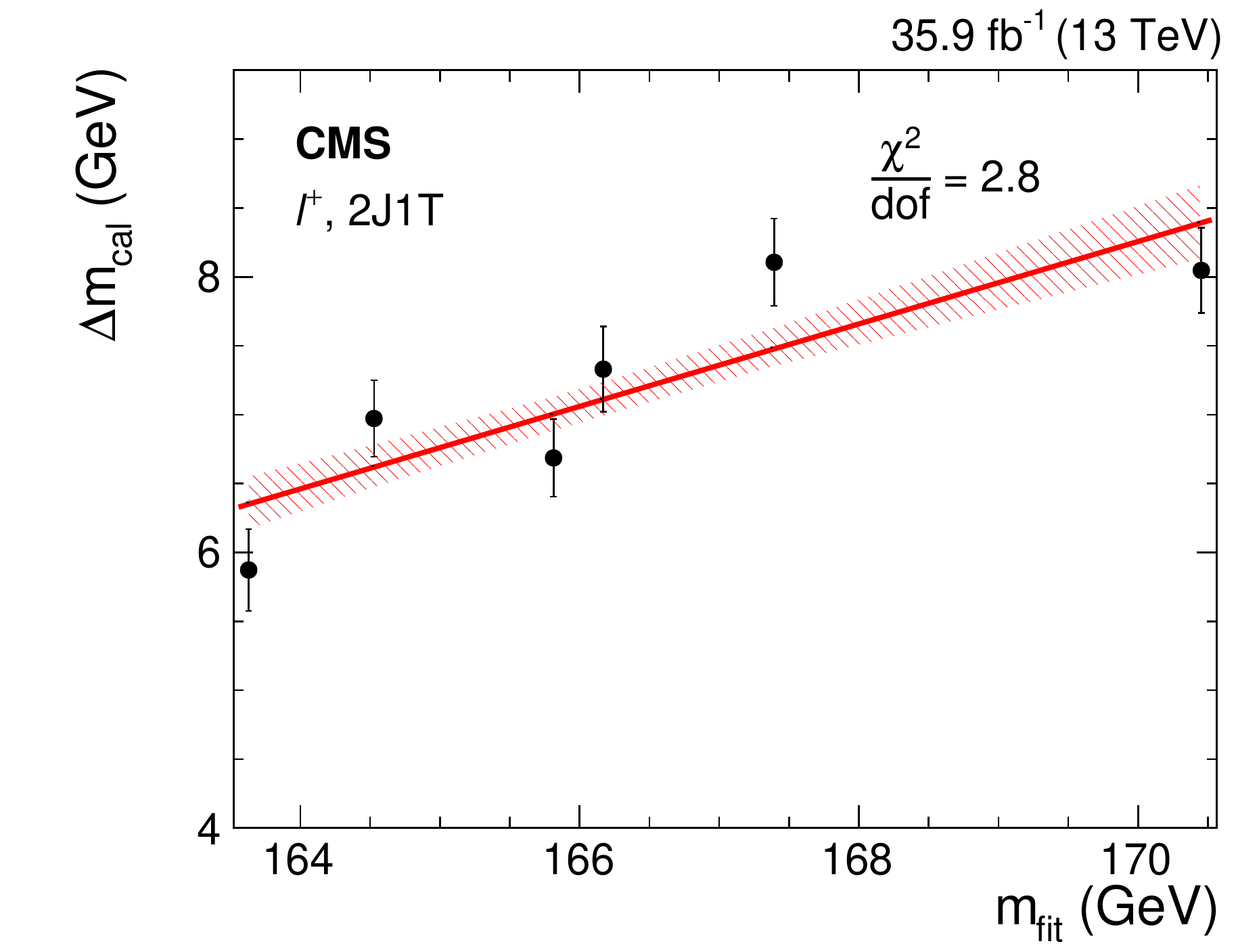}\\
\includegraphics[width=0.45\textwidth]{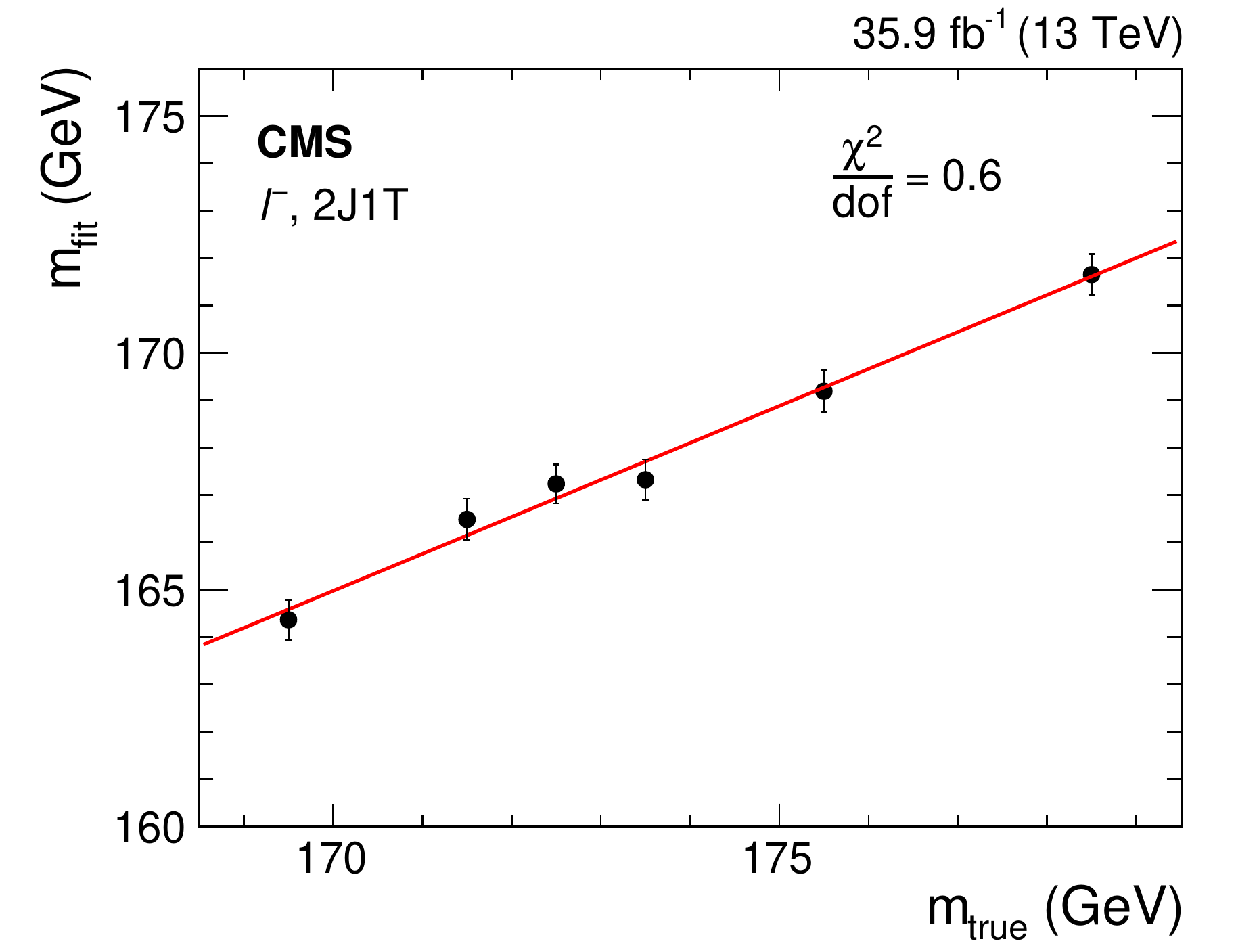}
\includegraphics[width=0.45\textwidth]{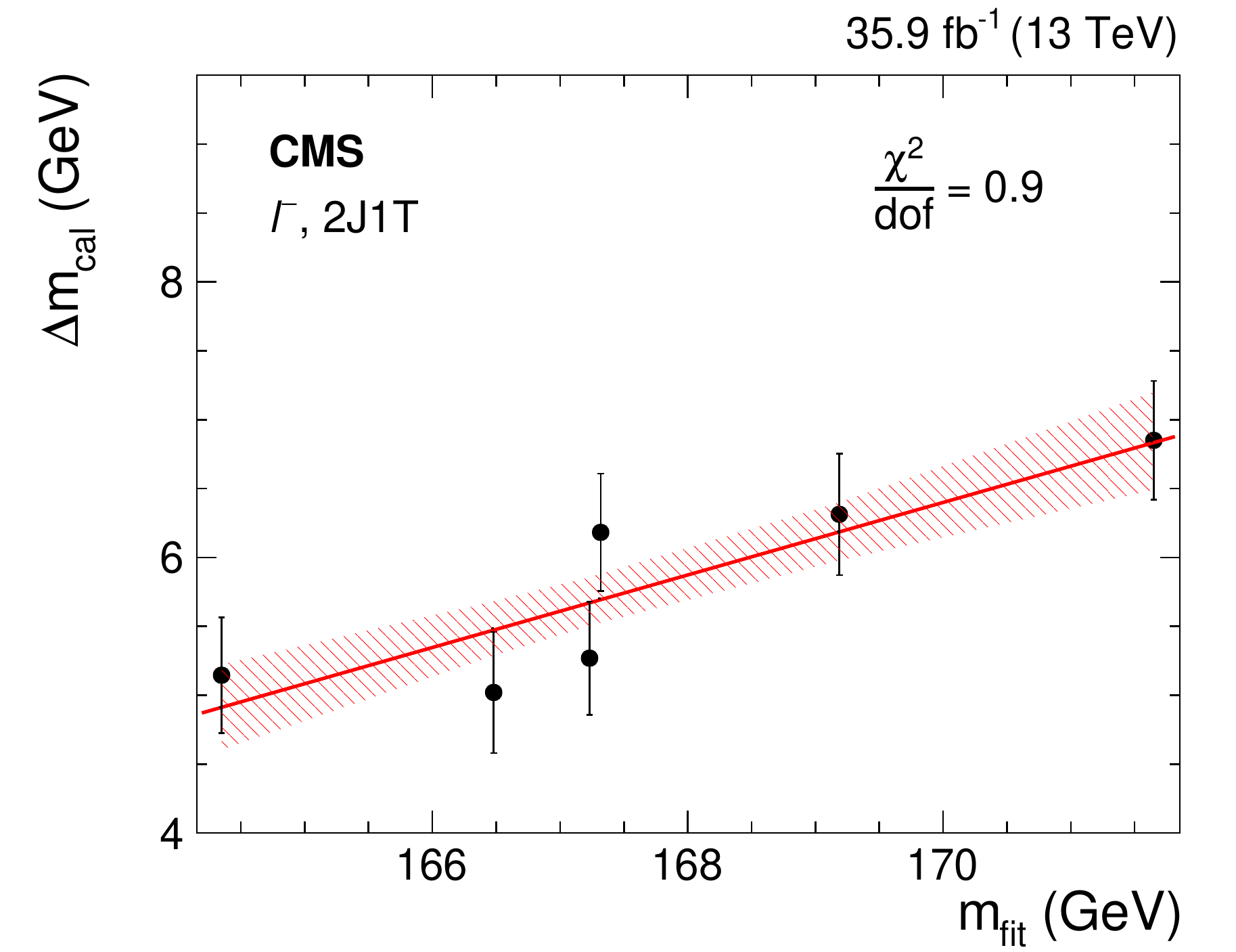}\\
\includegraphics[width=0.45\textwidth]{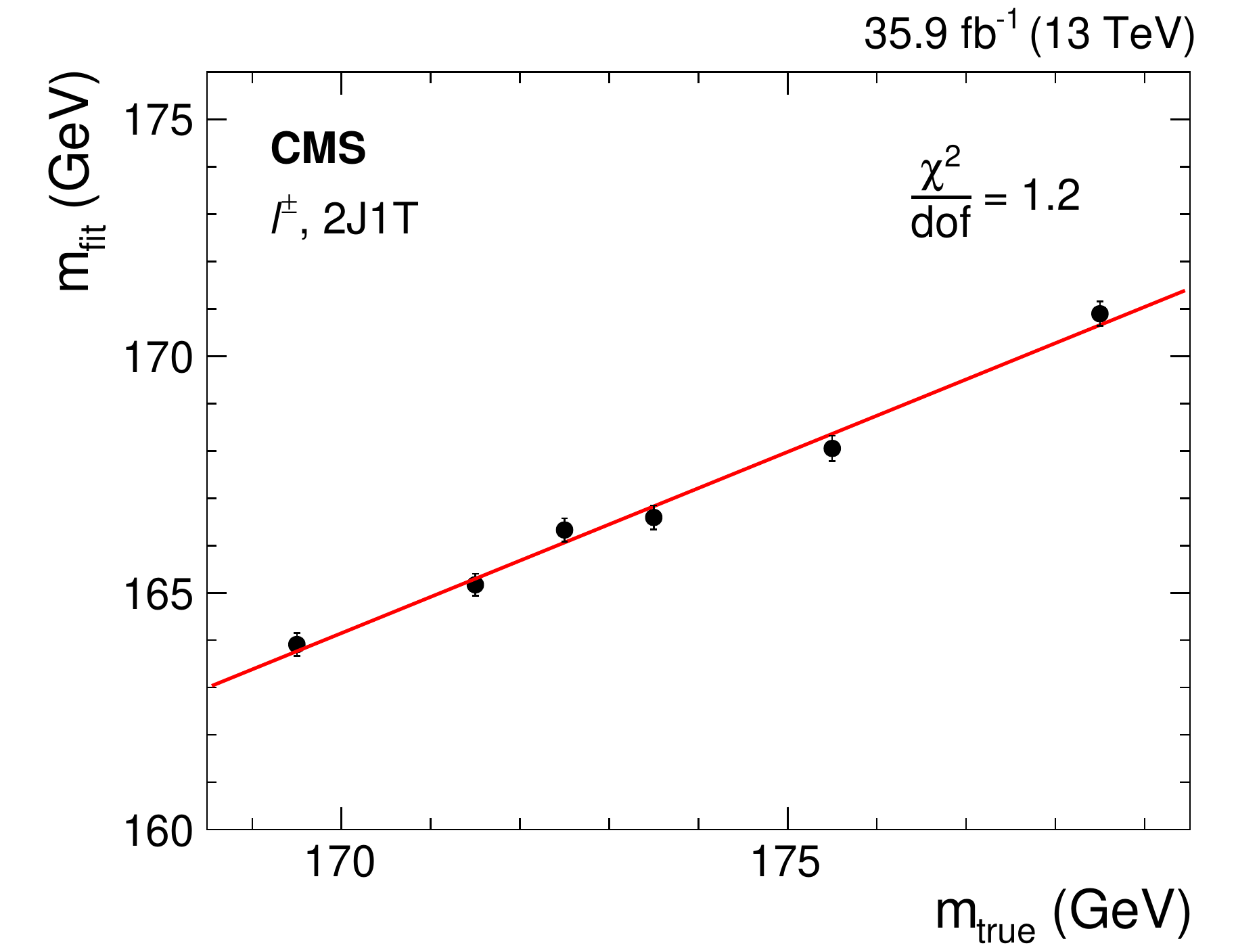}
\includegraphics[width=0.45\textwidth]{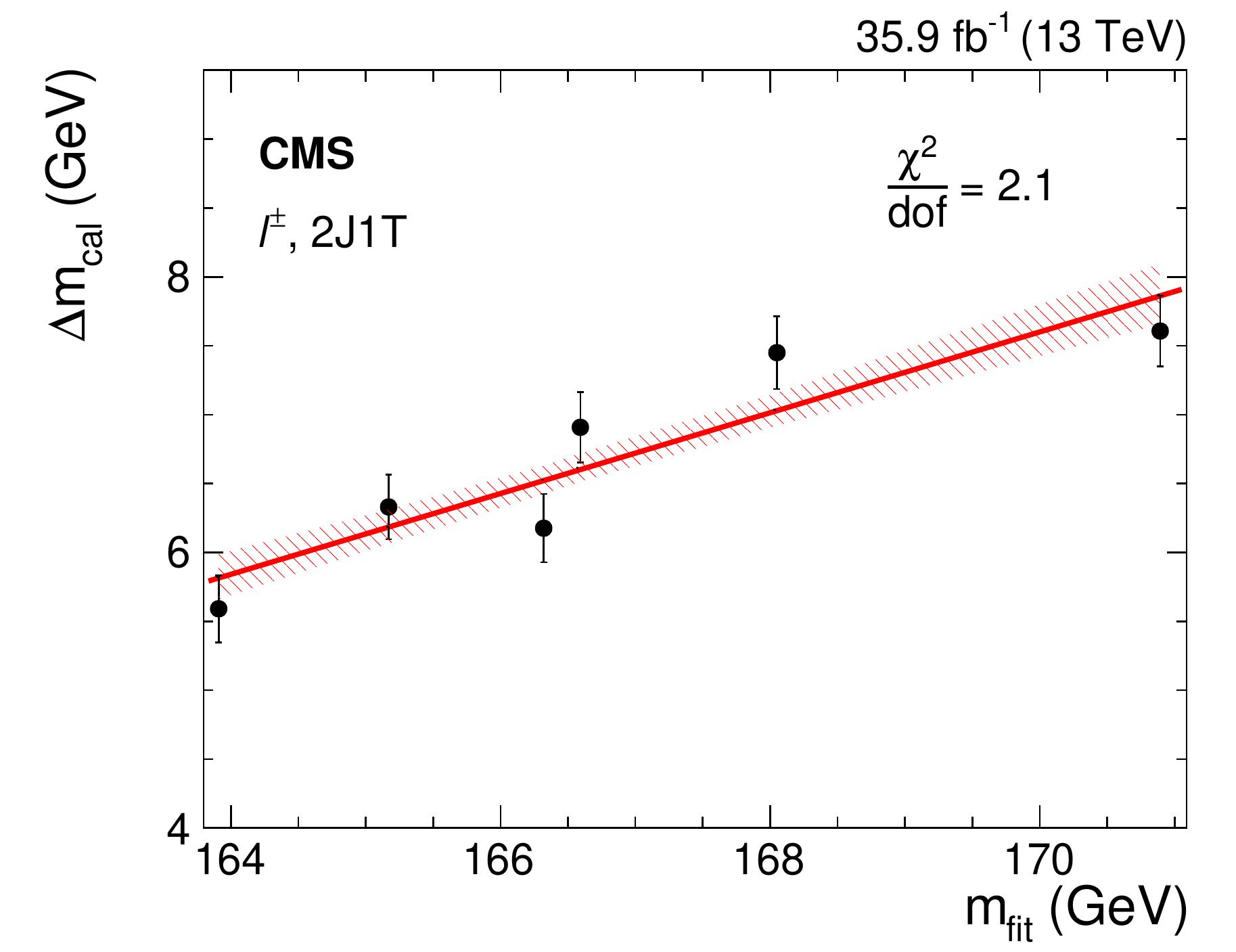}\\
\caption{\label{fig:massCalibration} Test of the linearity of the  $m_{\text{fit}}$ for different values of \textit{true} mass $m_{\text{true}}$ (left), and the resulting mass calibration $\Delta m_{\text{cal}}$ as a function of $m_{\text{fit}}$ (right) for events in the 2J1T category for the $l^{+}$ (upper), $l^{-}$ (middle), and $l^{\pm}$ (lower) cases.
The shaded regions indicate $\pm$1 standard deviations about the central values defined by the red line.
The value of the $\chi^{2}$ per degrees of freedom (dof) from the linear fit is shown in each plot.}
\end{figure*}

\begin{itemize}
\item{\textit{JES:} The energies of all reconstructed jets in simulated events are simultaneously scaled up and down according to their \pt- and $\eta$-dependent uncertainties~\cite{Chatrchyan:2011ds}, split into correlation groups, namely intercalibration, MPFInSitu, and uncorrelated according to the procedure described in Ref.~\cite{CMS-PAS-JME-15-001}.
A similar approach to subcategorize the JES uncertainties was used in Ref.~\cite{Sirunyan:2018gqx} for the \mtop measurement with \ttbar events in the \ljets final state.}
\item{\textit{Jet energy resolution (JER):} To account for the difference in JER between data and simulation, a dedicated smearing is applied~\cite{Chatrchyan:2011ds} in simulation that improves or worsens the resolutions within their uncertainties.}
\item{\textit{Unclustered energy:} The contributions of unclustered particles to \ptmiss are varied within their respective energy resolutions~\cite{Khachatryan:2014gga}.}
\item{\textit{Muon and electron efficiencies:} The lepton identification, isolation, and trigger efficiencies are determined with a ``tag-and-probe" method~\cite{Khachatryan:2010xn} from Drell--Yan events falling in the \PZ boson mass window.
The efficiency correction factors are applied to simulated events in order to match with data.
The uncertainties in these correction factors are varied in bins of \pt and $\abs{\eta}$.}
\item{\textit{Pileup:} The uncertainty in the expected distribution of pileup is propagated as a systematic uncertainty by varying the total inelastic cross section by $\pm$4.6\%~\cite{Sirunyan:2018nqx} about its central value of 69.2\unit{mb} by reweighting the simulated events.}
\item{\textit{\PQb tagging:} The scale factors used to calculate the efficiency corrections for the \PQb tagging algorithm are varied up and down within their uncertainties.
The efficiency corrections found from these variations are then applied to the simulation to estimate the corresponding systematic uncertainty.}
\item{\textit{QCD multijet background:} The contribution of the QCD multijet background is estimated based on data as discussed in Section~\ref{sec:bkgmodel}.
Its contribution is first subtracted from data before the final fit using the parametric model given in Eq.~(\ref{eq:Model}).
Each bin of the QCD multijet shape derived from SB data is varied independently by an uncertainty of 50\% and a new set of shapes is obtained.
The resulting new shapes are subtracted from data in the signal region one at a time and the fit is repeated.
In this method, a maximal variation in the rate and shape of the QCD multijet background is considered.
The resulting uncertainty is obtained from the difference in the mean value of the offset-corrected fit results relative to the nominal case.}
\item{\textit{Mass calibration:} The mass calibration ($\Delta m_{\text{cal}}$), i.e., the difference between $m_{\text{fit}}$ and $m_{\text{true}}$, is obtained as a function of $m_{\text{fit}}$, using dedicated MC samples with alternate $m_{\text{true}}$ hypotheses (Fig.~\ref{fig:massCalibration}). The linear behavior between $m_{\text{fit}}$ and $m_{\text{true}}$ dictates that $\Delta m_{\text{cal}}$ also has a linear dependence on $m_{\text{fit}}$.  
The band about the central line represents the $\pm$1 standard deviation owing to statistical fluctuations of the signal and \ttbar samples with different $m_{\text{true}}$ hypotheses.
The mass calibration is obtained from the central value, while its uncertainty is determined from the band and considered as an independent source of uncertainty.
This procedure has been implemented separately for the $l^{+}$, $l^{-}$, and $l^{\pm}$ cases.}
\item{\textit{Luminosity:} The relative uncertainty in the integrated luminosity is $\pm$2.5\%~\cite{CMS-PAS-LUM-17-001}.
This is propagated to the uncertainties in the expected rates of the signal and background processes except for the QCD multijet, which is determined from data.}
\end{itemize}

\subsection{\label{sec:syst-mod} Modeling uncertainties}

\begin{itemize}
\item{\textit{CR and early resonance decay (ERD):} The uncertainties due to ambiguities in modeling CR effects are estimated by comparing the default model in \PYTHIA 8 with two alternative CR models, one with string formation beyond leading color ("QCD inspired")~\cite{Christiansen:2015yqa}, and the other in which the gluons can be moved to another string ("gluon move")~\cite{Argyropoulos:2014zoa}.
In addition, CR effects due to the top quark decay products are assessed by switching off (default) and on ERD~\cite{Argyropoulos:2014zoa} in \PYTHIA 8.
In the first case, the lifetime of the top quark is assumed to be long enough to shield its decay products from color reconnecting with the rest of the event, whereas this restriction is lifted when we enable the ERD option.
All models are tuned to the UE measurements in Ref.~\cite{Sirunyan:2018avv}, and simultaneous variations of different CR models in signal and \ttbar simulations are considered.
The largest observed shift is quoted as the systematic uncertainty.}
\item{\textit{Flavor-dependent JES:} The Lund string fragmentation model of \PYTHIA 8~\cite{Sjostrand:2014zea} is compared with the cluster fragmentation of \HERWIGpp~\cite{Bahr:2008pv}.
Each model relies on a large set of tuning parameters that allows one to modify the individual fragmentation of jets initiated from gluons and light, charm, and bottom quarks.
Therefore, the difference in JES between \PYTHIA and \HERWIG is determined for each jet flavor~\cite{Chatrchyan:2011ds}.
The flavor uncertainties for jets from gluons and light, charm, and bottom quarks are evaluated separately and added linearly~\cite{Sirunyan:2018gqx} since these individual contributions are treated to be fully correlated.
This method ensures that the relevant systematic sources are varied simultaneously for all jet flavors.}
\item{\textit{\PQb quark hadronization:} This term accounts for flavor-dependent uncertainties arising from the simulation of parton fragmentation.
The fragmentation of \PQb quarks into \PQb hadrons is varied in simulation within uncertainties in the Bowler--Lund fragmentation function tuned to the ALEPH~\cite{Heister:2001jg}, DELPHI~\cite{DELPHI:2011aa}, and SLD~\cite{SLD:2002poq} data.
In addition, the difference between the Bowler--Lund~\cite{Bowler:1981sb} and Peterson~\cite{Peterson:1982ak} fragmentation functions is included in the uncertainty.
Lastly, the uncertainty due to the semileptonic \PQb hadron branching fraction is obtained by varying it by $-0.45$\% and $+0.77$\%, which is the range of measurements from $\PB^{0}$/$\PB^{\pm}$ decays and their uncertainties~\cite{Zyla:2020zbs}.}
\item{\textit{Signal modeling:} To determine the influence of possible mismodeling of the $t$-channel single top quark process, several sources are considered as listed below.
\begin{enumerate}
\item{\textit{Parton shower (PS) scale:} We compare the nominal signal shape with reweighted shapes obtained by using per-event weights corresponding to independent variations of initial- and final-state radiation (ISR and FSR) scales by a factor of 2 and $1/2$, respectively.
During estimation of the related uncertainties, the ISR scale is kept fixed at the nominal value while the FSR scale is varied and vice-versa.
The uncertainty is estimated from the difference in the fit results using reweighted shapes relative to the nominal one.}
\item{\textit{$\mu_{\mathrm{R}}$ and $\mu_{\mathrm{F}}$ scales:} The impacts of varying the $\mu_{\mathrm{R}}$ and $\mu_{\mathrm{F}}$ scales up and down by a factor of 2 relative to their respective nominal values (both set to $172.5\GeV$) are considered by applying per-event weights~\cite{Kalogeropoulos:2018cke} on the $\zeta$ distributions.
Two cases are considered for the evaluation of related uncertainties.
In the first case, one scale is varied while the other is kept fixed to its nominal value; in the other case, both scales are varied together in the same direction with respect to their nominal values.
The resulting uncertainties from each case are added in quadrature and quoted as the uncertainty due to the $\mu_{\mathrm{R}}$ and $\mu_{\mathrm{F}}$ scales.}
\item{\textit{PDF+\alpS:} The impact due to the choice of PDFs is studied using reweighted shapes that are derived from replicas of the NNPDF 3.0 NLO ($\alpS = 0.118$) PDF set~\cite{Ball:2014uwa}.
In addition, NNPDF3.0 sets with $\alpS = 0.117$ and 0.119 are evaluated and the observed difference is added in quadrature.}
\end{enumerate}
}
\item{\textit{\ttbar modeling:} The impacts due to variation of the ISR and FSR scales, the $h_{\text{damp}}$ parameter responsible for ME-PS matching~\cite{CMS-PAS-TOP-16-021} (where ME is the acronym for matrix element), the $\mu_{\mathrm{R}}$ and $\mu_{\mathrm{F}}$ scales, and PDF+\alpS in the \ttbar process are considered.
The uncertainties due to the ISR and FSR scales, $\mu_{\mathrm{R}}$ and $\mu_{\mathrm{F}}$ scales, and PDF+\alpS variations in \ttbar events are estimated by following exactly the same method as for the signal events.
Additionally, variation of \alpS in the UE tune for the \ttbar simulation sample is considered in order to cover the difference between the UE models used for simulated \ttbar and all other processes, as discussed in Section~\ref{sec:simulation}.   
We also take into account the mismodeling of the top quark \pt spectrum, which is harder in simulated \ttbar events than in the data~\cite{Khachatryan:2015oqa}.
The uncertainties due to the aforementioned sources are determined from the difference in fit results obtained from the varied or reweighted \ttbar shapes corresponding to each source relative to the nominal one.
The contributions from individual sources are then added in quadrature to obtain the total uncertainty due to \ttbar modeling.}
\item{\textit{Parametric shapes:} The impact from varying the shape parameters of the signal and background models is considered as a separate systematic uncertainty.
The shape parameters are varied by three standard deviations about their estimated values, derived using simulation.
The impacts due to individual sources are summed in quadrature to obtain the total uncertainty due to the parametric modeling of the signal and backgrounds.}
\end{itemize}

Table~\ref{tab:syst} summarizes the aforementioned sources of systematic uncertainty and their contributions.
The impacts due to alternative ME (\MGvATNLO), FS (5FS), PS (\HERWIGpp), and UE (CUETP8M2T4) modeling of the signal process are also evaluated for the $l^{\pm}$ final state using dedicated simulated event samples.
Their individual contributions range between $-0.36$ and $+0.16\GeV$.
As these values are covered by the total systematic uncertainty listed in Table~\ref{tab:syst}, no additional uncertainty is assigned to the measured \mtop value due to these sources.
\begin{table*}
\centering
\caption{\label{tab:syst} Summary of the \mtop uncertainties in \GeV for each final-state lepton charge configuration.
The statistical uncertainties are obtained by performing the fits again in each case while fixing the nuisance parameters to their estimated values from data. 
With the exception of the flavor-dependent JES sources, the total systematic uncertainty is obtained from the quadrature sum of the individual systematic sources.
The amount of statistical fluctuations in the systematic shifts are quoted within parentheses whenever alternative simulated samples with systematic variations have been used.
These are determined from 1000 pseudo-experiments in each case.
Entries with ${<}{ 0.01}$ denote that the magnitude of the systematic bias is less than $0.01$.}
\resizebox{\textwidth}{!}{
\begin{tabular}{ccccc}
Source &  & $\delta m_{l^{\pm}}$ & $\delta m_{l^{+}}$ & $\delta m_{l^{-}}$ \\
\hline
Statistical & & $\pm 0.19$ & $\pm 0.23$ & $\pm 0.33$ \\ [\cmsTabSkip]
Statistical + profiled systematic  & & $\pm 0.32$ & $\pm 0.37$ & $\pm 0.58$ \\ [\cmsTabSkip]
\multirow{4}{*}{JES} & Correlation group intercalibration & $\pm 0.09$ & $\pm 0.07$ & $\pm 0.12$  \\
 & Correlation group MPFInSitu & $\pm 0.02$ & $\pm 0.02$ & $\pm 0.01$ \\
 & Correlation group uncorrelated & $\pm 0.39$ & $\pm 0.17$ & $\pm 0.83$\\ [\cmsTabSkip]
 & Total (quadrature sum) & $\pm 0.40$ & $\pm 0.18 $ & $\pm 0.84$ \\ [\cmsTabSkip]
JER & & ${<}{ 0.01}$ & ${<}{ 0.01}$ & ${<}{ 0.01}$\\
Unclustered energy & & ${<}{ 0.01}$ & ${<}{ 0.01}$ & ${<}{ 0.01}$\\
Muon efficiencies & & ${<}{ 0.01}$ & ${<}{ 0.01}$ & ${<}{ 0.01}$ \\
Electron efficiencies & & $\pm 0.01$ & $\pm 0.01 $ & $\pm 0.01$\\
Pileup &  & $\pm 0.14$ & $\pm 0.04$ & $\pm 0.34$\\
\PQb tagging & & $\pm 0.20$ & $\pm 0.18 $ & $\pm 0.22$\\
QCD multijet background & & $\pm 0.02$ & $\pm 0.01$ & $\pm 0.02$ \\
Mass calibration &  & $\pm 0.11$ & $\pm 0.13$ & $\pm 0.20$ \\
Int. luminosity & & ${<}{ 0.01}$ & ${<}{ 0.01}$ & $\pm 0.01$\\ [\cmsTabSkip]
CR model and ERD & & $\pm 0.24\ (0.017)$ & $\pm 0.39\ (0.027)$ & $\pm 0.68\ (0.048)$ \\ [\cmsTabSkip]
\multirow{5}{*}{Flavor-dependent JES} & Gluon & $ + 0.52$ & $+ 0.75$ & $ -0.03$\\
 & Light quark (uds) & $ -0.18$ & $+0.18 $ & $ -0.23$ \\
 & Charm & $ + 0.01$ & $ +0.08 $ & $ +0.11 $ \\
 & Bottom & $ - 0.48$ & $ -0.29 $ & $ -0.31 $ \\ [\cmsTabSkip]
 & Total (linear sum) & $ -0.13 $ & $+0.72$ & $ -0.46 $ \\ [\cmsTabSkip]
\multirow{4}{*}{\PQb quark hadronization model} & \PQb frag. Bowler--Lund & $\pm 0.03$ & $\pm 0.06 $ & $\pm 0.08 $ \\
 & \PQb frag. Peterson & $+0.14 $ & $+0.11 $ & $+0.19 $ \\
 & Semileptonic \PQb hadron decays & $\pm 0.18$ & $\pm 0.17 $ & $\pm 0.19$ \\ [\cmsTabSkip]
 & Total (quadrature sum) &$+0.23$\, $-0.18$ & $+0.21$\, $-0.18 $ & $+0.28$\, $-0.21$ \\ [\cmsTabSkip]
\multirow{5}{*}{Signal modeling} & ISR & $\pm 0.01$ & $\pm 0.01 $ & ${<}{ 0.01}$ \\
 & FSR & $\pm 0.28$ & $\pm 0.31$ & $\pm 0.20$\\
 & $\mu_{\mathrm{R}}$ and $\mu_{\mathrm{F}}$ scales & $\pm 0.09$ & $\pm 0.13$ & $\pm 0.03$ \\
 & PDF+\alpS & $\pm 0.06$ & $\pm 0.06$ & $\pm 0.07$ \\ [\cmsTabSkip]
 & Total (quadrature sum) & $\pm 0.30$ & $\pm 0.34$ & $\pm 0.21$ \\  [\cmsTabSkip]
\multirow{8}{*}{\ttbar modeling} & ISR & $\pm 0.11\ (0.008)$ & $\pm 0.02\ (0.001)$ & $\pm 0.22\ (0.016)$ \\
 & FSR & $\pm 0.10\ (0.007)$ & $\pm 0.14\ (0.010)$ & $\pm 0.40\ (0.028)$ \\
 & ME-PS matching scale & $\pm 0.10\ (0.007)$ & $\pm 0.10\ (0.006)$ & $\pm 0.10\ (0.008)$ \\
 & $\mu_{\mathrm{R}}$ and $\mu_{\mathrm{F}}$ scales & $\pm 0.03$ & $\pm 0.03$ & $\pm 0.01$\\
 & PDF+\alpS & ${<}{ 0.01}$ & ${<}{ 0.01}$ & ${<}{ 0.01}$ \\
 & Top quark \pt reweighting & $-0.04$ & $-0.08$ & $-0.04$\\
 & UE & $\pm 0.07\ (0.005)$ & $\pm 0.04\ (0.003)$ & $\pm 0.17\ (0.012)$ \\ [\cmsTabSkip]
 & Total (quadrature sum) & $\pm 0.20$ & $+0.18$\ $-0.20$ & $\pm 0.50$ \\ [\cmsTabSkip]
\multirow{4}{*}{Parametric shapes} & Signal shape & $\pm 0.05$ & $\pm 0.03$ & $\pm 0.04$ \\
 & \ttbar bkg. shape & $\pm 0.07$ & $\pm 0.04$ & $\pm 0.05$ \\
 & EW bkg. shape & $\pm 0.03$ & $\pm 0.01$ & $\pm 0.02$ \\ [\cmsTabSkip]
 & Total (quadrature sum) & $\pm 0.09$ & $\pm 0.05$ & $\pm 0.07$ \\ [\cmsTabSkip]
Total externalized systematic & & $+0.69$\, $-0.71$ & $+0.97$\, $-0.65$ & $+1.32$\, $-1.39 $\\ [\cmsTabSkip]
Grand total & & $+0.76$\, $-0.77$ & $+1.04$\, $-0.75$ & $+1.44$\, $-1.51$\\
\end{tabular}}
\end{table*}

In the case of the signal process, top quarks are produced more abundantly relative to their antiquark partners due to the charge asymmetry of the \PW boson radiated from the initial-state quark in \Pp{}\Pp collisions at the LHC. 
This leads to a higher relative background contamination in the $l^{-}$ final state arising from top antiquark decay compared to the $l^{+}$ final state from top quark decay, as shown in Fig.~\ref{fig:massfit}.
As a result, the measurement in the $l^{-}$ final state is more sensitive to the sources that significantly alter the background contributions along with the signal, compared to the ones that impact the signal contribution only.
This is reflected in Table~\ref{tab:syst} where the uncertainties from the signal modeling are lower for the $l^{-}$ case; whereas other sources, except for the ones listed under flavor-dependent JES, that alter the background contributions along with the signal have a larger impact on the total uncertainty.
In the case of the flavor-dependent JES uncertainty sources, the uncertainty is primarily dictated by the untagged jet.
The presence of a light quark in the final state is a salient feature of the signal process.
This light quark, or the FSR gluon radiated from it, is often detected as the untagged jet in the endcap region.
The untagged-jet kinematic properties are heavily exploited by the BDTs in order to achieve a better separation between the signal and backgrounds.
Hence, it has a large impact on the final acceptance, as well as on the $\zeta$ shapes obtained after the BDT selection.
The energy calibration of the endcap detector is known to have larger uncertainties compared to the barrel and hence it has a larger impact on the untagged jet found in the endcap region.
In the $l^{-}$ final state, the relative contribution of the flavor-dependent JES uncertainty is smaller owing to a lower signal-to-background ratio.

\section{\label{sec:result} Results}

The \mtop value is measured with events dominated by $t$-channel single top quark process, inclusive of the lepton charge in the final state, as
\begin{linenomath}
\begin{equation}
\label{eq:mass_incl}
\mtop = 172.13 \pm 0.32\statprof {}^{+0.69}_{-0.71}\ (\text{ext})\GeV = 172.13^{+0.76}_{-0.77}\GeV.
\end{equation}
\end{linenomath} 
The masses of the top quark and antiquark are determined separately by requiring positively and negatively charged leptons in the final state, respectively.
We find
\begin{linenomath}
\begin{eqnarray}
\label{eq:mass_top}
\mtop = 172.62 \pm 0.37\statprof {}^{+0.97}_{-0.65}\ (\text{ext})\GeV = 172.62^{+1.04}_{-0.75}\GeV, \\
\label{eq:mass_antitop}
m_{\PAQt} = 171.79 \pm 0.58\statprof {}^{+1.32}_{-1.39}\ (\text{ext})\GeV = 171.79^{+1.44}_{-1.51}\GeV.
\end{eqnarray}
\end{linenomath} 

\begin{figure*}
\centering
\includegraphics[width=0.6\textwidth]{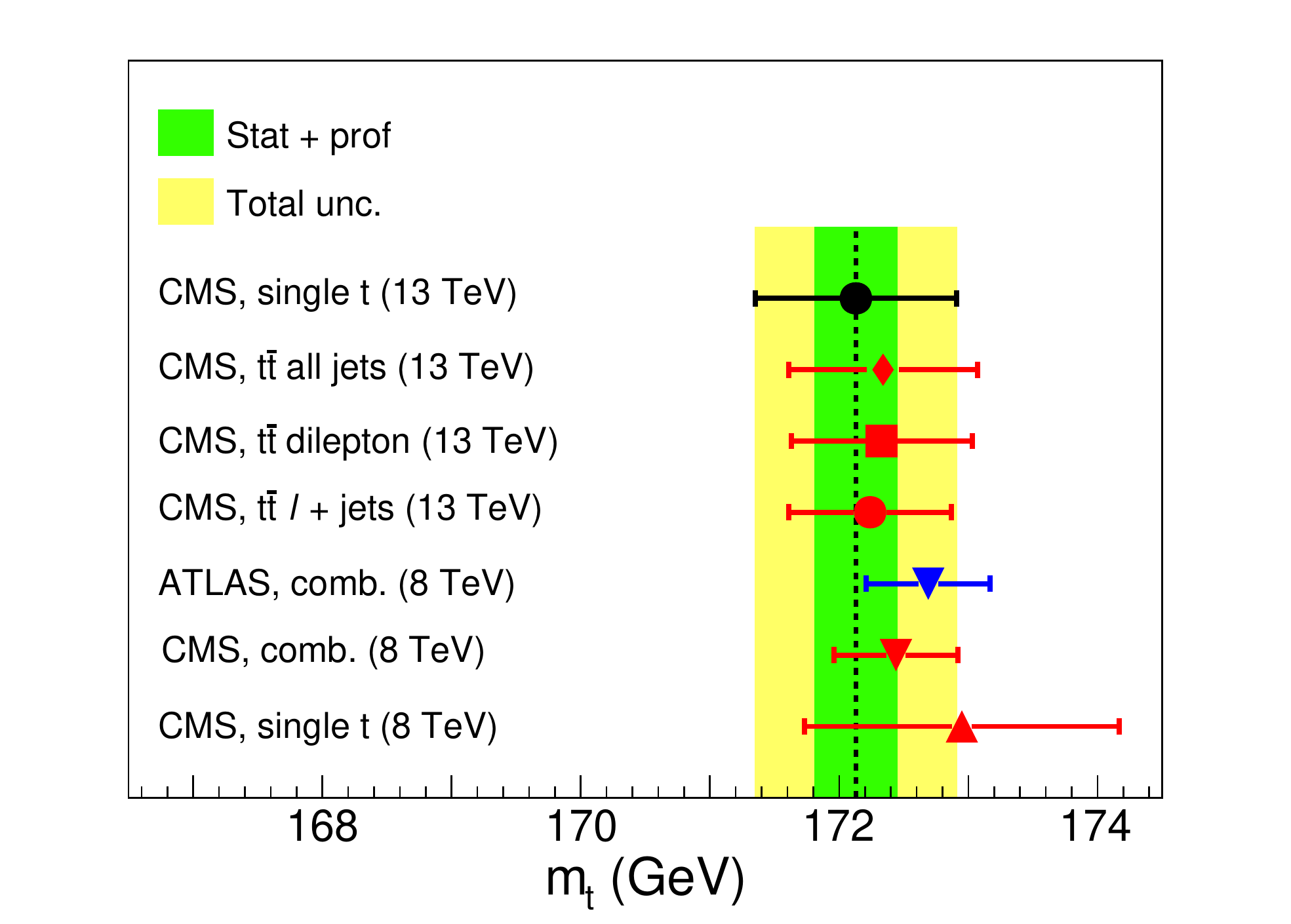}\\
\caption{\label{fig:mt_Comp} A comparison of measured \mtop values from this analysis (black circle), from previous CMS results in \ttbar events at $\sqrt{s} = 13\TeV$ for fully hadronic~\cite{Sirunyan:2018mlv}, dileptonic~\cite{Sirunyan:2018goh}, and semileptonic~\cite{Sirunyan:2018gqx} final states, and from ATLAS~\cite{Aaboud:2018zbu} and CMS~\cite{Khachatryan:2015hba,CMS_topMass_SingleTop_8TeV} analyses at $\sqrt{s} = 8\TeV$.
The horizontal bars on the points show the combined statistical and systematic uncertainties in each measurement.
The vertical dashed black line indicates the central value obtained from this measurement in the $l^{\pm}$ final state.
The green band represents the combined statistical and profiled systematic uncertainties in the present result, whereas the yellow band shows the total uncertainty.}
\end{figure*}

The first uncertainty is the combination of the statistical and profiled systematic uncertainties, whereas the second denotes the uncertainty due to the externalized systematic sources.
The total uncertainty is obtained by adding the two values in quadrature. 
The measured masses of the top quark and antiquark are consistent with each other, as well as with the $l^{\pm}$ result, within uncertainties.
The measured value corresponding to the charge-inclusive final states agrees with previous measurements by the ATLAS~\cite{Aaboud:2018zbu} and CMS~\cite{Khachatryan:2015hba,Sirunyan:2018gqx,Sirunyan:2018mlv,Sirunyan:2018goh,CMS_topMass_SingleTop_8TeV} Collaborations at different center-of-mass energies using various final states, within the uncertainties.
Comparisons of the previous measurements and the result from this analysis are shown in Fig.~\ref{fig:mt_Comp}.

\begin{figure*}
\centering
\includegraphics[width=0.6\textwidth]{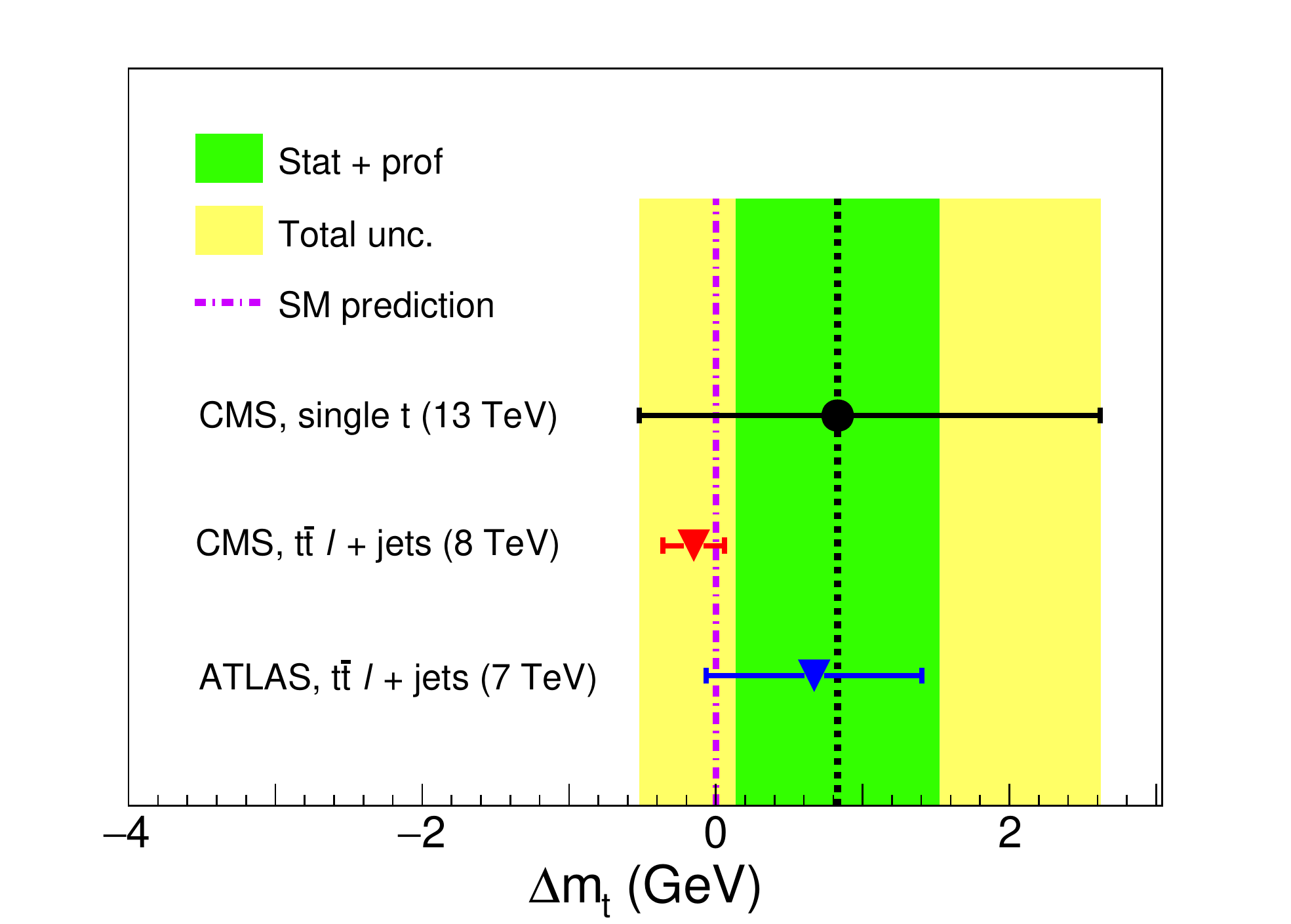}\\
\caption{\label{fig:delM_Comp} A comparison of the $\Delta\mtop$ measurement from this analysis (black circle) with the previous ATLAS~\cite{Aad:2013eva} and CMS~\cite{Chatrchyan:2016mqq} results  in \ttbar events at 7 and 8\TeV, respectively.
The horizontal bars on the points show the combined statistical and systematic uncertainties in each measurement.
The vertical dashed black line indicates the central value obtained from this measurement, and the vertical dash-dotted magenta line is the SM prediction.
The green band represents the combined statistical and profiled systematic uncertainties in the present result, whereas the yellow band shows the total uncertainty.}
\end{figure*}

The mass ratio of the top antiquark to quark is determined to be
\begin{linenomath}
\begin{equation}
\label{eq:ratio}
R_{\mtop} = \frac{m_{\PAQt}}{\mtop}=0.9952 \pm 0.0040\statprof {}^{+0.0068}_{-0.0096}\,(\text{ext}) = 0.9952^{+0.0079}_{-0.0104},
\end{equation}
\end{linenomath} 
and the mass difference between the top quark and antiquark is
\begin{linenomath}
\begin{equation}
\label{eq:diff}
\Delta\mtop = \mtop - m_{\PAQt} = 0.83 \pm 0.69\statprof {}^{+1.65}_{-1.16}\,(\text{ext})\GeV = 0.83^{+1.79}_{-1.35}\GeV.
\end{equation}
\end{linenomath}
The uncertainties in the mass ratio and difference are obtained by combining the uncertainties from the individual measurements, as indicated in Eqs.~(\ref{eq:mass_top}) and (\ref{eq:mass_antitop}).
Here, it is assumed that the combined statistical and profiled systematic uncertainties in the top quark and antiquark masses are uncorrelated, while the externalized systematic uncertainties are fully correlated source by source.
The estimated values of R$_{\mtop}$ and $\Delta \mtop$ are consistent with unity and zero, respectively, within uncertainties, showing no evidence for violation of $CPT$ invariance.
Figure~\ref{fig:delM_Comp} compares $\Delta \mtop$ from this analysis with the previous measurements by the ATLAS~\cite{Aad:2013eva} and CMS~\cite{Chatrchyan:2016mqq} Collaborations in \ttbar events at $\sqrt{s} = 7$ and $8\TeV$, respectively.
All results agree with each other and are consistent with zero, within the uncertainties.

The precision of the \mtop measurement presented here shows about a 30\% improvement over the previous CMS result~\cite{CMS_topMass_SingleTop_8TeV} from single top quark events. 
The inclusion of the electron final state improves the overall signal yield, thus reducing the statistical component of the total uncertainty. 
The MVA discriminant and the nuisance parameters in the ML fit constrain the background contamination to a level where the impact of dominant systematic uncertainty sources including JES can be brought under control. 
These improved strategies are responsible for reducing the overall uncertainty in the measured mass.
The statistical uncertainty plays a minor role in the achieved precision, which is limited by the systematic uncertainties due to JES, CR, and FSR modeling in the signal process.
A deeper understanding of these effects would be needed to further improve the precision.

\section{\label{sec:summary} Summary}

Measurements of the top quark and antiquark masses, as well as their ratio and difference, are performed using a data sample enriched with single top quark events produced in proton-proton collisions at $\sqrt{s} = 13\TeV$.
The analyzed data correspond to an integrated luminosity of 35.9\fbinv recorded by the CMS experiment at the LHC.
Events containing an isolated muon or electron and two jets, of which one is \PQb tagged, in the final state are used in the study.
From the inclusive measurement the top quark mass is found to be $172.13^{+0.76}_{-0.77}\GeV$, where the uncertainty includes both the statistical and systematic components.
The masses of the top quark and antiquark are separately determined as $172.62^{+1.04}_{-0.75}$ and $171.79^{+1.44}_{-1.51}\GeV$, respectively.
These quantities are used to determine the mass ratio of the top antiquark to top quark of $0.9952^{+0.0079}_{-0.0104}$, along with the difference between the top quark and antiquark masses of $0.83^{+1.79}_{-1.35}\GeV$, both for the first time in single top quark production.
The obtained mass ratio and difference agree with unity and zero, respectively, within the uncertainties, and are consistent with the conservation of $CPT$ symmetry.
This is the first measurement of the top quark mass in this particular final state to achieve a sub-GeV precision.

\begin{acknowledgments}
We congratulate our colleagues in the CERN accelerator departments for the excellent performance of the LHC and thank the technical and administrative staffs at CERN and at other CMS institutes for their contributions to the success of the CMS effort. In addition, we gratefully acknowledge the computing centers and personnel of the Worldwide LHC Computing Grid and other centers for delivering so effectively the computing infrastructure essential to our analyses. Finally, we acknowledge the enduring support for the construction and operation of the LHC, the CMS detector, and the supporting computing infrastructure provided by the following funding agencies: BMBWF and FWF (Austria); FNRS and FWO (Belgium); CNPq, CAPES, FAPERJ, FAPERGS, and FAPESP (Brazil); MES (Bulgaria); CERN; CAS, MoST, and NSFC (China); MINCIENCIAS (Colombia); MSES and CSF (Croatia); RIF (Cyprus); SENESCYT (Ecuador); MoER, ERC PUT and ERDF (Estonia); Academy of Finland, MEC, and HIP (Finland); CEA and CNRS/IN2P3 (France); BMBF, DFG, and HGF (Germany); GSRI (Greece); NKFIA (Hungary); DAE and DST (India); IPM (Iran); SFI (Ireland); INFN (Italy); MSIP and NRF (Republic of Korea); MES (Latvia); LAS (Lithuania); MOE and UM (Malaysia); BUAP, CINVESTAV, CONACYT, LNS, SEP, and UASLP-FAI (Mexico); MOS (Montenegro); MBIE (New Zealand); PAEC (Pakistan); MSHE and NSC (Poland); FCT (Portugal); JINR (Dubna); MON, RosAtom, RAS, RFBR, and NRC KI (Russia); MESTD (Serbia); SEIDI, CPAN, PCTI, and FEDER (Spain); MOSTR (Sri Lanka); Swiss Funding Agencies (Switzerland); MST (Taipei); ThEPCenter, IPST, STAR, and NSTDA (Thailand); TUBITAK and TAEK (Turkey); NASU (Ukraine); STFC (United Kingdom); DOE and NSF (USA).
    
\hyphenation{Rachada-pisek} Individuals have received support from the Marie-Curie program and the European Research Council and Horizon 2020 Grant, contract Nos.\ 675440, 724704, 752730, 758316, 765710, 824093, 884104, and COST Action CA16108 (European Union); the Leventis Foundation; the Alfred P.\ Sloan Foundation; the Alexander von Humboldt Foundation; the Belgian Federal Science Policy Office; the Fonds pour la Formation \`a la Recherche dans l'Industrie et dans l'Agriculture (FRIA-Belgium); the Agentschap voor Innovatie door Wetenschap en Technologie (IWT-Belgium); the F.R.S.-FNRS and FWO (Belgium) under the ``Excellence of Science -- EOS" -- be.h project n.\ 30820817; the Beijing Municipal Science \& Technology Commission, No. Z191100007219010; the Ministry of Education, Youth and Sports (MEYS) of the Czech Republic; the Deutsche Forschungsgemeinschaft (DFG), under Germany's Excellence Strategy -- EXC 2121 ``Quantum Universe" -- 390833306, and under project number 400140256 - GRK2497; the Lend\"ulet (``Momentum") Program and the J\'anos Bolyai Research Scholarship of the Hungarian Academy of Sciences, the New National Excellence Program \'UNKP, the NKFIA research grants 123842, 123959, 124845, 124850, 125105, 128713, 128786, and 129058 (Hungary); the Council of Science and Industrial Research, India; the Latvian Council of Science; the Ministry of Science and Higher Education and the National Science Center, contracts Opus 2014/15/B/ST2/03998 and 2015/19/B/ST2/02861 (Poland); the Funda\c{c}\~ao para a Ci\^encia e a Tecnologia, grant CEECIND/01334/2018 (Portugal); the National Priorities Research Program by Qatar National Research Fund; the Ministry of Science and Higher Education, project no. 14.W03.31.0026 (Russia); the Programa Estatal de Fomento de la Investigaci{\'o}n Cient{\'i}fica y T{\'e}cnica de Excelencia Mar\'{\i}a de Maeztu, grant MDM-2015-0509 and the Programa Severo Ochoa del Principado de Asturias; the Stavros Niarchos Foundation (Greece); the Rachadapisek Sompot Fund for Postdoctoral Fellowship, Chulalongkorn University and the Chulalongkorn Academic into Its 2nd Century Project Advancement Project (Thailand); the Kavli Foundation; the Nvidia Corporation; the SuperMicro Corporation; the Welch Foundation, contract C-1845; and the Weston Havens Foundation (USA).
\end{acknowledgments}

\bibliography{auto_generated}
\numberwithin{figure}{section}
\clearpage
\appendix
\section{\label{sec:suppl_MVA_Inputs} MVA input variables and their correlations}

\begin{figure*}[!htb]
\centering
\includegraphics[width=0.45\textwidth]{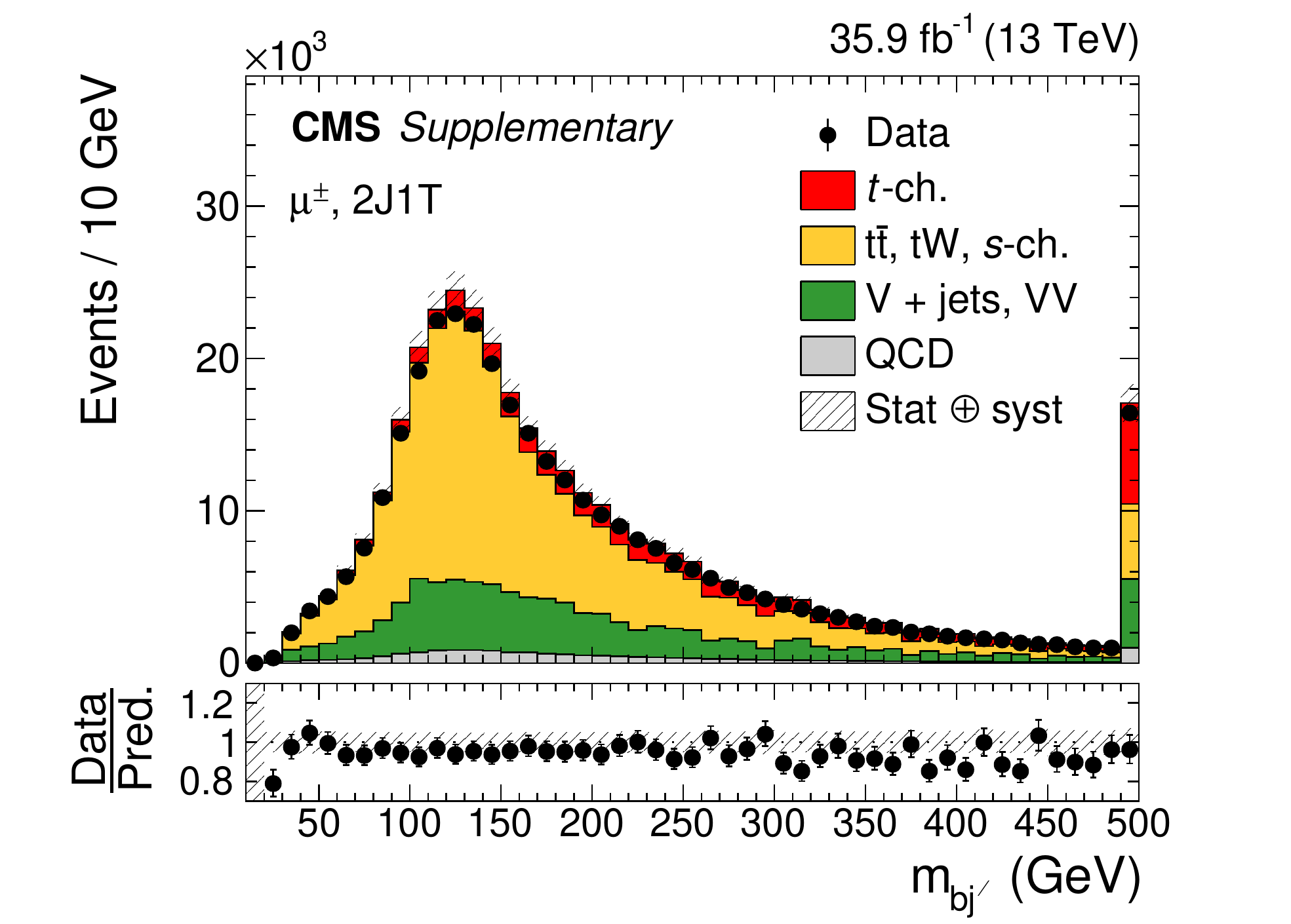}
\includegraphics[width=0.45\textwidth]{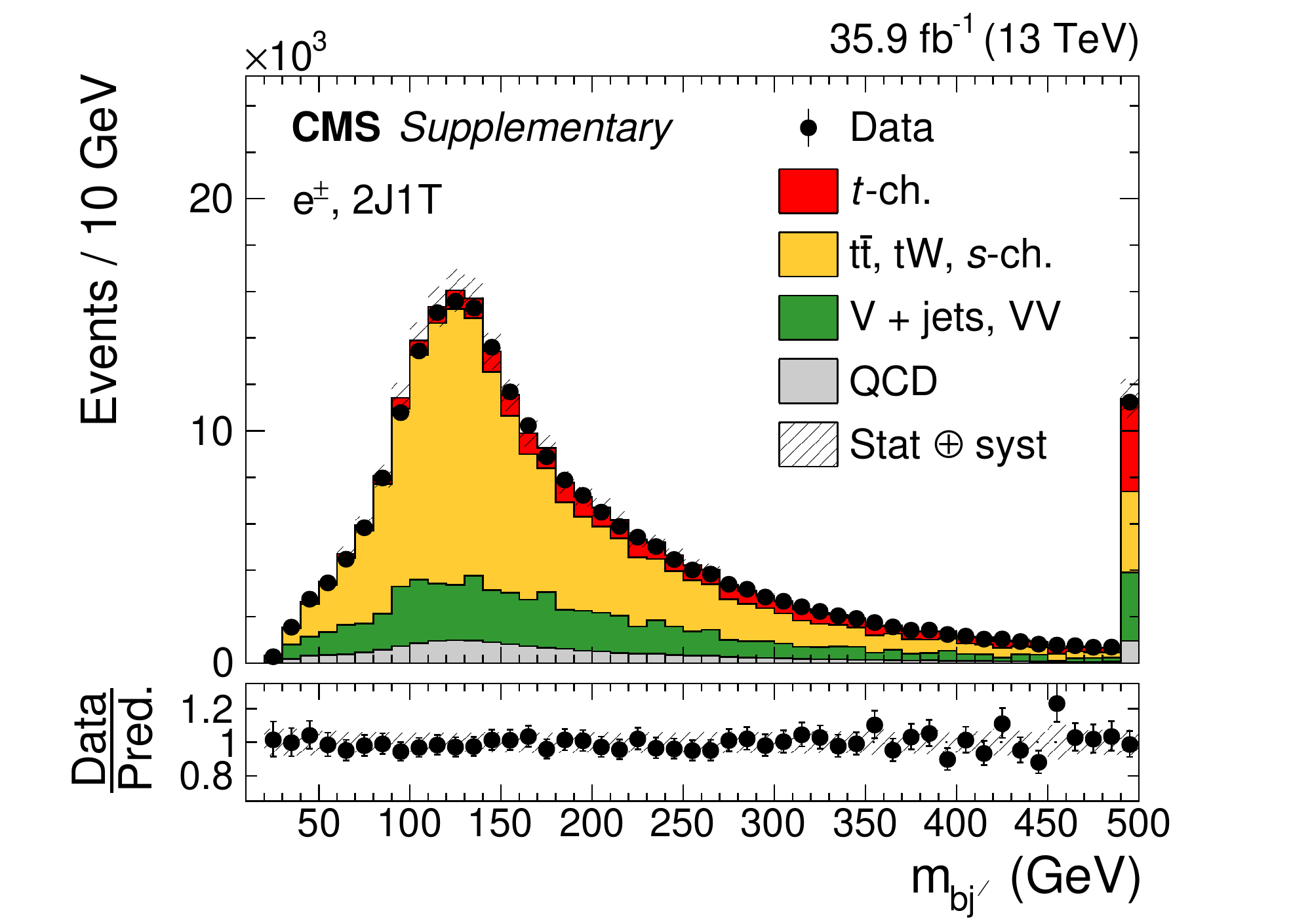}\\
\includegraphics[width=0.45\textwidth]{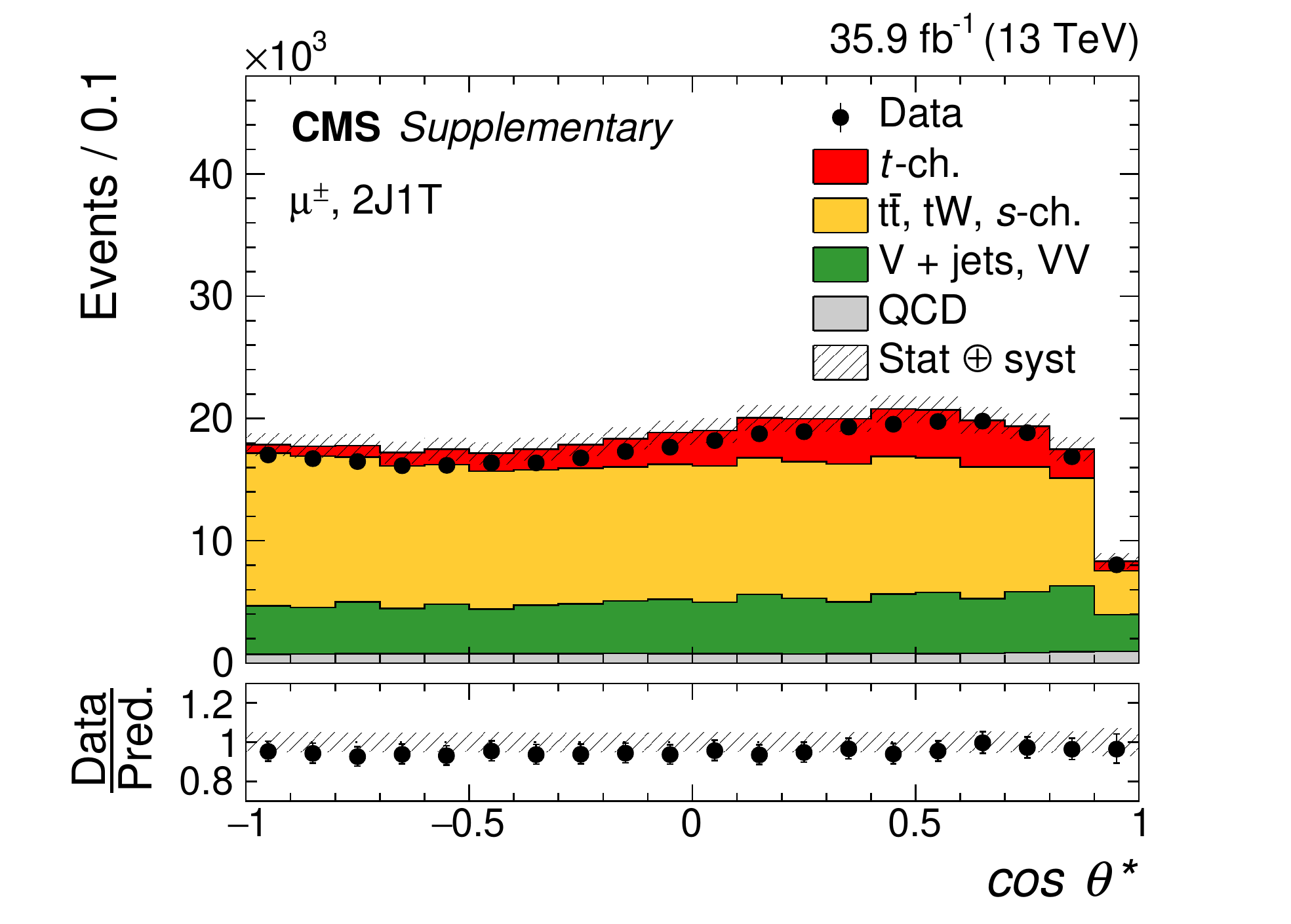}
\includegraphics[width=0.45\textwidth]{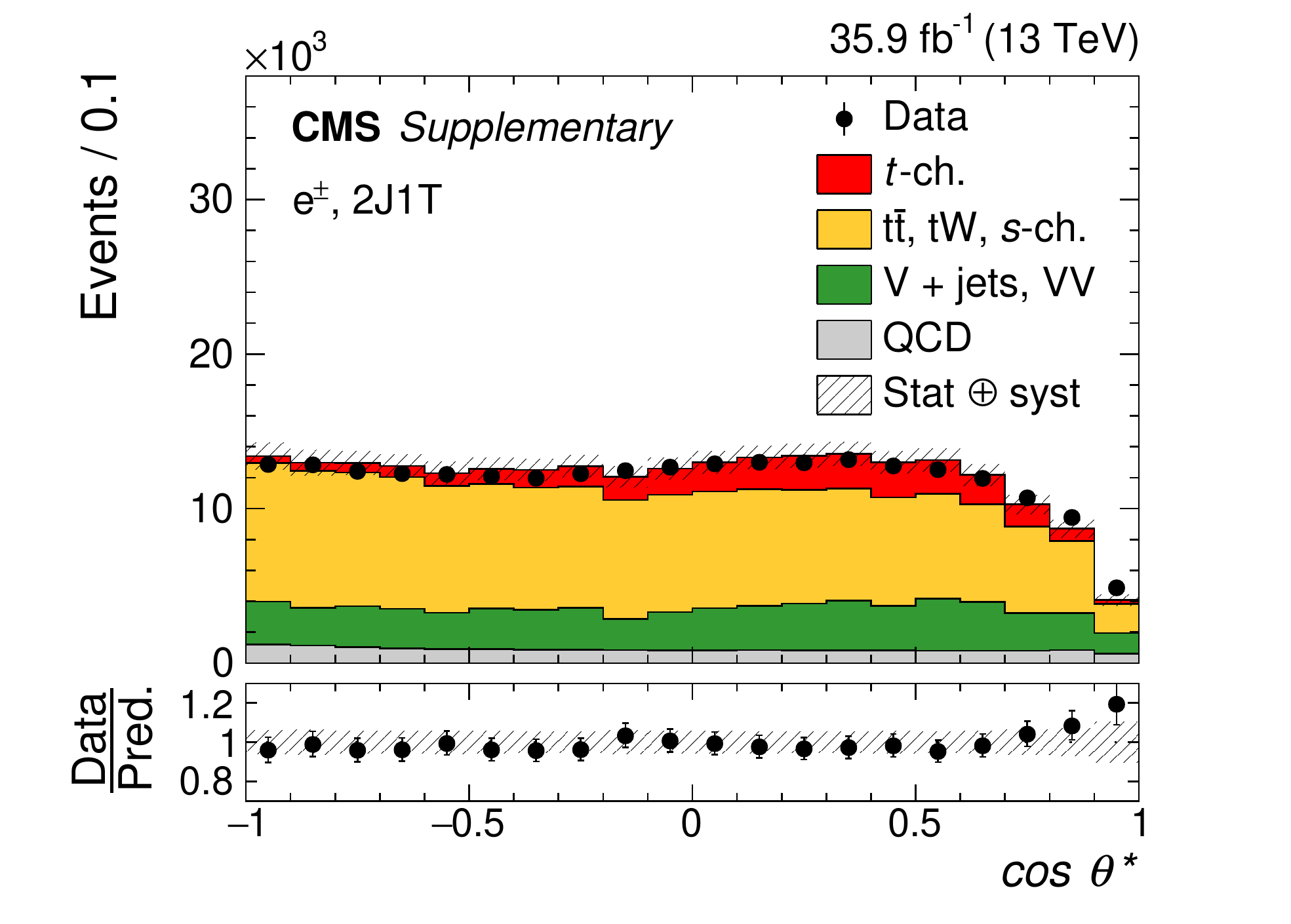}\\
\includegraphics[width=0.45\textwidth]{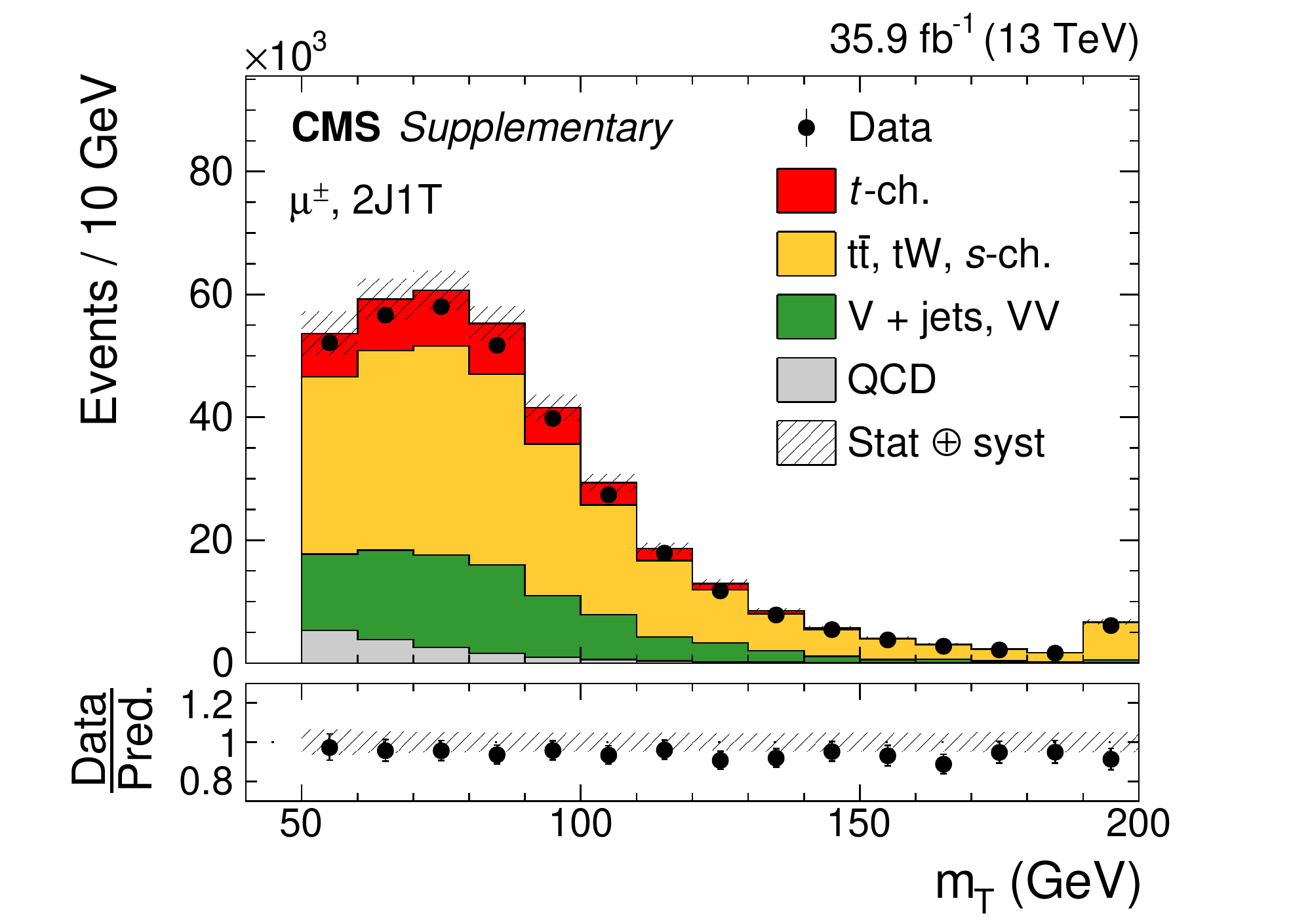}
\includegraphics[width=0.45\textwidth]{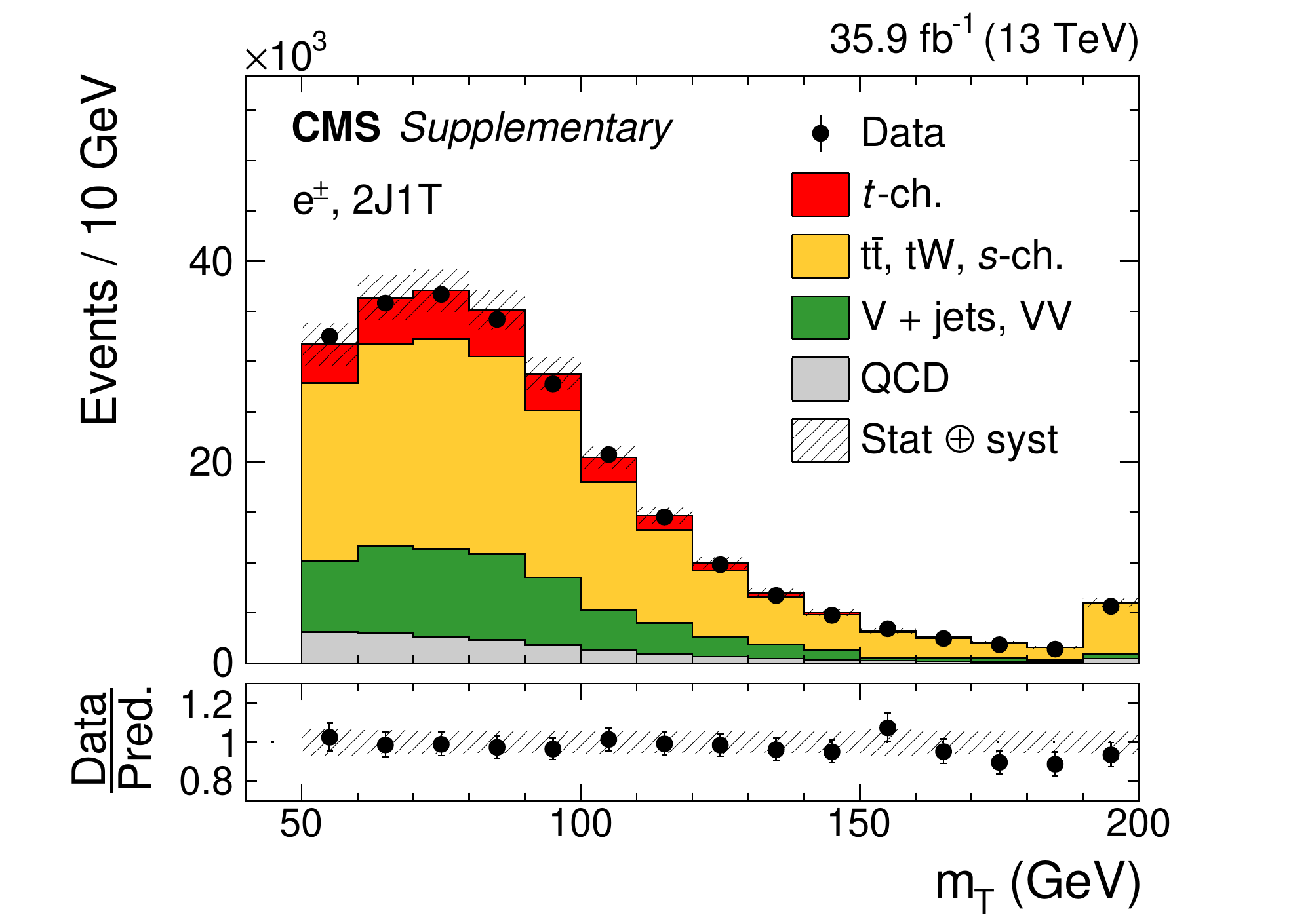}\\
\caption{\label{fig:inputVarsuppl1} Distributions of $m_{\PQb\mathrm{j^{\prime}}}$ (upper row), $\cos{\theta^{*}}$ (middle row), and \mT (lower row) for the muon (left) and electron (right) final states in the 2J1T category for data (points) and simulation (colored histograms).
The lower panel in each plot shows the ratio of the data to the predictions.
The bands indicate the statistical and systematic uncertainties added in quadrature.
The last bin in each of the upper- and lower-row plots includes the overflow.}
\end{figure*}

\begin{figure*}
\centering
\includegraphics[width=0.45\textwidth]{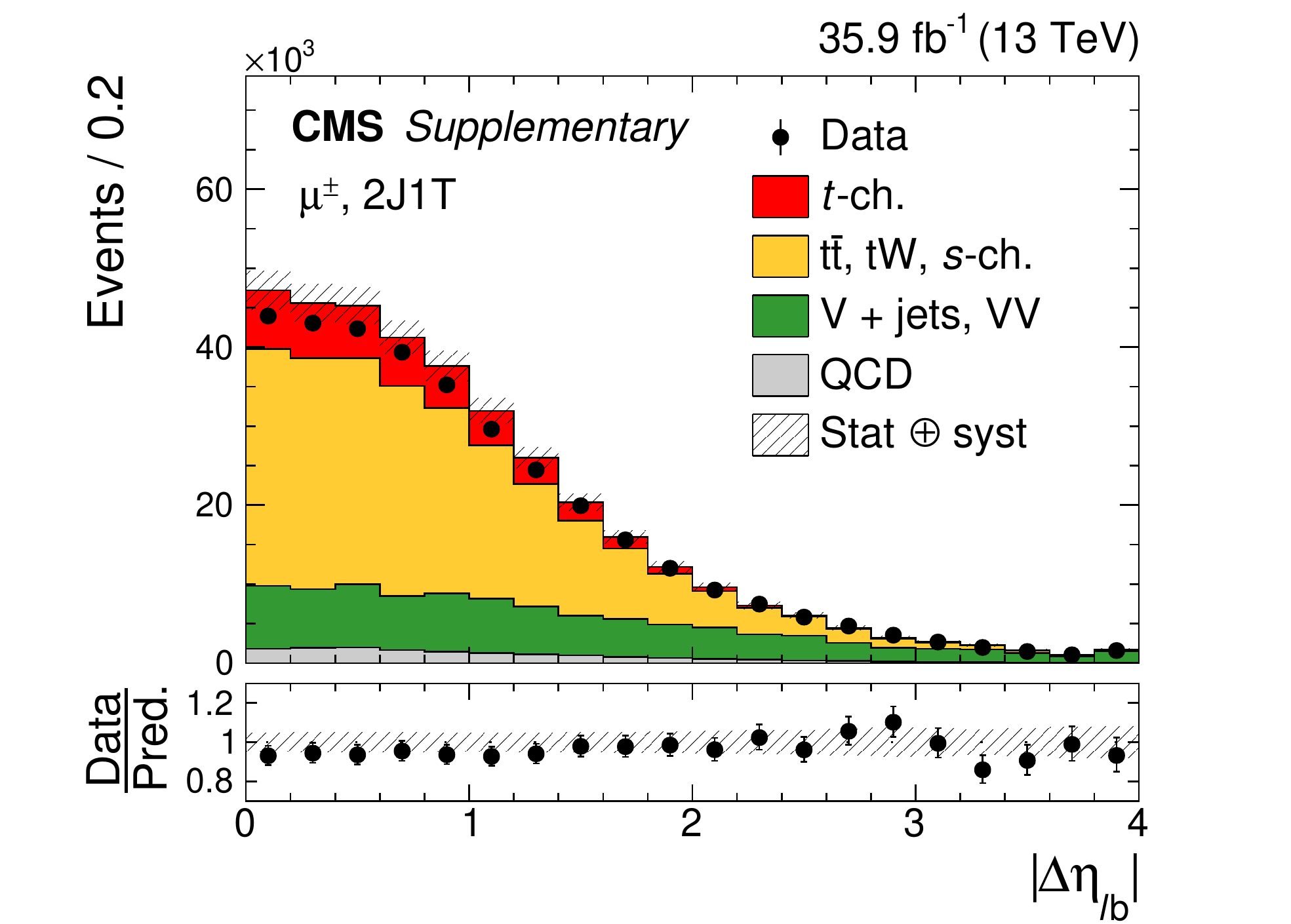}
\includegraphics[width=0.45\textwidth]{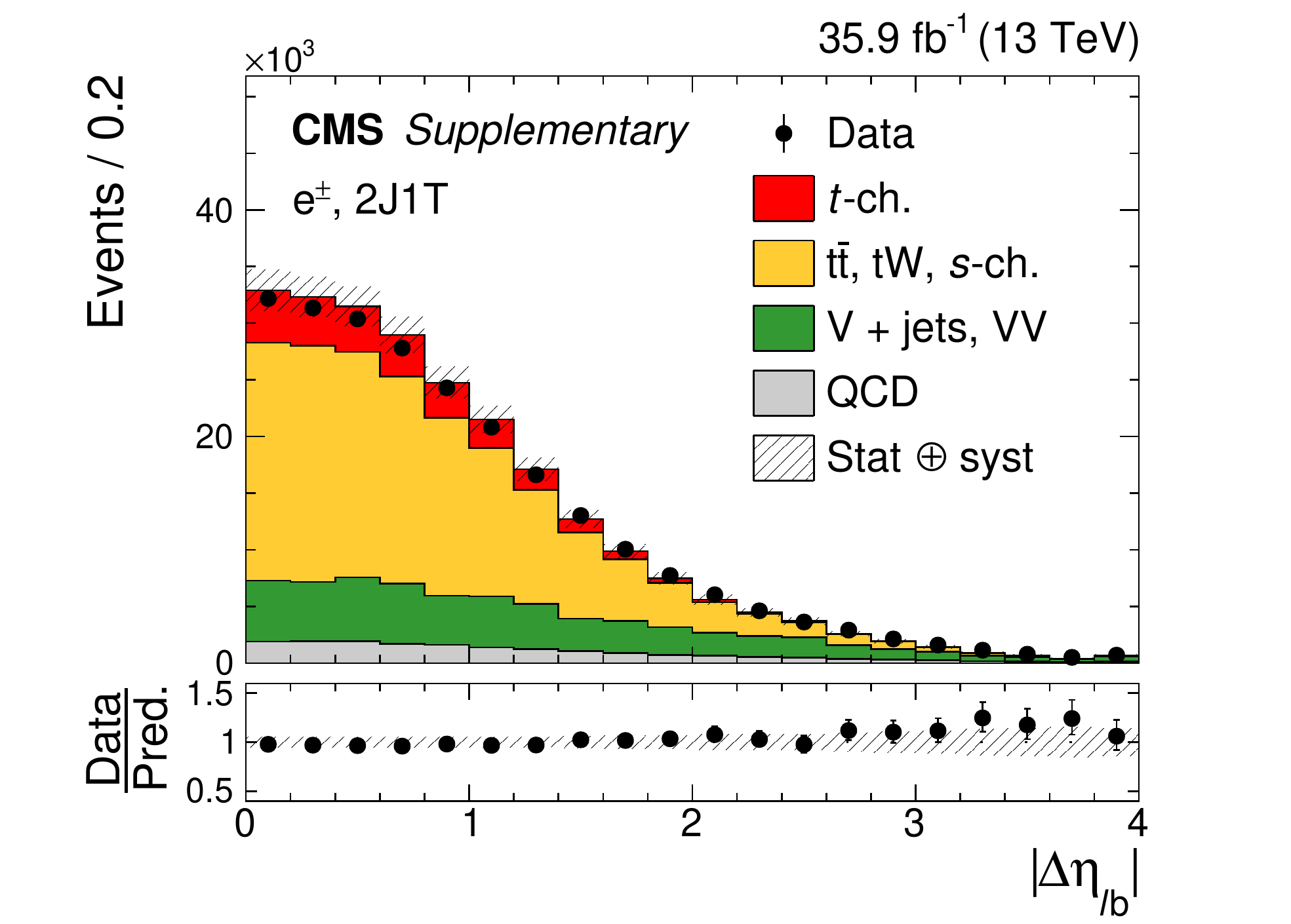}\\
\includegraphics[width=0.45\textwidth]{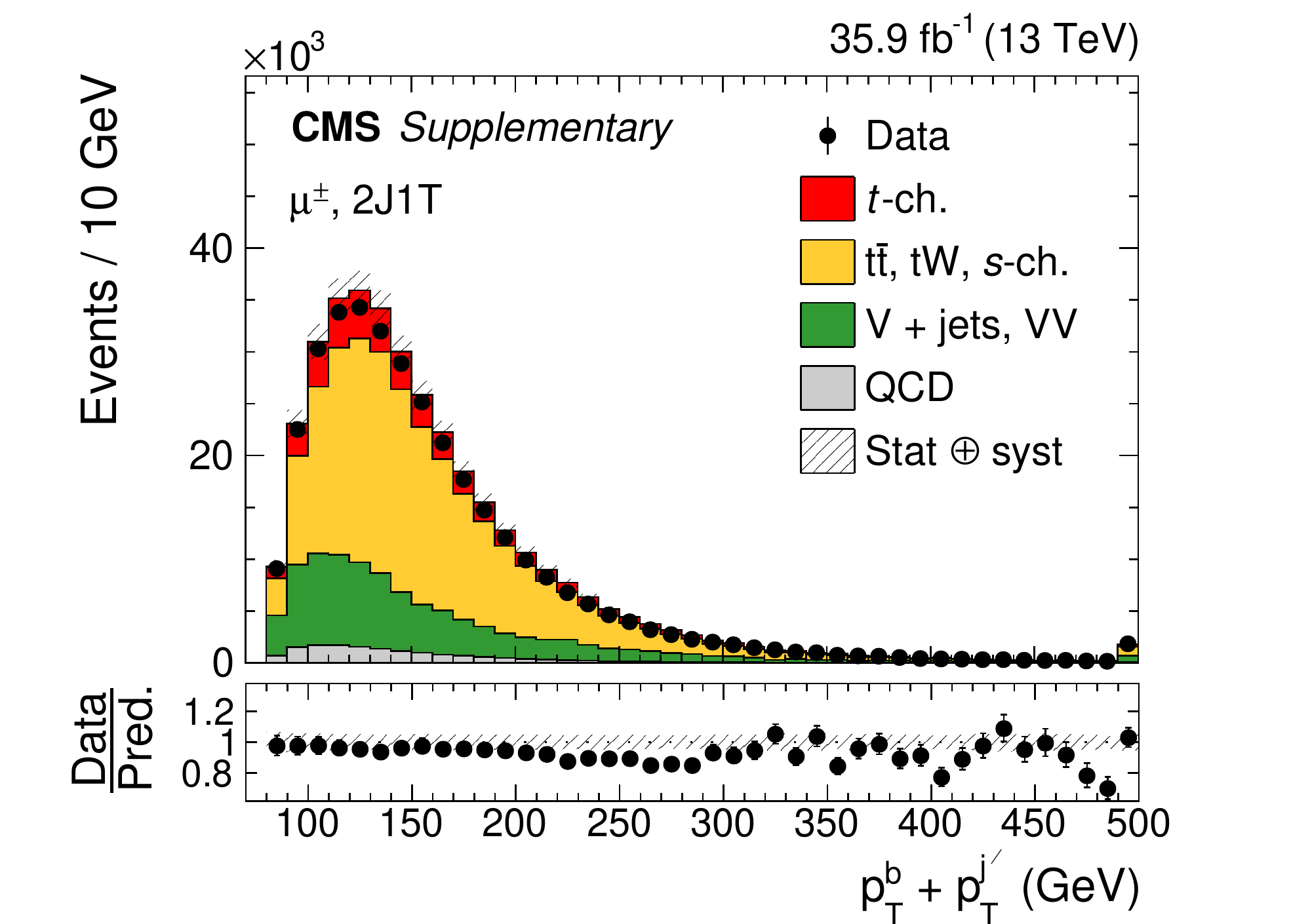}
\includegraphics[width=0.45\textwidth]{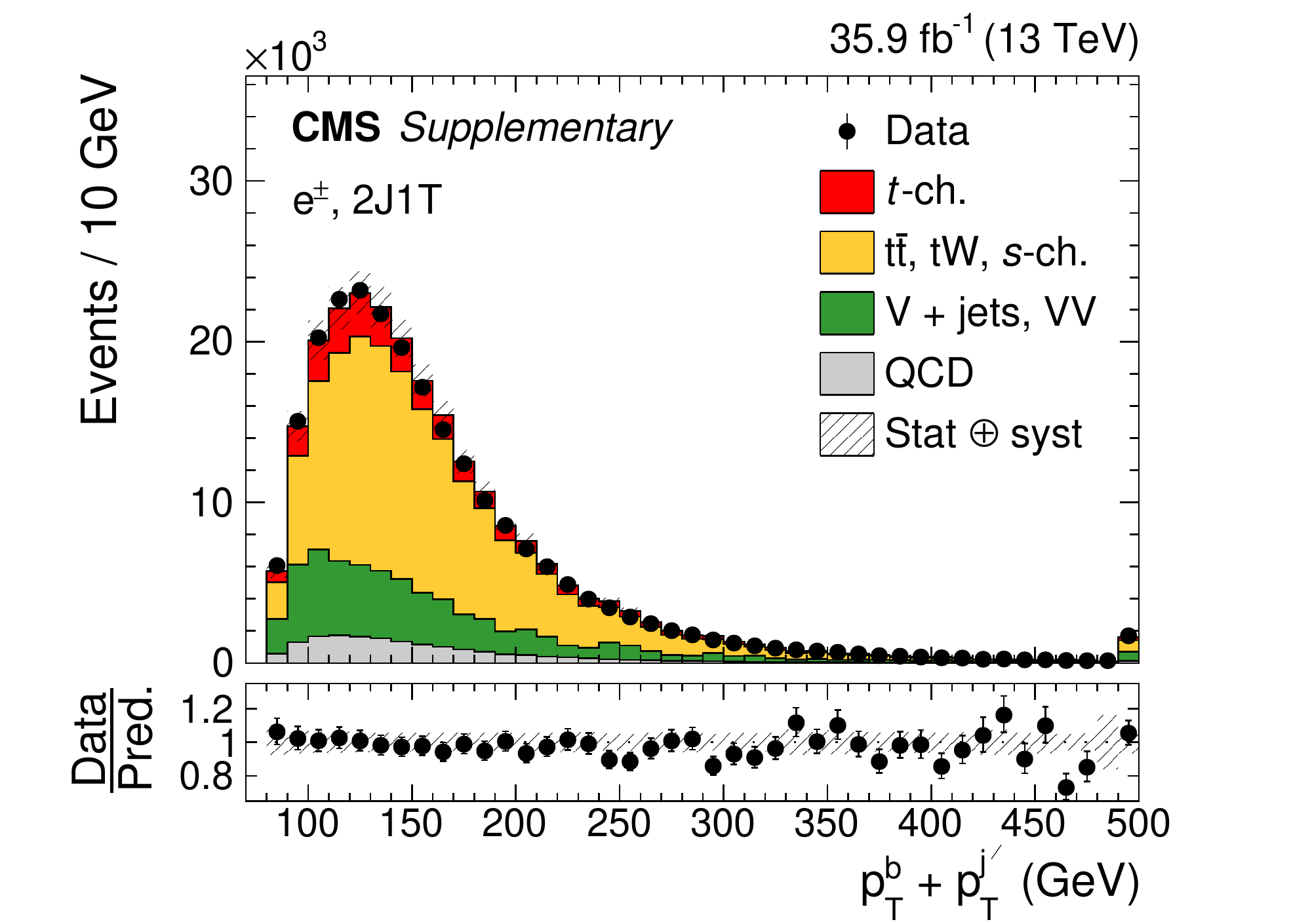}\\
\includegraphics[width=0.45\textwidth]{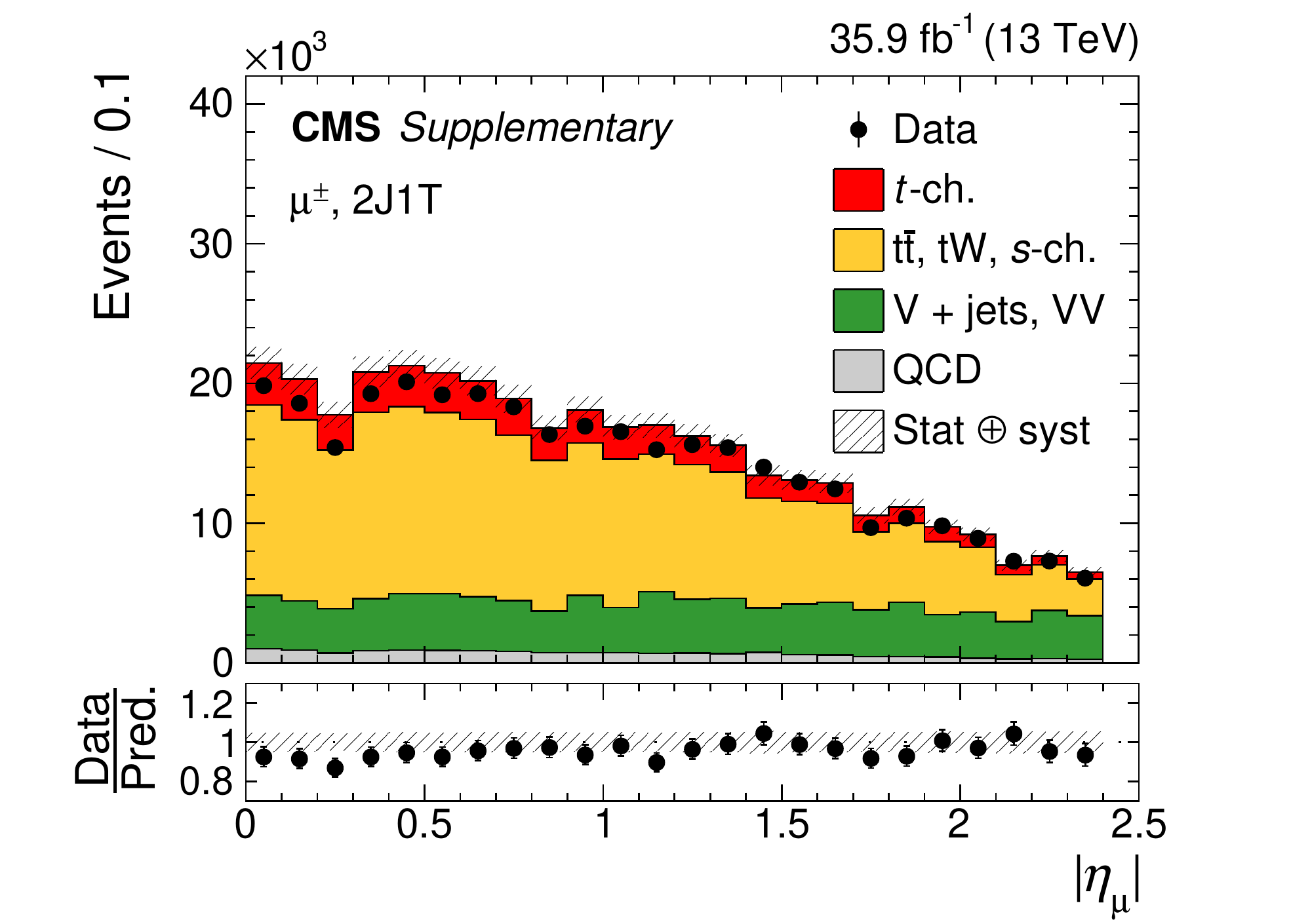}
\includegraphics[width=0.45\textwidth]{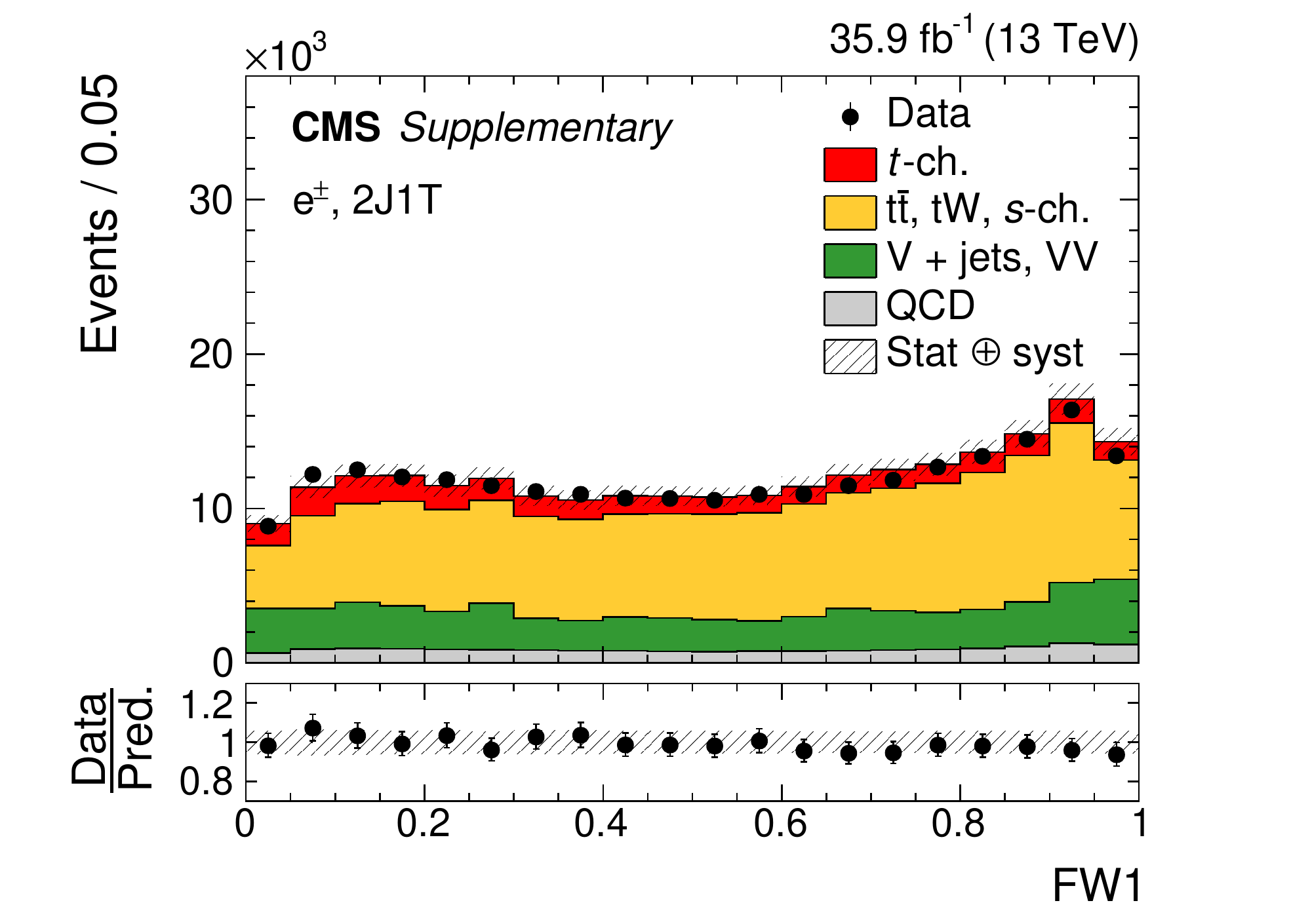}\\
\caption{\label{fig:inputVarsuppl2} Distributions of $\abs{\Delta\eta_{l\PQb}}$ (upper row) and $\pt^{\PQb} + \pt^{\mathrm{j}^{\prime}}$ (middle row) for the muon (left) and electron (right) final states in the 2J1T category for data (points) and simulation (colored histograms).
The lower row shows the similar data-to-simulation comparison for $\abs{\eta_{l}}$ (left) and FW1 (right) for the muon (left) and electron (right) final states in the 2J1T category.
The lower panel in each plot shows the ratio of the data to the predictions.
The bands indicate the statistical and systematic uncertainties added in quadrature.
The last bin in each plot except for the lower-right one includes the overflow.}
\end{figure*}

\begin{figure*}
\centering
\includegraphics[width=0.45\textwidth]{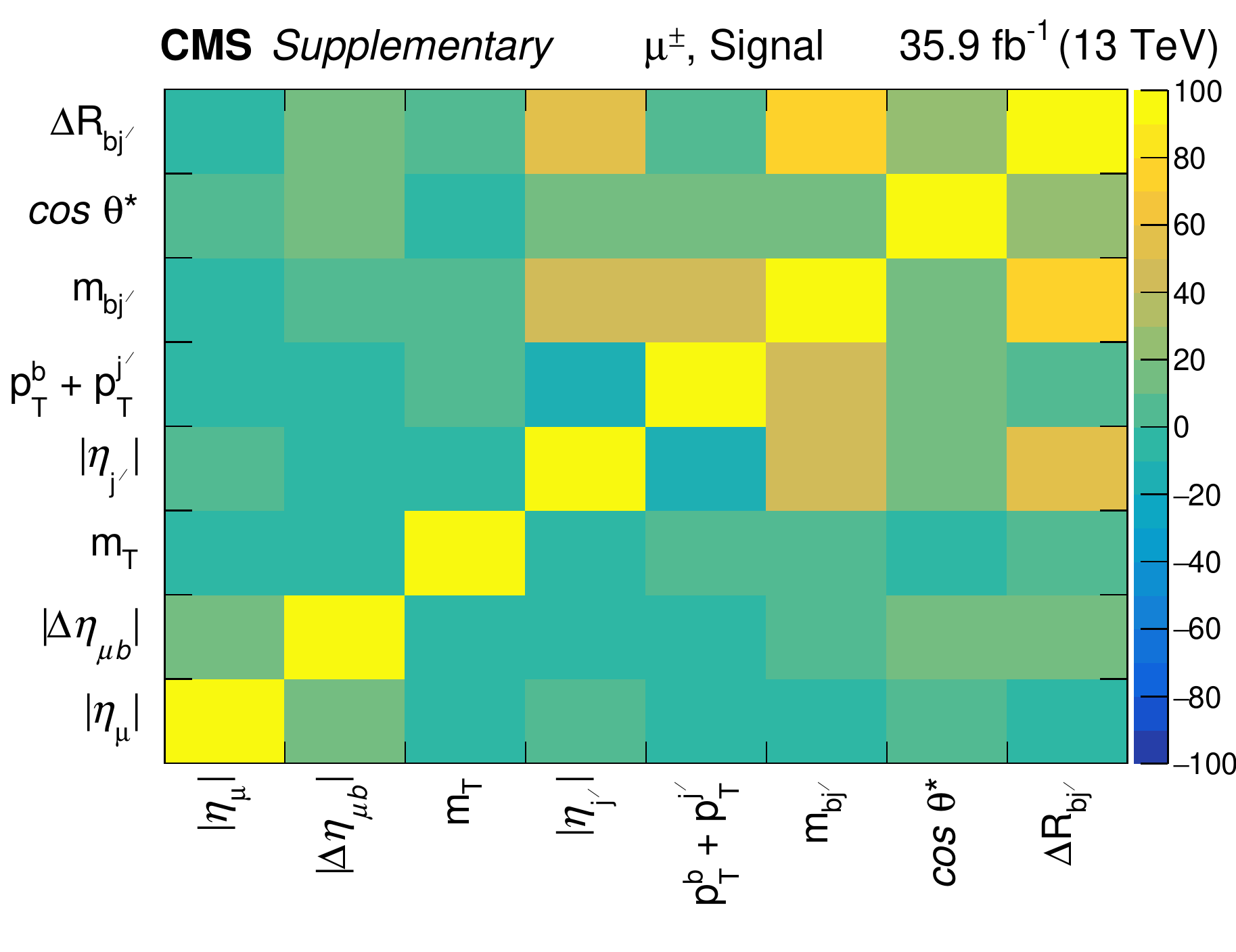}
\includegraphics[width=0.45\textwidth]{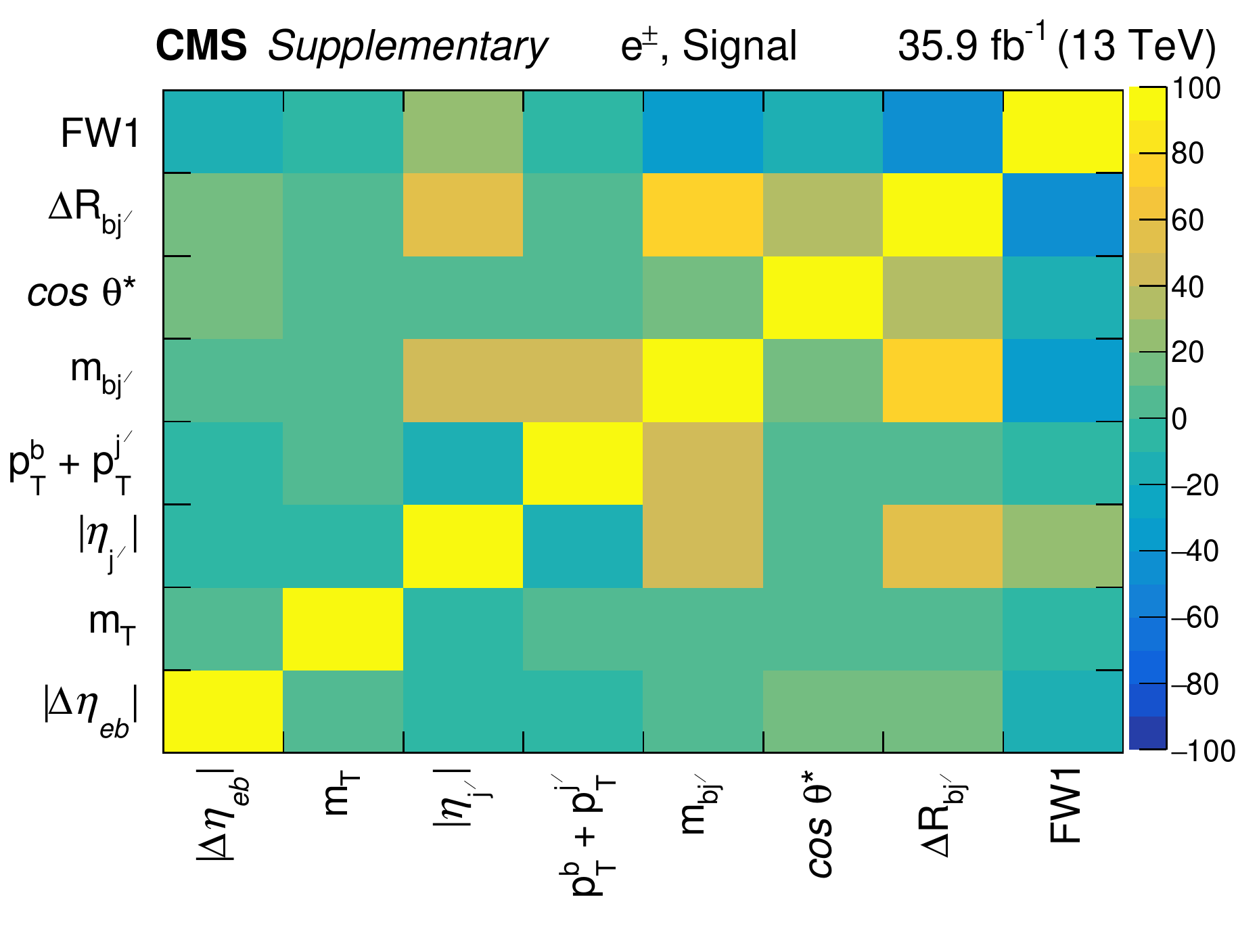}\\
\includegraphics[width=0.45\textwidth]{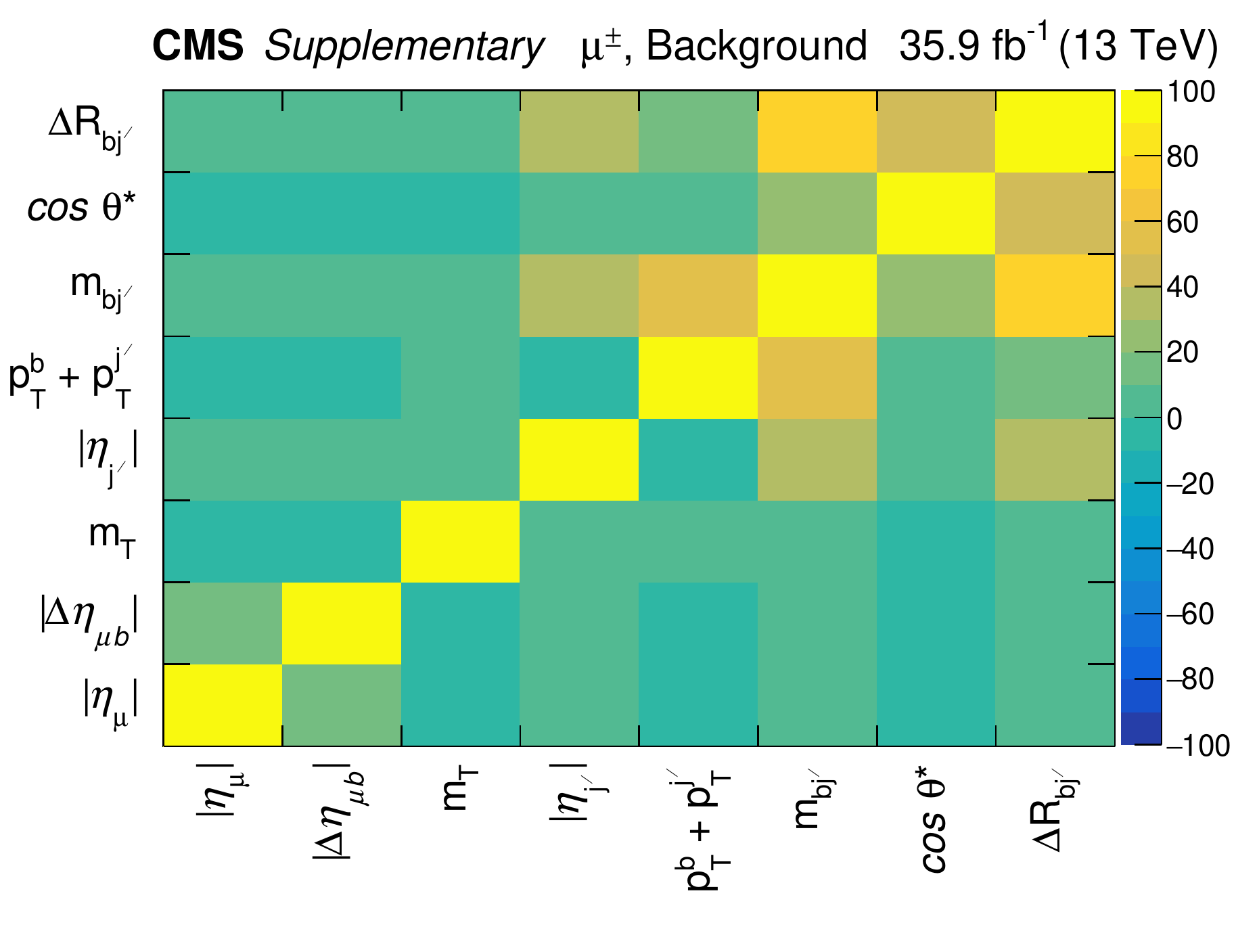}
\includegraphics[width=0.45\textwidth]{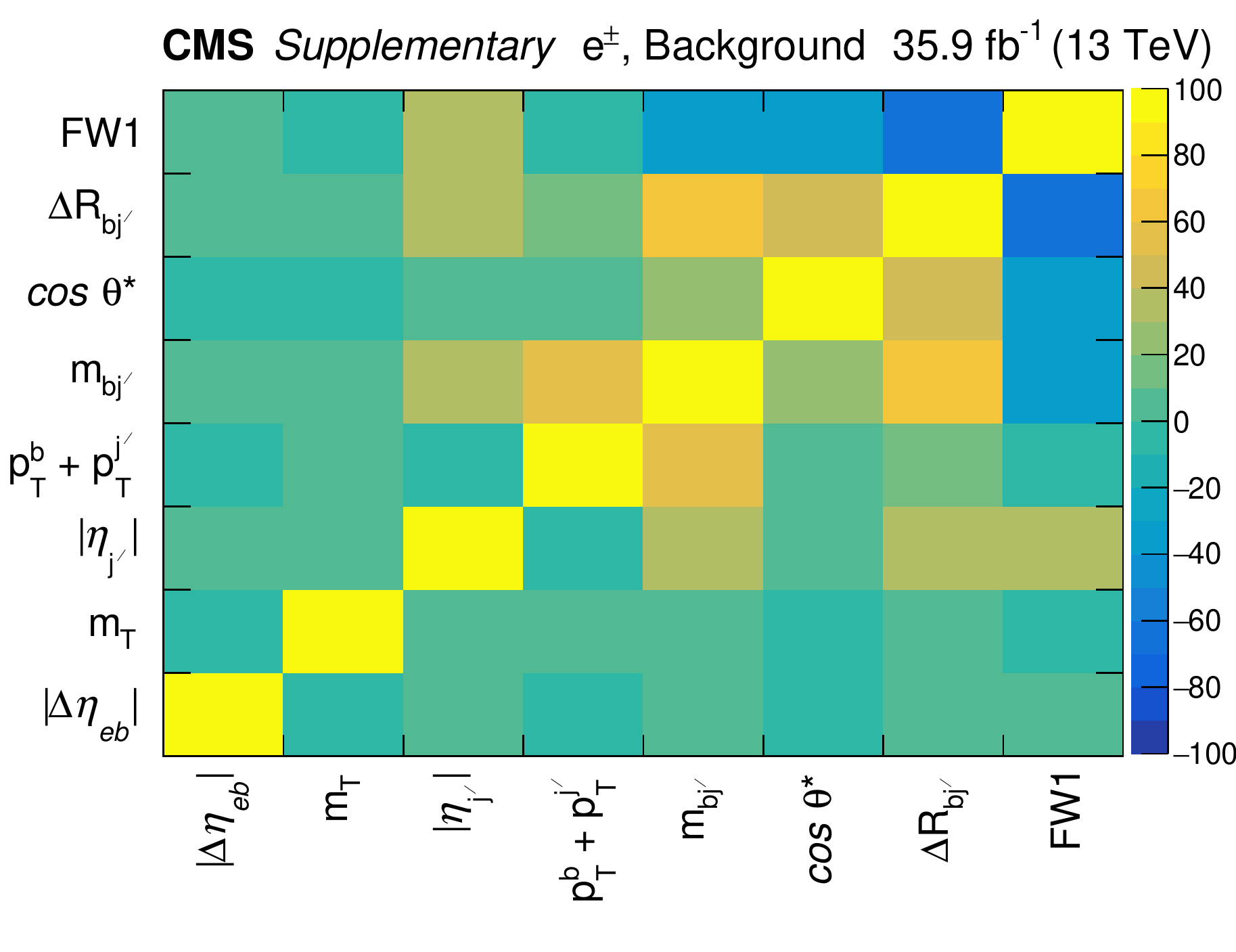}\\
\includegraphics[width=0.45\textwidth]{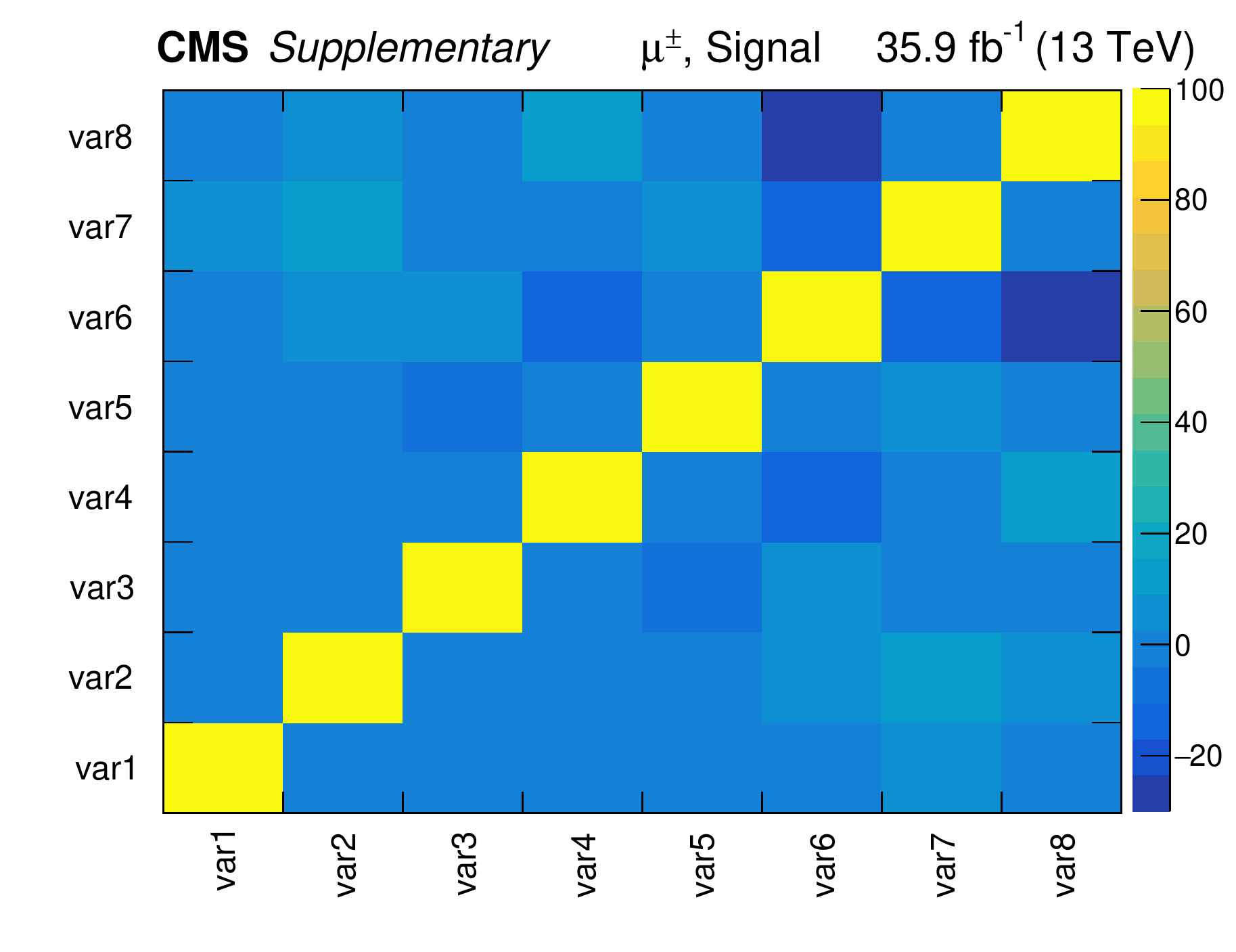}
\includegraphics[width=0.45\textwidth]{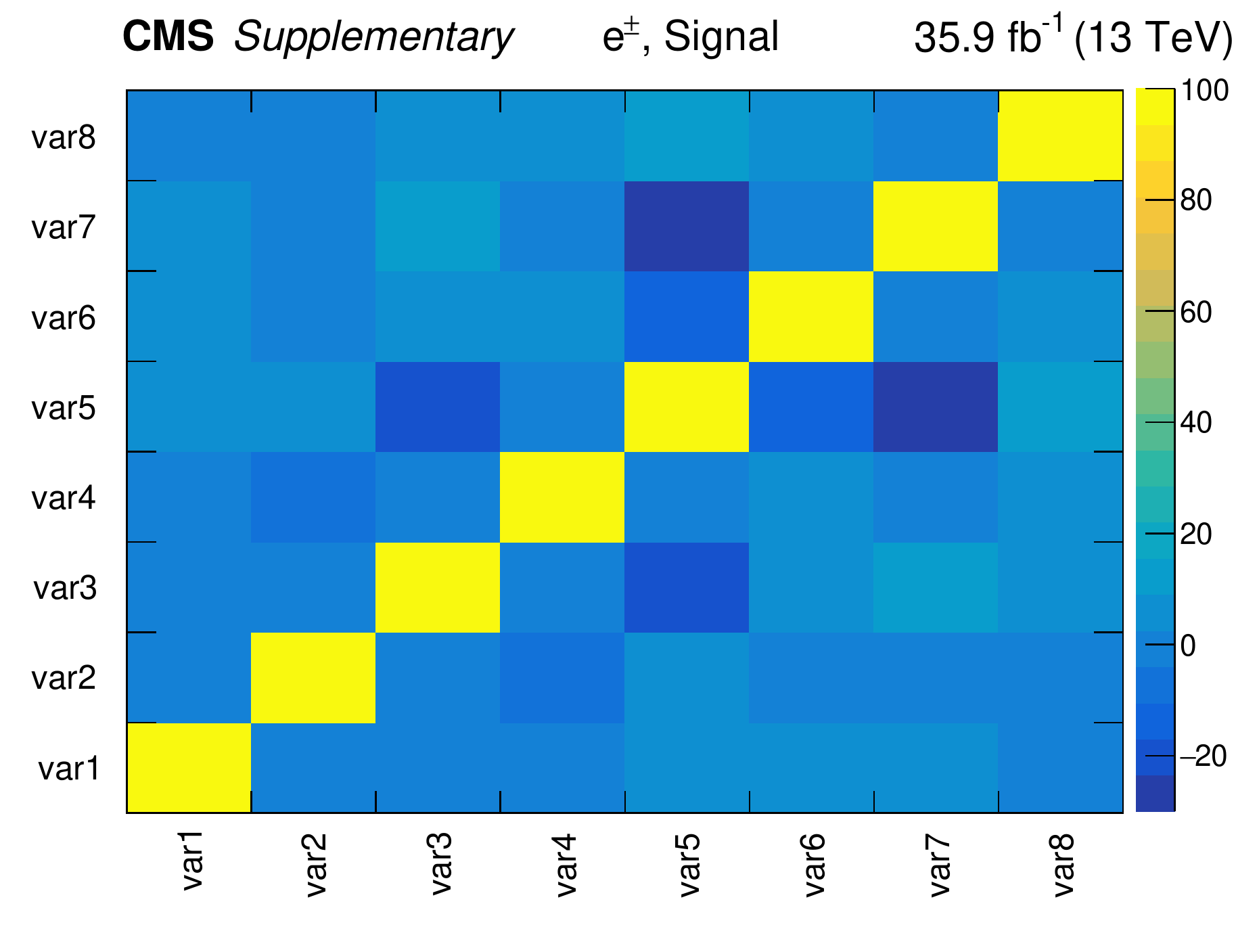}\\
\includegraphics[width=0.45\textwidth]{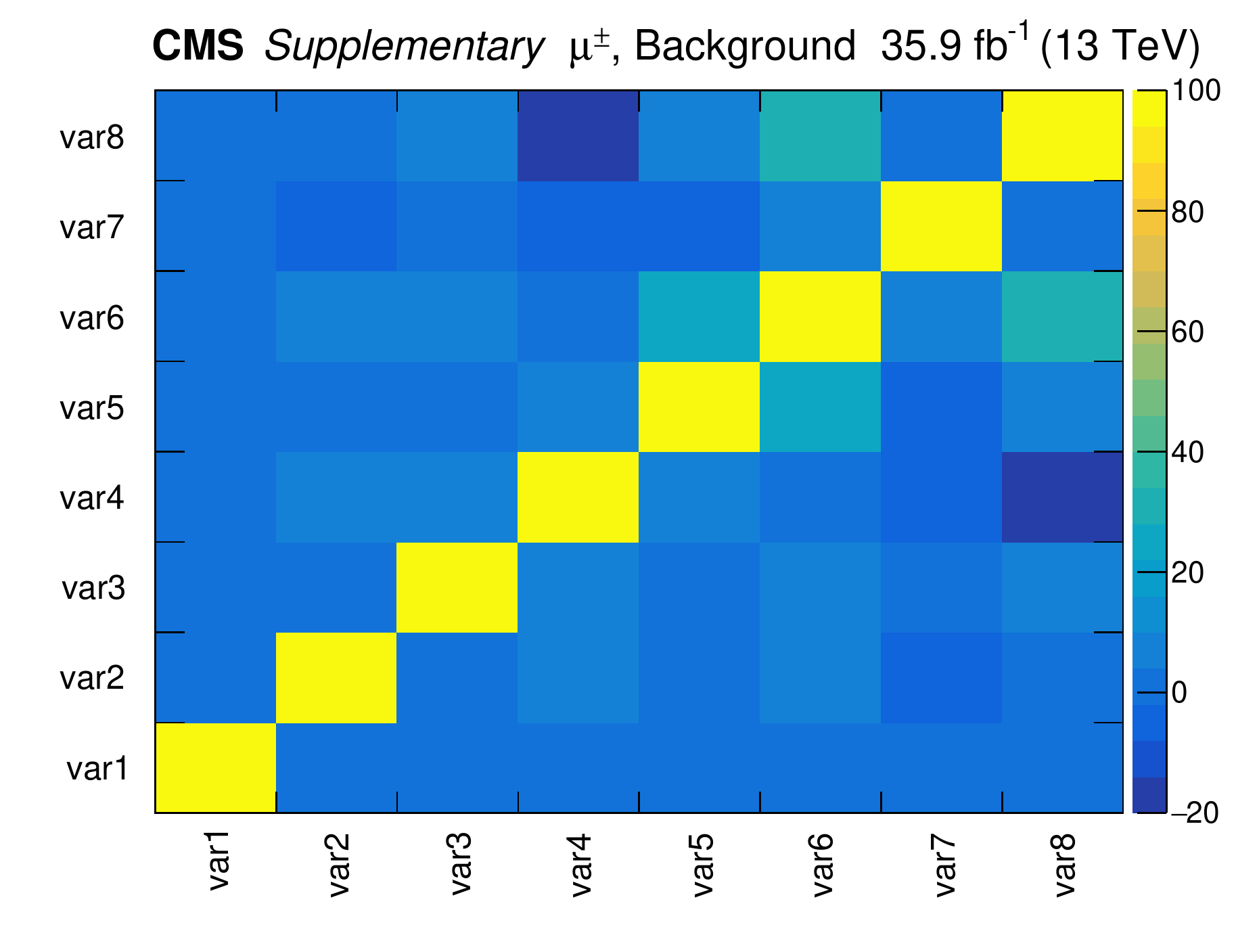}
\includegraphics[width=0.45\textwidth]{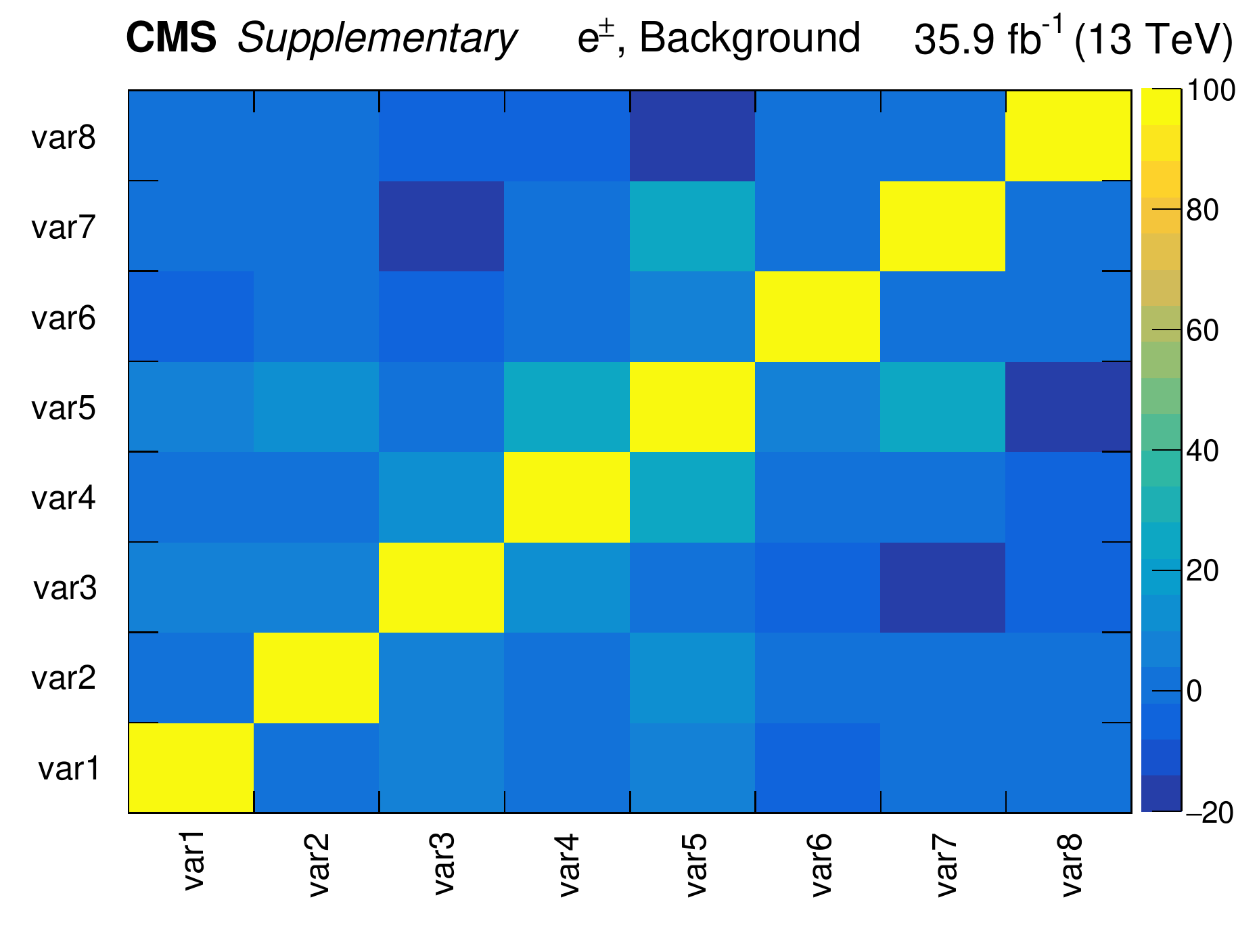}\\
\caption{\label{fig:inputVarCorrMuon} Correlations in \% among the BDT input variables used for the muon (left) and electron (right) final states in the signal and background events of the 2J1T category before (upper-two rows) and after (lower-two rows) decorrelation.}
\end{figure*}

Figures~\ref{fig:inputVarsuppl1} and \ref{fig:inputVarsuppl2} show the distributions of the other six BDT input variables in data and simulation for the muon and electron final states.
Figure~\ref{fig:inputVarCorrMuon} presents the correlations among all BDT input variables for these final states before and after applying the decorrelation method available in \textsc{tmva}.
As expected, the correlations are significantly reduced after decorrelation.

\clearpage

\section{\label{sec:suppl_fit} Likelihood scan and correlations among fit parameters}

Figure~\ref{fig:likelihood} (left) shows the scan of the profile likelihood ratio as a function of the POI for the ML fit model used to determine the \mtop value in the charge-inclusive muon and electron final states.
The scan results are presented both for data and simulation.
Figure~\ref{fig:likelihood} (right) presents correlations among the POI and three nuisance parameters corresponding to the ML fit applied to data in the same final states.

\renewcommand{\thefigure}{B.\arabic{figure}}
\begin{figure*}[!htb]
\centering
\includegraphics[width=0.45\textwidth]{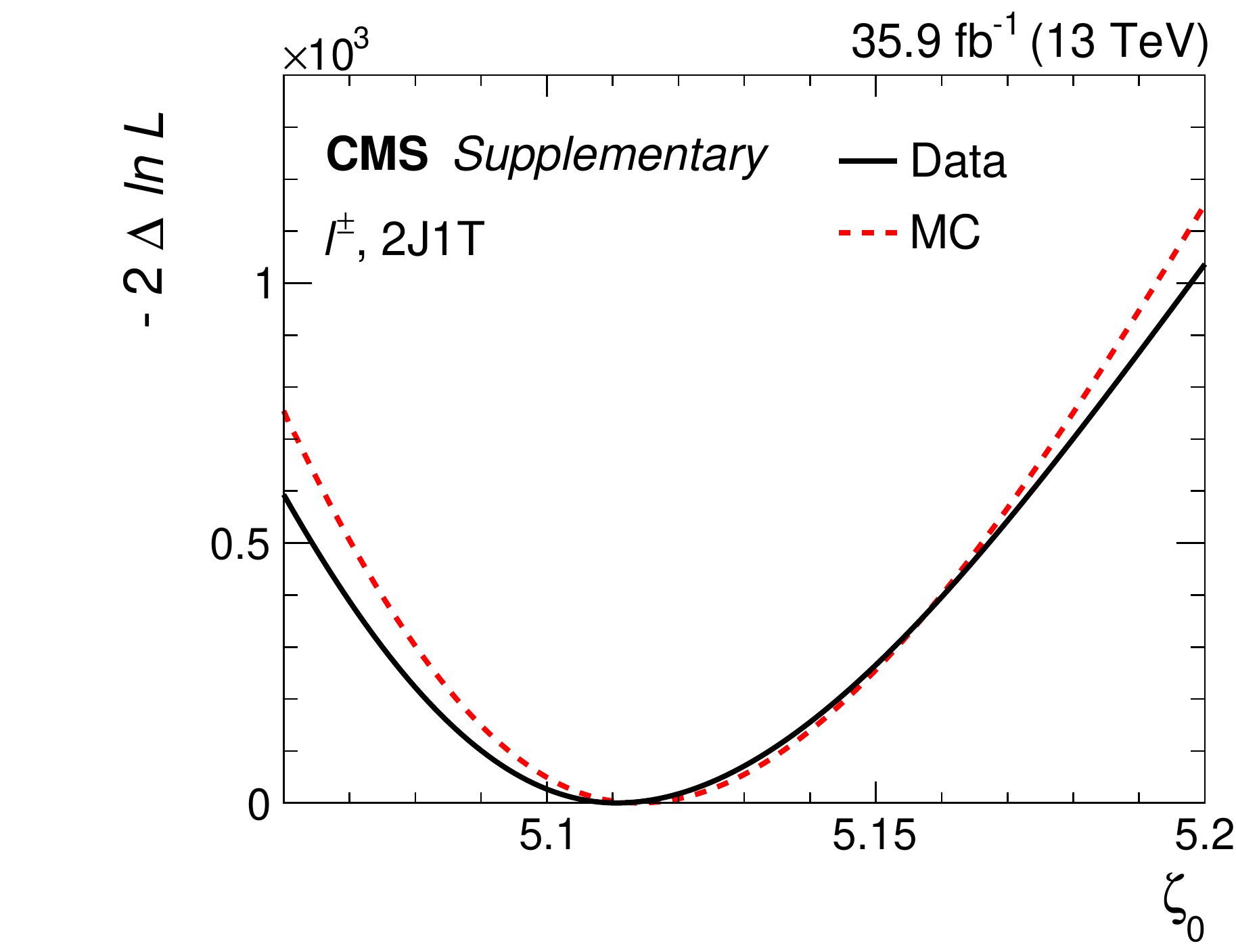}
\includegraphics[width=0.45\textwidth]{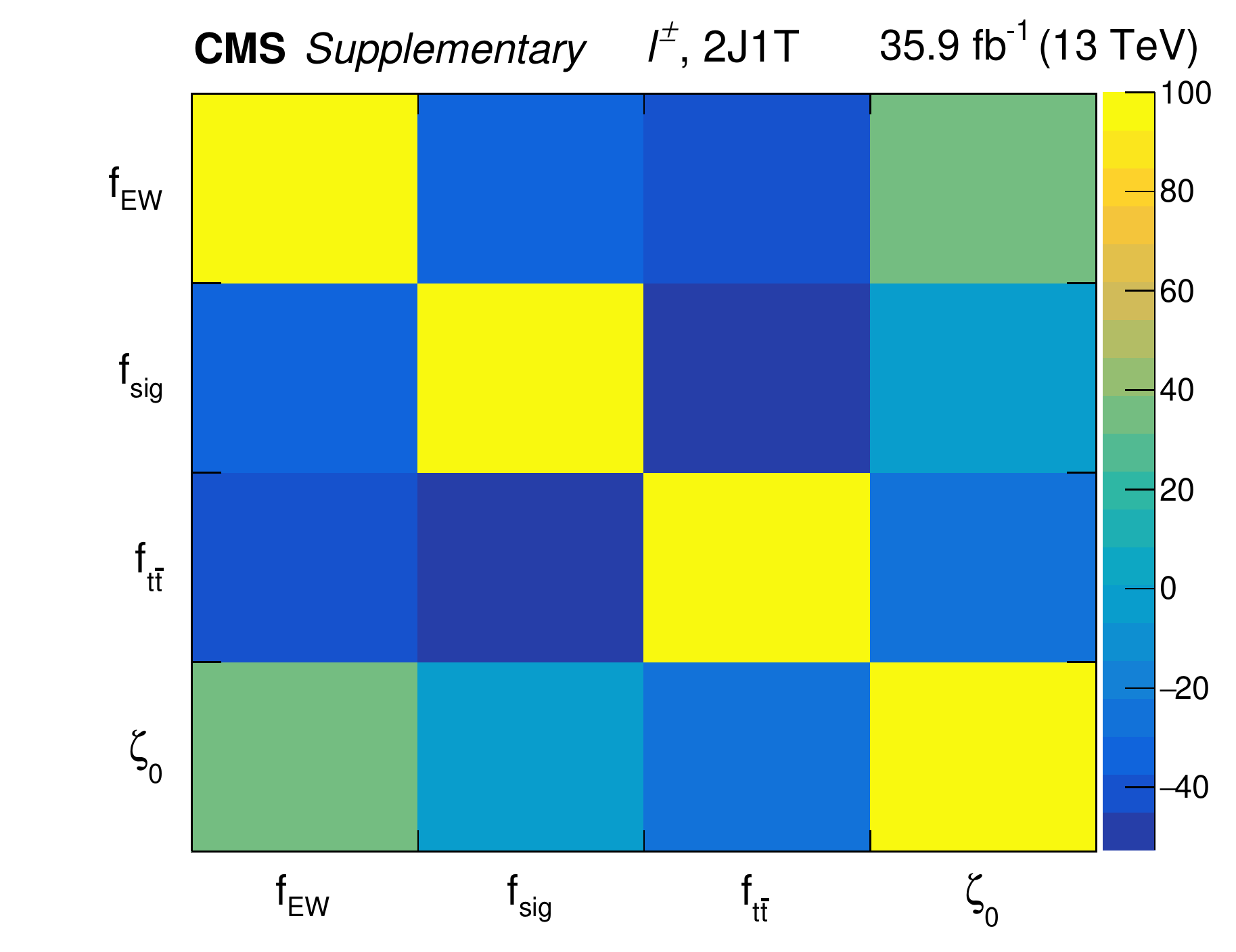}\\
\caption{\label{fig:likelihood} Scan of the profile likelihood ratio as a function of the POI for the parametric fit model used in the $l^{\pm}$ final state of the 2J1T category in data and simulated events (left).
Correlations in \% among the POI and nuisance parameters corresponding to the fit to data for the $l^{\pm}$ final state of the 2J1T category (right).}
\end{figure*}
\clearpage
\cleardoublepage \section{The CMS Collaboration \label{app:collab}}\begin{sloppypar}\hyphenpenalty=5000\widowpenalty=500\clubpenalty=5000\input{TOP-19-009-authorlist.tex}\end{sloppypar}
\end{document}

%% file: TOP-19-009-authorlist.tex
\vskip\cmsinstskip
\textbf{Yerevan Physics Institute, Yerevan, Armenia}\\*[0pt]
A.~Tumasyan
\vskip\cmsinstskip
\textbf{Institut f\"{u}r Hochenergiephysik, Wien, Austria}\\*[0pt]
W.~Adam, J.W.~Andrejkovic, T.~Bergauer, S.~Chatterjee, M.~Dragicevic, A.~Escalante~Del~Valle, R.~Fr\"{u}hwirth\cmsAuthorMark{1}, M.~Jeitler\cmsAuthorMark{1}, N.~Krammer, L.~Lechner, D.~Liko, I.~Mikulec, P.~Paulitsch, F.M.~Pitters, J.~Schieck\cmsAuthorMark{1}, R.~Sch\"{o}fbeck, M.~Spanring, S.~Templ, W.~Waltenberger, C.-E.~Wulz\cmsAuthorMark{1}
% ...
\vskip\cmsinstskip
% ...
1:  Also at TU Wien, Wien, Austria\\
% ...
